%% file: paper_smj.tex
\documentclass[publish]{smj}

\usepackage{tikz}
\usepackage{bm} 
\usepackage{dsfont}
\usepackage{comment}
\usepackage{theorem}
\usepackage{algpseudocode}
\usepackage{algorithmicx} % algorithm and algorithmic environment
\usepackage{algorithm}    % \listofalgorithms command 
\usepackage{booktabs}
\usepackage{multirow}
\usepackage{enumitem}
\usepackage{subcaption}
\newtheorem{theorem}{Theorem}[section]
\newtheorem{lemma}[theorem]{Lemma}

\usepackage[T1]{fontenc}% http://ctan.org/pkg/fontenc
\usepackage[outline]{contour}% http://ctan.org/pkg/contour

\usepackage{rotating}

\input{macros}

\newcommand{\epochs}{101}
\newcommand{\bestmodel}{RW_normal}

\allowdisplaybreaks

\Author{Jan V\'avra\Affil{1}\ORCID{0000-0003-0068-5356},\\
        Bettina Gr\"un\Affil{2}\ORCID{0000-0001-7265-4773}, 
        \and\\ Paul Hofmarcher\Affil{3}\ORCID{0000-0002-9206-6386}
}
\AuthorRunning{Jan V\'avra \textrm{et al.}}

\Affiliations{
    \item   %Department of Probability and Mathematical Statistics, 
            Faculty of Mathematics and Physics, 
            Charles University,
            Prague,
            Czech Republic
    \item   % Institute for Statistics and Mathematics,
            WU Vienna University of Economics and Business,
            Wien,
            Austria
    \item   Paris Lodron University of Salzburg,
            Salzburg,
            Austria

}   %% end \Affiliations

\CorrAddress{Jan V\'avra, 
             Department of Probability and Mathematical Statistics,
             Charles University Prague,
             Sokolovsk\'a 83, 
             186 75, Praha 8, 
             Czech Republic}
\CorrEmail{vavraj@karlin.mff.cuni.cz}
\CorrPhone{(+420)\;731\;066\;832}

\Title{Evolving voices based on temporal Poisson factorisation
}
\TitleRunning{Temporal Poisson factorisation}

\Abstract{
The world is evolving and so is the vocabulary used to discuss topics in speech. Analysing political speech data from more than 30 years requires the use of flexible topic models to uncover the latent topics and their change in prevalence over time as well as the change in the vocabulary of the topics. We propose the temporal Poisson factorisation (TPF) model as an extension to the Poisson factorisation model to model sparse count data matrices obtained based on the bag-of-words assumption from text documents with time stamps. We discuss and empirically compare different model specifications for the time-varying latent variables consisting either of a flexible auto-regressive structure of order one or a random walk. Estimation is based on variational inference where we consider a combination of coordinate ascent updates with automatic differentiation using batching of documents. Suitable variational families are proposed to ease inference. We compare results obtained using independent univariate variational distributions for the time-varying latent variables to those obtained with a multivariate variant. We discuss in detail the results of the TPF model when analysing speeches from 18 sessions in the U.S.\ Senate (1981--2016).
}

\Keywords{auto-regressive process; Poisson factorisation; time-varying; topic model; variational inference
}

\begin{document}

%%% Title page
%%% -------------------------------------------------------------------------------
%%% Use command\maketitle to produce the title page.
\maketitle

\renewcommand{\arraystretch}{0.4}

\input{section01_final}
\input{section02_final}

\input{section03_final}
\input{section04_final}
\input{section05_final}

%%% Supplementary materials (if any)
%%% ------------------------------------------
%\section*{Supplementary materials}
%The supplementary materials include [the list of supplementary materials and their brief description]. They are available from the journals page at \url{https://journals.sagepub.com/home/smj}. [At the submission stage, all supplementary materials (or the links to access them) should be submitted jointly with the main paper.]

%%% Acknowledgements (if any)
%%% ------------------------------------------
%\section*{Acknowledgements}
%We want to thank\ldots 

%%% Declaration of conflicting interests (should always be included)
%%% -----------------------------------------------------------------
\section*{Declaration of Conflicting Interests}
The authors declared no potential conflicts of interest with respect to the research, authorship and/or publication of this article.
   %%% Alternatively, please, disclose here potential conflicting interests. 

%%% Funding (if any)
%%% ------------------------------------------
\section*{Funding}
This research was supported by the Jubil\"aumsfonds of the Oesterreichische Nationalbank (OeNB, grant no.~18718).

%%% Appendix (if any)
%%% ------------------------------------------
\input{appendix_final}

\clearpage

%%% Supplementary materials
%%% ------------------------------------------
\section*{Supplementary Materials}
%In the supplement, we provide additional technical details about the variational approach we took. In particular, we properly introduce the latent auxiliary counts, show how to obtain the variational means for quadratic forms under an auto-regressive prior, and write down an explicit formula for the ELBO. Moreover, we present details of a simulation study we preformed to assess and compare our proposed modelling approach and we provide additional empirical results of the analysis of the U.S.\ Senate speech data. 
% Within Appendix~\ref{appsec:CAVI} of this paper, we keep only the details on how the CAVI updates are performed within Algorithm~\ref{alg:TVPF_CAVI_and_SVI}.

\renewcommand\thesection{S.\arabic{section}}
\renewcommand\thesubsection{S.\arabic{section}.\arabic{subsection}}
\setcounter{section}{0}

This section provides more details complementing the derivations and results in the main paper ``Evolving voices based on temporal Poisson factorisation''. Each section within this section starts with the letter ``S'' for Supplement to distinguish the section numbering within this document and the original paper. 

Section~\ref{supsec:tpf} provides an overview on the notation used within our main paper and briefly summarises the hierarchy of the proposed temporal Poisson factorisation (TPF) model. Section~\ref{supsec:auxiliary_variables} introduces the latent auxiliary counts that decompose the observed counts into counts for each topic. These variables allow for straightforward updates within our algorithm. Some of the updates require computation of variational means of quadratic forms which are not trivial under auto-regressive prior (see Section~\ref{supsec:Eq_quadratic_form}). The $\mathsf{ELBO}$ function consists of contributions from the model based on the auxiliary counts, the prior distributions and the entropies of the variational families. Therefore, Section~\ref{supsec:ELBO} writes down its full extent while rearranging the terms so that is clearer to see where the CAVI updates (listed in the Appendix of the main paper) come from. 
Section~\ref{supsec:simulation} outlines the design and provides the results of a simulation study where the main focus is the identification of the true value of the auto-regressive coefficient $\delta$ under 7 different models, including the competing Dynamic Poisson Factorisation (DPF) model that requires a certain data structure. Models are compared in terms of variational information criteria and computational  demands. Section~\ref{supsec:processing_senate_data} provides more information on the pre-processing of the U.S.\ Senate speeches. The final Section~\ref{supsec:additional_outputs} provides additional plots and a table complementing the empirical results included in the main paper. In particular, we inspect the same plots included in the main paper 
for all four models to demonstrate that in fact similar conclusions are drawn with only marginal differences across the four choices.  

\clearpage

\input{supplement01_TPF_summary}
\input{supplement02_auxiliary_counts}
\input{supplement03_Eq_quadratic}

\input{supplement04_ELBO}

%% \input{supplement05_CAVI}
\input{supplement06_simulation}
\input{supplement07_senate_preprocessing}

\input{supplement08_senate_outputs}

%%% References if bibTeX is used
%%%
%%% Please, do not specify any \bibliographystyle{} command!
%%%
%%% It is already specified in the smj.cls and its
%%% second specification here causes error.
%%% ------------------------------------------------------------
%\clearpage
\bibliography{literature}

%%% References (if created by hand).
%%% -----------------------------------------------------------------------------------

\end{document}

%% file: macros.tex
\definecolor{red2}{rgb}{0.9333333, 0, 0}
\definecolor{mediumblue}{rgb}{0, 0, 0.8039216}
\definecolor{magenta4}{rgb}{0.545098, 0, 0.545098}
\definecolor{orcidlogocol}{rgb}{0.651, 0.8078, 0.2235}
\newcommand{\R}{\mathbb{R}}

\DeclareMathOperator{\E}{\mathsf{E}}

\DeclareMathOperator{\var}{\mathsf{var}}
\DeclareMathOperator{\cov}{\mathsf{cov}}

\DeclareMathOperator{\diag}{\mathsf{diag}}
\DeclareMathOperator{\diagzero}{\mathsf{diag}_0}
\DeclareMathOperator{\superdiag}{\overline{\mathsf{diag}}}
\DeclareMathOperator{\subdiag}{\underline{\mathsf{diag}}}

\DeclareMathOperator{\Tr}{\mathsf{Tr}}

\newcommand{\T}[1]{#1^\top}        

\newcommand{\indicator}{\mathds{1}} %% tied with package dsfont 
\newcommand{\ones}{\bm 1}
\newcommand{\norm}[2]{\mathsf{N}\left(#1, \,#2 \right)} %normal 1=E,2=var
\newcommand{\knorm}[3]{\mathsf{N}_{#1}\left(#2, \,#3 \right)} %MV normal 1=dim k, 2=E, 3=var
\newcommand{\gammadist}[2]{\Gamma\left(#1, #2 \right)} %Gamma distribution 1=shape alpha, 2=rate beta
\newcommand{\Mult}[3]{\mathsf{Mult}_{#1}\left(#2, #3 \right)} %multinomial 1=K (ngroups), 2=n(observations), 3=probs
\newcommand{\Pois}[1]{\mathsf{Pois}\left(#1 \right)} %Poisson 1=lambda
\newcommand{\AR}[1]{\mathsf{AR}\left(#1\right)} % Autoregressive sequence of order #1
\newcommand{\sAR}[4]{\mathsf{AR}_{#1}\left(#2, #3, #4\right)} % shifted Autoregressive sequence of order 1 of length #1, mean #2, ARcoef #3, ARprec #4
\newcommand{\ELBO}[1]{\mathsf{ELBO}\left(#1\right)} %ELBO function of parameter 1
\newcommand{\Eq}{\E_{q_{\bm\phi}}}

\newcommand{\shp}{\mathrm{shp}}
\newcommand{\rte}{\mathrm{rte}}
\newcommand{\loc}{\mathrm{loc}}
\newcommand{\scl}{\mathrm{scl}}
\newcommand{\sclsq}{\mathrm{var}}
\newcommand{\covm}{\mathrm{cov}}
\DeclareMathAlphabet{\mathbbb}{U}{bbold}{m}{n}

%% file: section01_final.tex
%%%!!!!!!!!!!!!!!!!!!!!!!!!!!!!!!!!!!!!!!!!!!!!!!!!!!!!!!!!!!!!!!!!!!!!!!!%%%
%%%%%%%%%%%%%%%%%%%%%%%%%%%%%%%%%%%%%%%%%%%%%%%%%%%%%%%%%%%%%%%%%%%%%%%%%%%%%
%%%%%%%%%%%%               NEW SECTION                      %%%%%%%%%%%%%%%%%
%%%%%%%%%%%%%%%%%%%%%%%%%%%%%%%%%%%%%%%%%%%%%%%%%%%%%%%%%%%%%%%%%%%%%%%%%%%%%
%%%!!!!!!!!!!!!!!!!!!!!!!!!!!!!!!!!!!!!!!!!!!!!!!!!!!!!!!!!!!!!!!!!!!!!!!!%%%
\section{Introduction} \label{sec:introduction}

Probabilistic topic models represent the state-of-the-art method for analysing a corpus of text documents based on the bag-of-words assumption to infer the content based on latent topics \citep{blei:2011}. The vanilla topic model provides a data generating process for the document-term matrix (DTM) which contains the frequency counts of the terms specified by a vocabulary for all documents in the corpus. The co-occurrence structure observed for terms within documents is explained by latent topics which differ in their term intensities and the document intensities which reflect how prevalent topics are for each document. 

Two main approaches are considered in topic models: latent Dirichlet allocation \citep[LDA;][]{blei_etal_2003} and Poisson factorisation \citep[PF;][]{Gopalan:2014}. LDA assumes that the frequency counts for a given document are drawn from a multinomial distribution conditional on the length of the document. The success probabilities of the multinomial distribution correspond to a convex combination of term intensities weighted by the document intensities. Alternatively, PF directly models the counts based on independent Poisson distributions. The rate parameter for a document-term combination is obtained by the convex combination of the document intensities weighted by the term intensities.

If the documents in the corpus have time stamps as meta-information and the documents are collected for an extensive period of time, it is natural to include time in the modelling of the text documents and infer how the prevalence of topics as well as the content of topics evolve over time. Previous approaches including time information in topic models for text data focused on the LDA approach  \citep{Blei+Lafferty:2006, wang2015continuous, glynn_etal:2015_bayesian_DLTM}. Extensions of the PF model to include a time-varying component were only proposed in the context of a different application where the number of interactions for user-item combinations were modelled with user-item matrices being available for several time points  \citep{charlin2015DPF, hosseini_etal:2017_RPF_for_temporal_recommendation}.

In this paper, we propose the temporal Poisson factorisation (TPF) model which extends PF to modelling a DTM where the documents have associated time stamps. The TPF model thus represents a dynamic topic model for text data 
%not relying on LDA 
where both the topical content as well as topical prevalence may vary over time.
Section~\ref{sec:tvpf} provides the model specification together with suitable choices for the prior distributions. 
To model the time dependence, we consider an auto-regressive specification where the coefficient induces a stationary behaviour of the $\AR{1}$ process in addition to the random walk parameterisation usually considered in the context of dynamic topic models. 
Section~\ref{sec:inference} describes the use of variational inference (VI) for estimation (with details given in Appendix~\ref{appsec:CAVI}). We challenge the use of independent univariate variational families only, as usually pursued in mean-field VI \citep{charlin2015DPF}, and consider also a multivariate specification. In contrast to \citet{Blei+Lafferty:2006} and \citet{wang2015continuous} who account for the dependence on the previous time period in the variational distribution, we use a multivariate specification covering all time points. In Section~\ref{sec:inference} we also present different model selection criteria proposed for VI \citep{McGrory_Titterington:2007_VIC,You+Ormerod+Muller:2014_VIC} and suitable post-processing tools to summarise the estimated models and infer insights.
Further details, including the results of a simulation study with synthetic data, are provided in the Supplementary Material (sections labelled with ``S'').
We apply the TPF model to 36 years of political discourse in the U.S.\ Senate in Section~\ref{sec:senate_speech_data}. We empirically investigate the performance of the auto-regressive model specification compared to using a random walk and assess the difference in the estimation when using independent univariate variational families compared to a multivariate specification. The suitability of the model to uncover topical variation over time as 
well as the change in prevalence is assessed. 
We conclude in Section~\ref{sec:conclusion}.  

%% file: section02_final.tex
%%%!!!!!!!!!!!!!!!!!!!!!!!!!!!!!!!!!!!!!!!!!!!!!!!!!!!!!!!!!!!!!!!!!!!!!!!%%%
%%%%%%%%%%%%%%%%%%%%%%%%%%%%%%%%%%%%%%%%%%%%%%%%%%%%%%%%%%%%%%%%%%%%%%%%%%%%%
%%%%%%%%%%%%               NEW SECTION                      %%%%%%%%%%%%%%%%%
%%%%%%%%%%%%%%%%%%%%%%%%%%%%%%%%%%%%%%%%%%%%%%%%%%%%%%%%%%%%%%%%%%%%%%%%%%%%%
%%%!!!!!!!!!!!!!!!!!!!!!!!!!!!!!!!!!!!!!!!!!!!!!!!!!!!!!!!!!!!!!!!!!!!!!!!%%%
\section{Temporal Poisson factorisation (TPF)} \label{sec:tvpf}
In this section we define our TPF model. 
The model is in principle applicable to any time-varying modelling of count matrices where rows are associated with different time periods. For ease of understanding, we present the notation having the empirical application in mind where we apply the TPF model to a corpus of text documents collected over time transformed to a DTM using the bag-of-words assumption.

The fundamental unit of our model is the \emph{document}, indexed by $d = 1, \ldots, D$, where $D$ is the total number of documents. Under the bag-of-words assumption, each document is represented by a sparse vector $\bm Y_d$ of counts (non-negative integers) of the \emph{terms}. The set of all considered terms, indexed by $v = 1, \ldots, V$, is referred to as the \emph{vocabulary}. Terms can consist of unigrams, bigrams, or other $n$-grams. 
Stacking the individual document vectors $\bm Y_d$ row-wise results in the DTM $\mathbb{Y} = \left(\bm Y_d\right)_{d=1}^D = \left(y_{dv}\right)_{d, v = 1}^{D, V}$.

We assume that the documents are collected over an extended period of time. Consequently, we split time into $T$ \emph{time periods} and assume in the model that temporal variations occur only between distinct time periods $t = 1, \ldots, T$, but parameters remain constant within each individual time period. A suitable choice of splitting the time into these time periods is crucial to justify this assumption. To fix notation, let $t_d$ denote the time period in which document~$d$ is written and $\mathcal{D}_t$ denote the set of all documents written in time period~$t$.

%%%!!!!!!!!!!!!!!!!!!!!!!!!!!!!!!!!!!!!!!!!!!!!!!!!!!!!!!!!!!!!!!!!!!!!!!!%%%
%%%%%%%%%%%%               NEW SUBSECTION                   %%%%%%%%%%%%%%%%%
%%%!!!!!!!!!!!!!!!!!!!!!!!!!!!!!!!!!!!!!!!!!!!!!!!!!!!!!!!!!!!!!!!!!!!!!!!%%%
\subsection{Model specification} \label{subsec:tvpf_def}
The TPF model is defined as a generalisation of the classical Poisson factorisation model with $K$ underlying latent topics:
\begin{equation} \label{eq:tvpf_rates}
y_{dv} \sim \Pois{\lambda_{dv}} 
\qquad \text{with rate} \qquad 
\lambda_{dv} 
= \sum \limits_{k=1}^K \lambda_{dkv}
= \sum \limits_{k=1}^K \theta_{dk} \exp 
\left\{ 
    h_{kv,t_d}
\right\}.
\end{equation}

Each element $y_{dv}$ of the DTM $\mathbb{Y}$ 
is assumed to be drawn from a Poisson distribution, independent from the other counts, with a rate $\lambda_{dv}$ that sums up the contributions from different topics $ k = 1, \ldots, K $. The topic-specific Poisson rates $\lambda_{dkv}$  combine the \emph{document intensities} $\theta_{dk}$ for topic $k$ and the \emph{term intensities} $h_{kv,t_d}$ of topic $k$ at time point $t_d$. 
%According to \citep{ShikunOleaNesbit2024RobustMLAlgforTextAnalysis} topic models are prone to identifiability issues. Here we only anchor the permutation of topic labels, but a robust .
%The problem of un-identifiability of topic models \citep{ShikunOleaNesbit2024RobustMLAlgforTextAnalysis} is overcomed by prior distributions.

The \emph{document intensities} $ \theta_{dk}, \theta_{dk} > 0 $, capture the contribution of each topic $k$ to a given document $d$. Since the document label $d$ also includes the information about the time period when it was written ($t_d$), there is no need to explicitly specify $\theta_{dk}$ as time-varying. Changes in intensity of a topic over time can be derived by aggregating over the document intensities of documents written in the specific time periods. 
%Potentially, we could capture some temporal evolution through a more elaborate prior structure than proposed in \eqref{eq:tvpf_prior_theta}. \textcolor{red}{ but in section~\ref{subsec:topic_prevalence} ... we present an approach to account for the time variation in the topical prevalence. simply averaging at each time t over the associated theta}.

The \emph{term intensities} $h_{kv,t}$  
capture how the importance of a term for a topic varies over time 
on the logarithmic scale. To link the term intensities across adjacent time periods, we make use of a suitable prior setup mirroring auto-regressive processes. 
We use the raw, non-centred term intensities $h_{kv,t}$ without explicitly separating out the overall mean. Thus, adjustments related to the average rate are incorporated into the prior structure of $h_{kv,t}$. 
This differs from the specification of the dynamic Poisson factorisation (DPF) model \citep[][see Section~\ref{subsec:tvpf_vs_dpf} for a detailed comparison]{charlin2015DPF}.
%This differs from the specification in DPF \citep{charlin2015DPF} where the model separates out the average rate of document-topic associations across all time periods and directly includes this overall mean in the formula for the Poisson rates $\lambda_{dkv}$. 

%%%!!!!!!!!!!!!!!!!!!!!!!!!!!!!!!!!!!!!!!!!!!!!!!!!!!!!!!!!!!!!!!!!!!!!!!!%%%
%%%%%%%%%%%%               NEW SUBSECTION                   %%%%%%%%%%%%%%%%%
%%%!!!!!!!!!!!!!!!!!!!!!!!!!!!!!!!!!!!!!!!!!!!!!!!!!!!!!!!!!!!!!!!!!!!!!!!%%%
\subsection{Prior specification} \label{subsec:tvpf_prior}
Priors need to be selected for the time-varying term intensities $h_{kv,t}$ and the document intensities $\theta_{kd}$. 
First, we discuss the prior choices for the time-varying term intensities $h_{kv,t}$.
We use a~shifted auto-regressive prior for $h_{kv,t}$ parameterised as follows:
\begin{equation}\label{eq:ar_prior}
    \left. h_{t}- \mu \;\right|\; h_{t-1} \sim \norm{\delta
    \left(h_{t-1} - \mu\right)}{\tau^{-1}}
    \quad \text{and} \quad 
    h_{1} - \mu \sim \norm{0}{\tau^{-1}},
\end{equation}
where we drop for clarity all indices except time. The sequence is centred around the mean $\mu$. The centred sequence then follows an $\AR{1}$ process where $\delta$ is the auto-regressive coefficient. The precision $\tau$ controls the variability of the noise of the centred sequence including the first deviation from an artificial observation $h_0 - \mu = 0$. 

We write prior~\eqref{eq:ar_prior} in vectorised form using:
\begin{equation} \label{eq:tvpf_prior_ar_kv}
    \bm h \sim \sAR{T}{\mu}{\delta}{\tau} 
    \Longleftrightarrow
    \bm h \sim \knorm{T}{\mu \ones}{
    \tau^{-1}
    \Delta_T(\delta)^{-1}
    },
\end{equation}
where the precision matrix $\Delta_T(\delta)$ is a tridiagonal matrix
%decomposed as:
%$$
%\Delta_T\left(\delta\right) 
%=
%\mathbb{I}_T
%+ \delta^2 \mathbb{I}_{T,0}
%- \delta \left(\mathbb{I}_{T,-1} + \mathbb{I}_{T,1}\right)
%$$
%where $\mathbb{I}_T$ is identity matrix, $\mathbb{I}_{T,0} = \diagzero(1, \ldots, 1) = \diag(1, \ldots, 1, 0)$ is identity matrix where the last element is zero, $\mathbb{I}_{T,-1} = \subdiag(1, \ldots, 1)$ is the subdiagonal and $\mathbb{I}_{T,1} = \superdiag(1, \ldots, 1)$ is the superdiagonal.
with $\left(1+\delta^2, \ldots, 1+\delta^2, 1\right)$ on the main diagonal and $-\delta$ on subdiagonal and superdiagonal. 
%$$
%\Delta_T(\delta)
%=
%\begin{pmatrix}
%\textcolor{red}{1+\delta^2} & - \delta & 0 & \ldots & 0 & 0\\
%-\delta & 1+\delta^2 & - \delta & \ldots & 0 & 0\\
%0 & -\delta & 1+\delta^2 & \ldots & 0 & 0\\
%\vdots & \vdots & \vdots & \vdots & \vdots & \vdots \\
%0 & 0 & 0 & \ldots & 1+\delta^2 & -\delta \\
%0 & 0 & 0 & \ldots & -\delta & 1 
%\end{pmatrix}
%$$
In the complete notation including all the indices, this implies 
for each combination of topic~$k$ and term~$v$ independently that
$\bm h_{kv} \sim \sAR{T}{\mu_{kv}}{\delta_{kv}}{\tau_{kv}}$.

This specification allows for a flexible auto-regressive coefficient $\delta$ and thus differs from the traditional, more rigid random-walk parameterisation ($\delta = 1$), as for example used by \citet{Blei+Lafferty:2006}  or \citet{charlin2015DPF}. 
We assign the prior  $\delta_{kv} \sim \norm{\mu^{\delta}}{\left(\sigma^{\delta}\right)^2}$ with fixed hyperparameters, enabling direct updates during VI. Alternatively, a truncated version with support restricted to $[-1, 1]$ could be used to enforce stationarity while retaining the simplicity of the update steps.
    % We assign the prior $\delta_{kv} \sim \norm{\mu^{\delta}}{\left(\sigma^{\delta}\right)^2}$ with fixed hyperparameters as it allows for direct updates when performing VI. Alternatively, one could also consider using a truncated version with support limited to $[-1, 1]$ to ensure the stationarity of the process while preserving the straightforward updates. 
    % However, as \citet{Stan_AR:2024} recommends, we do not constrain this parameter. Especially when we set $\delta_{kv} \sim \norm{0.5}{0.2^2}$ in applications, as it assigns a lot of mass to the interval $(0,1)$ anyway.
For values of $\delta$ close to zero, the sequence of term intensities varies around its mean $\mu$ without any time dependence. For values of $\delta$ close to $1$, the intensity of the current term depends heavily on the value from the previous time period.  

The means $\mu_{kv}$ of the auto-regressive sequences are also given a normal prior to take advantage of conjugacy: $\mu_{kv} \sim \norm{\mu^{\mu}}{\left(\sigma^{\mu}\right)^2}$. The use of a large variance together with a mean of zero, e.g., $\norm{0}{100^2}$, induces a flat prior and reduces the shrinkage of $\mu$ towards zero to yield a flexible fit of $\mu$ to the sequence $\bm h$. 
Finally, the precisions $\tau$ are given a~gamma prior to preserve conjugacy: $\tau_{kv} \sim \gammadist{a^{\tau}}{b^{\tau}}$, in applications we prefer $\gammadist{0.3}{0.3}$. 

For the document intensities, we follow \citet{gopalanhofmanblei2015scalableHPoisF} and assign a hierarchical gamma prior with additional gamma layer for the document-specific rates to account for differences in document length: $\theta_{dk} \sim \gammadist{a^\theta}{\xi_d}$ and $\xi_d \sim \gammadist{a^{\xi}}{b^{\xi}}$.
%\begin{equation*} %\label{eq:tvpf_prior_theta}
%    \theta_{dk} \sim \gammadist{a^\theta}{\xi_d}
%    \qquad \text{and} \qquad
%    \xi_d \sim \gammadist{a^{\xi}}{b^{\xi}}.
%\end{equation*}
This hierarchical gamma structure provides us with closed-form updates contrary to the non-conjugate alternative of a log-normal distribution used, e.g., by~\citet{charlin2015DPF}. 
%In order to capture the time evolution of $\bm \theta$ intensities we would have to either change the rate from document specific to time period specific (possibly also topic specific) or add one more additional gamma layer for the auxiliary rates to aggregate documents over time periods. 
%In the paper at hand, we follow a different strategy and make
%\textcolor{red}{.. and make what?}
%\textcolor{red}{However, we do not find it worth of implementation as we later provide different tools for capturing the evolution in time.} 
As $\gammadist{0.3}{0.3}$ is widely used in the literature as a prior for $\theta_{dk}$, we use $a^\theta = 0.3$ and $a^{\xi} = 0.3$ with $b^{\xi} = 1$ to fix $\E \xi_d = 0.3$ a~priori. 
%The whole structure of the TPF model is summarised and depicted in plate notation in Figure~\ref{fig:hierarchy_TVPF}.

Figure~\ref{fig:hierarchy_TVPF} depicts the whole structure of the TPF model in plate notation.
\input{diagram_tvpf}
%\FloatBarrier

%%%!!!!!!!!!!!!!!!!!!!!!!!!!!!!!!!!!!!!!!!!!!!!!!!!!!!!!!!!!!!!!!!!!!!!!!!%%%
%%%%%%%%%%%%               NEW SUBSECTION                   %%%%%%%%%%%%%%%%%
%%%!!!!!!!!!!!!!!!!!!!!!!!!!!!!!!!!!!!!!!!!!!!!!!!!!!!!!!!!!!!!!!!!!!!!!!!%%%
\subsection{Comparing TPF to DPF} \label{subsec:tvpf_vs_dpf}

The most comparable approach to our model is dynamic Poisson factorisation (DPF) by \citet{charlin2015DPF}, which also extends PF by incorporating a time-varying component. However, its data structure requirements differ significantly from ours. Specifically, DPF assumes document-term matrices (DTMs) with fixed dimensions across all time periods:
% The closest competitor to our modelling approach is the DPF by \citet{charlin2015DPF} which also extends PF to include a time-varying component. However, the data structure required crucially differs from our  approach. In particular, the DPF specification requires DTMs with the same dimensions for each time period:
\begin{equation*} %\label{eq:dpf_rates}
y_{av,t} \sim \Pois{\lambda_{av,t}}
\quad \text{where} \quad 
\lambda_{av,t} 
= \sum \limits_{k=1}^K \exp 
\left\{
    g_{ak,t} + \overline{g}_{ak}
    +
    h_{kv,t} + \overline{h}_{kv}
\right\}.
\end{equation*}
In this case, the first index does not indicate the document~$d$ but rather the author~$a \in \{1, \ldots, A\}$, in contrast to~\eqref{eq:tvpf_rates} for TPF. Eventually, the index does not have to be an author, but it needs to refer to a~quantity that appears once per time period and repeatedly across all time periods. Hence, a~single document is not a~suitable index. 

This specification allows the intensity $\theta$ to also be directly time-varying by using $\exp\{g_{ak,t} + \overline{g}_{ak}\}$ where $\overline{g}_{ak}$ captures the author-specific mean topic intensity and $g_{ak,t}$ is a centred random walk ($\delta = 1$) sequence. Different from our specification of the mean through the prior, see~\eqref{eq:tvpf_rates} and \eqref{eq:tvpf_prior_ar_kv}, the mean values $\overline{g}_{ak}$ and $\overline{h}_{kv} \approx \mu_{kv}$ are separated from the random walk sequence and directly included within the formula for the Poisson rates. This induces identifiability issues when both $g_{ak,t}$ and $\overline{g}_{ak}$ are estimated with their own unrestricted location parameter.

In Section~\ref{sec:senate_speech_data}, we only apply TPF to the U.S.\ Senate speeches, because their data structure does not comply with the one required by DPF. Data with a suitable structure could eventually be obtained by aggregating documents from the same author within the same session into one very long document. However, then one could no longer determine the topic composition of individual speeches. Such an aggregation would reduce the dimension of the DTM to $AT \ll D$, drastically reducing sparsity. Moreover, the presence of an author within a specific time period depends on that person having a seat in the Senate in this session. This implies that the vast majority of Senators would have only observations for a few sessions and missing observations for sessions where they were not members of the Senate.
%An alternative aggregation step -- which would avoid missing observations -- could consist of using a different index for aggregation, e.g., federal states, and focus on the rhetorical evolution of these units of measurement. However, these units might not be the units of interest for  analysis and also might undergo drastic changes in their rhetoric due to switches between different individuals. 

We conduct an empirical comparison of DPF and TPF through a simulation study using synthetic data that aligns with the structural requirements of both modelling approaches. The results demonstrate that TPF outperforms DPF in terms of both goodness-of-fit metrics and computational efficiency. The study design and detailed results are presented in 
%Section~S.5.
Section~\ref{supsec:simulation}.
% We provide an empirical comparison of DPF and TPF based on a simulation study using synthetic data meeting the data structure requirements of both modelling approaches. The results show that TPF outperforms DPF both with respect to relevant goodness-of-fit criteria as well as computational costs. The design as well as the detailed results of the simulation study are given in Section~S.5. 

%These issues lead us to reinvent the model and adapt it to our situation. 

%% file: diagram_tvpf.tex
\begin{figure}[t!]
	\centering
\resizebox{0.5\columnwidth}{!}{%
    \begin{tikzpicture}[x=0.9cm,y=0.9cm,
        datanode/.style={draw = black, rectangle, inner sep = 10pt},
        primarynode/.style={draw = black, rectangle, rounded corners, inner sep = 7 pt},
        paramnode/.style={draw = black, circle, inner sep = 5pt}](9.9,7.3)
        \linethickness{0.075mm}
        % borders of the picture
        \clip (-2.6,-3) rectangle (7.3,4.3);
        %% Plates - D, V, K
        \filldraw[fill=green!80!white, draw=green!40!black, fill opacity=0.2] (-2.5,4.3) -- (1,4.3) -- (1,1.5) -- (0,0.5) -- (0,-3) -- (-2.5,-3) -- cycle;
        \filldraw[fill=blue!60!white, draw=blue!60!black, fill opacity=0.2] (6,4.3) -- (-1,4.3) -- (-1,1.5) -- (0,0.5) -- (0,-3) -- (6,-3) -- cycle;
        \filldraw[fill=red!60!white, draw=red!60!black, fill opacity=0.2] (-2,1) -- (7.2, 1) -- (7.2,-3) -- (0,-3) -- (0,-1) -- (-2,-1) -- cycle;
        %% Labels of the plates
        \draw (-1.8, 3.7) node {$D$};
        \draw (5.5, 3.7) node {$V$};
        \draw (6.6, -2.4) node {$K$};
        %% Nodes
        % outcomes Y + regressors X
        \draw (0,3.4) node(Y)[datanode] {$\mathbb{Y}$};
        % Poisson intensities lambda
        \draw (0,1.8) node(lambda)[primarynode] {$\bm \lambda$};
        % Primary parameters
        \draw (-1,0) node(theta)[primarynode] {$\bm \theta$};
        \draw (0.7,0) node[primarynode, anchor=west] (h) {$\bm h_1 \rightarrow \bm h_2 \rightarrow \cdots \rightarrow \bm h_T$};
        %\draw (1,0) node(h1)[primarynode] {$h_1$};
        %\draw (2.5,0) node(h2)[primarynode] {$h_2$};
        %\draw (4,0) node(hdot) {$\cdots$};
        %\draw (5.5,0) node(hT)[primarynode] {$h_T$};
        %\node [draw=black!50, fit={(h1) (h2) (hdot) (hT)}] {};
        % Other auxiliary parameters
        \draw (-1,-2) node(xi)[paramnode] {$\bm \xi$};
        \draw (1.5,-2) node(mu)[paramnode] {$\bm \mu$};
        \draw (3,-2) node(delta)[paramnode] {$\bm \delta$};
        \draw (4.5,-2) node(tau)[paramnode] {$\bm \tau$};
        %% Directions
        \draw [->] (lambda) -- (Y);
        \draw [->] (theta) -- (lambda);
        \draw [->] (h) -- node[above right] {$\exp$} (lambda);
        %\draw [->] (h1) -- (h2);
        %\draw [->] (h2) -- (hdot);
        %\draw [->] (hdot) -- (hT);
        \draw [->] (xi) -- (theta);
        \draw [->] (mu) -- (h);
        \draw [->] (delta) -- (h);
        \draw [->] (tau) -- (h);
    \end{tikzpicture}
    }
    \caption{\label{fig:hierarchy_TVPF} 
		The TPF model in plate notation. Observed data (rectangles), primary model parameters (rounded corners), additional parameters (circles). Coloured planes indicate dimension sizes with respect to documents ($D$), the vocabulary ($V$) and topics ($K$).
	}
\end{figure} 

%% file: section03_final.tex
%%%!!!!!!!!!!!!!!!!!!!!!!!!!!!!!!!!!!!!!!!!!!!!!!!!!!!!!!!!!!!!!!!!!!!!!!!%%%
%%%%%%%%%%%%%%%%%%%%%%%%%%%%%%%%%%%%%%%%%%%%%%%%%%%%%%%%%%%%%%%%%%%%%%%%%%%%%
%%%%%%%%%%%%               NEW SECTION                      %%%%%%%%%%%%%%%%%
%%%%%%%%%%%%%%%%%%%%%%%%%%%%%%%%%%%%%%%%%%%%%%%%%%%%%%%%%%%%%%%%%%%%%%%%%%%%%
%%%!!!!!!!!!!!!!!!!!!!!!!!!!!!!!!!!!!!!!!!!!!!!!!!!!!!!!!!!!!!!!!!!!!!!!!!%%%
\section{Inference, model evaluation and interpretation} \label{sec:inference}

We aim to infer the document intensities and the time-varying term intensities. In a Bayesian model, this requires estimating the posterior distribution of the model's parameters and latent variables. While exact analytical solutions for such posterior distributions are typically limited to simpler models, computational methods offer effective approximations. Markov chain Monte Carlo (MCMC) techniques, for instance, 
producing samples from the posterior
%construct a Markov chain that samples from the posterior 
%and 
are widely regarded as the gold standard due to their strong theoretical guarantees. However, variational methods offer a more computationally efficient and scalable alternative, albeit at the cost of some approximation error. Given the high dimensionality of the latent variables and parameter space, we opted for VI to leverage its efficiency and scalability.
% and are considered a gold standard due to their 
% strong theoretical guarantees. Alternatively, variational methods are computationally more efficient and scalable, while introducing some approximation error. Due to the size of the  latent variables and parameter vector, we decided to estimate the model based on  VI to exploit this computational efficiency and scalability. }
%However, due to their high computational demands, especially for large models with many latent variables, VI has emerged as the preferred approximation method in these cases. We thus use VI to approximate the posterior of the TPF model.

%%%!!!!!!!!!!!!!!!!!!!!!!!!!!!!!!!!!!!!!!!!!!!!!!!!!!!!!!!!!!!!!!!!!!!!!!!%%%
%%%%%%%%%%%%               NEW SUBSECTION                   %%%%%%%%%%%%%%%%%
%%%!!!!!!!!!!!!!!!!!!!!!!!!!!!!!!!!!!!!!!!!!!!!!!!!!!!!!!!!!!!!!!!!!!!!!!!%%%
\subsection{Variational inference} \label{subsec:variational_inference}

Variational inference \citep[VI; for a review, see][]{BleiVI:2017} frames the problem of posterior approximation as an optimisation problem. Instead of directly finding the posterior distribution, a suitable (variational) family of simply structured known distributions is constructed, and the member of that family that minimises the Kullback-Leibler (KL) divergence from the posterior is used for final inference. Minimising the KL divergence between the variational distribution and the true posterior is equivalent to maximising the so-called evidence lower bound (ELBO), which represents a lower bound on the log-likelihood of the observed data \citep[see, e.g.,][]{BleiVI:2017}.

For PF, latent variables $\widetilde{\mathbb{Y}} = (\widetilde{y}_{dkv})_{d,k,v=1}^{D,K,V}$ are included in the inference which correspond to the word-document counts $\widetilde{y}_{dkv}$ assigned to each topic $k$ with the sum across all topics being equal to $y_{dv}$.
Latent variables are kept separately from the vector $\bm \zeta$ that combines the model parameters $\bm \zeta = \{\bm \theta, \bm \xi, \bm \tau, \bm \delta, \bm \mu, \bm h\}$. We denote by $q_{\bm \phi} (\widetilde{\mathbb{Y}}, \bm \zeta) = q_{\bm \phi} (\widetilde{\mathbb{Y}}) q_{\bm \phi} (\bm \zeta)$ the probability density function of the variational distribution given by a collection of variational parameters $\bm \phi$.
%$$
%\bm \phi = \left\{ \bm \phi_{\theta}^\shp, \bm \phi_{\theta}^\rte, \bm \phi_{\xi}^\shp, \bm \phi_{\xi}^\rte, \bm \phi_{\tau}^\shp, \bm \phi_{\tau}^\rte, \bm \phi_{\delta}^\loc, \bm \phi_{\delta}^\sclsq, \bm \phi_{\mu}^\loc, \bm \phi_{\mu}^\sclsq, \bm \phi_{h}^\loc, \bm \phi_{h}^\sclsq\right\}.
%$$
We then maximise the  $\mathsf{ELBO}$ function with respect to the variational parameters $\bm \phi$:
\begin{equation} \label{eq:ELBO_general}
\mathsf{ELBO}(\bm \phi) = \Eq \left[ \log p(\mathbb{Y} | \widetilde{\mathbb{Y}}) + \log p(\widetilde{\mathbb{Y}} | \bm \zeta) - \log q_{\bm \phi} (\widetilde{\mathbb{Y}}) + \log p (\bm \zeta) - \log q_{\bm \phi} (\bm \zeta) \right],
\end{equation}
where $\Eq$ denotes expectation with respect to the variational family $q_{\bm \phi}$.

A common approach is mean-field VI, which uses independent variational distributions for each variable and parameter. However, for auto-regressive sequences where strong dependencies among parameters are expected, this specification might be too restrictive. Therefore, we consider a variational family consisting of several independent blocks, each tailored to capture different aspects of the model:
%\begin{compactitem}
\begin{itemize}
\item univariate gamma distributions 
%$\gammadist{\phi_\bullet^\shp}{\phi_\bullet^\rte}$
$\gammadist{\phi_{\theta dk}^\shp}{\phi_{\theta dk}^\rte}
, 
\gammadist{\phi_{\xi d}^\shp}{\phi_{\xi d}^\rte}
, 
\gammadist{\phi_{\tau kv}^\shp}{\phi_{\tau kv}^\rte}
$
for $\theta_{dk}$, $\xi_d$, and $\tau_{kv}$ to 
match the gamma-distributed variables;
%capture the shape and rate parameters of document and topic intensities;
\item univariate normal distributions 
%$\norm{\phi_\bullet^\loc}{\phi_\bullet^\sclsq}$ 
$\norm{\phi_{\delta kv}^\loc}{\phi_{\delta kv}^\sclsq}
, 
\norm{\phi_{\mu kv}^\loc}{\phi_{\mu kv}^\sclsq}$
for $\delta_{kv}$ and $\mu_{kv}$ to model the 
coefficient and mean of the AR sequence;
%auto-regressive coefficients and their means;
\item multivariate normal distributions $\knorm{T}{\bm \phi_{hkv}^\loc}{\bm \phi_{hkv}^\covm}$ for $\bm h_{kv}$ of dimension $T$ to handle the auto-regressive sequences with their inherent dependencies;
\item multinomial distributions $\Mult{K}{y_{dv}}{\bm \phi_{dv}^y}$ for the latent variables $\widetilde{y}_{dkv}$ if $y_{dv} > 0$.
\end{itemize}
%\end{compactitem}
This leads to the following set of variational parameters $\bm \phi$:
$$
\resizebox{\hsize}{!}{
	$
	\left\{ 
	\underset{\in \R_{>0}^{DK}}{\bm \phi_{\theta}^\shp}, 
	\underset{\in \R_{>0}^{DK}}{\bm \phi_{\theta}^\rte}, 
	\underset{\in \R_{>0}^{D}}{\bm \phi_{\xi}^\shp}, 
	\underset{\in \R_{>0}^{D}}{\bm \phi_{\xi}^\rte}, 
	\underset{\in \R_{>0}^{KV}}{\bm \phi_{\tau}^\shp}, 
	\underset{\in \R_{>0}^{KV}}{\bm \phi_{\tau}^\rte}, 
	\underset{\in \R^{KV}}{\bm \phi_{\delta}^\loc}, 
	\underset{\in \R_{>0}^{KV}}{\bm \phi_{\delta}^\sclsq}, 
	\underset{\in \R^{KV}}{\bm \phi_{\mu}^\loc}, 
	\underset{\in \R_{>0}^{KV}}{\bm \phi_{\mu}^\sclsq}, 
	\underset{\in \R^{KVT}}{\bm \phi_{h}^\loc},
	\underset{\in \R^{KVT(T+1)/2}}{\bm \phi_{h}^\covm},
	\underset{\in \R_{>0}^{D(K-1)V}}{\bm \phi^y}
	\right\}.
	$
}
$$
Using the coordinate ascent principle, we update the individual blocks of $\bm \phi$ to maximise $\ELBO{\bm \phi}$ 
while keeping the other parameters fixed. Due to our careful choice of prior distributions and variational families, we can update the parameters of the variational families for $\bm \theta$, $\bm \xi$, $\bm \tau$, $\bm \mu$, $\bm \delta$ and $\widetilde{\mathbb{Y}}$ in a closed-form manner using coordinate ascent variational inference (CAVI) updates. For a detailed discussion of these updates, we refer to  Appendix~\ref{appsec:CAVI}. Only the variational parameters $\bm \phi_{hkv}^\loc$ and $\bm \phi_{hkv}^\covm$ require a gradient-based approach based on automatic differentiation to obtain updates (ADVI updates;
\citealt{Kucukelbir+Tran+Ranganath:2017}). 

Due to the high memory demands when using objects of dimension
$D\times K\times V$, we reduce the first dimension to a~batch of documents (of size $|\mathcal{B}| = 512$ in applications) and pursue a stochastic gradient approach. One epoch of the algorithm consists of one random split of the corpus of documents into batches and each batch corresponds to one iteration of updates. Given a~batch of documents, we first update \emph{locally} the document-specific parameters for the documents in the current batch. The remaining ones are updated during the same epoch but in a different iteration. The other parameters updated with CAVI are updated \emph{globally} for each batch. We pursue a similar approach than in stochastic optimisation and move the previous value in direction of the CAVI update with step size $\rho_s$. To satisfy the Robbins-Monro condition we set $\rho_s = (s+\tau)^{-\kappa}$ and use $\kappa = 0.51$ and $\tau = 0$, which increases the weight for the new direction for later iterations.
%This results in a stochastic gradient approach being used for $\bm \phi_{hkv}^\loc$ and $\bm \phi_{hkv}^\covm$.
Finally, the \emph{global} ADVI updates for $\bm \phi_{hkv}^\loc$ and $\bm \phi_{hkv}^\covm$ employ the Adam algorithm \citep{kingma_ba2015Adam} with a learning rate of $\alpha = 0.01$ \citep[cp.][]{Vavra_etal_2024} based on an (appropriately scaled) approximation of $\ELBO{\bm \phi}$ determined for the documents in the current batch. Unlike in other similar estimation approaches \citep[e.g.,][]{Vafa_etal_2020}, we avoid replacing expectations $\Eq$ with Monte Carlo integrals when calculating the $\mathsf{ELBO}$ since all terms are tractable in closed form 
%(see Section~S.4). 
(see Section~\ref{supsec:ELBO}). 
Therefore, the only source of randomness for the stochastic gradients stems from the selection of documents in each batch. 
%The step size for global updates is determined by $\kappa = 0.51$ and $\tau = 0$ to determine the step size for the global updates, see Algorithm~\ref{alg:TVPF_CAVI_and_SVI}.

Algorithm~\ref{alg:TVPF_CAVI_and_SVI} sketches how we combine batching, CAVI updates (denoted by $\widehat{\bullet}$, always constructed based on the most recent parameter values) and ADVI updates. We initialise $\bm \phi_{\theta}$ and $\bm \phi_{\mu}^\loc$ with (transformed) estimates obtained for non-time-varying Poisson factorisation \citep{gopalanhofmanblei2015scalableHPoisF} to anchor the meaning of topic labels across different TPF settings and attenuate identifiability problems; other parameters are initialised reasonably at random. Algorithm~\ref{alg:TVPF_CAVI_and_SVI} is run for as many epochs $E$ as needed for the sequence of the approximated $\mathsf{ELBO}$ values to converge. More details about the inference algorithm can be found in Appendix~\ref{appsec:CAVI} and 
%Sections~S.2--S.4. 
Sections~\ref{supsec:auxiliary_variables}--\ref{supsec:ELBO}. 
Algorithm~\ref{alg:TVPF_CAVI_and_SVI} also points out the optional steps which can be included to calculate the exact $\mathsf{ELBO}$ values (using all documents) and $\mathsf{VAIC}$ and $\mathsf{VBIC}$ for model selection (see Sections~\ref{subsec:variational_criteria} and S.4). Our implementation using the Tensorflow environment is available in a  Github repository
(\url{https://github.com/vavrajan/TPF}).
%~\citep{Github_TPF}.

\begin{algorithm}[t!]
	\caption{Estimation combining CAVI and ADVI.}
	\label{alg:TVPF_CAVI_and_SVI}
	\begin{algorithmic} [1]
		\State \textbf{Input:} $\mathbb{Y}$, initial $\bm \phi^0$, $E$, $|\mathcal{B}|$, $\kappa \in (0.5, 1]$, $\tau \geq 0$, $\alpha$, hyperparameter values.
        \State Initialise the step counter $s:= 0$.
		\For{$e$ in $1:E$}
		\State \textbullet~Divide $D$ documents into $B$ batches $\mathcal{B}_b, b = 1, \ldots, B,$ of size $|\mathcal{B}_b| \approx |\mathcal{B}|$.
            \For{$b$ in $1:B$}
            %\State \textbullet~Only the documents $d\in\mathcal{B}_b$ are available.
            %\State \textbullet~Use sums $\sum\limits_{d \in \mathcal{B}_b} \cdots$ scaled by $\frac{D}{|\mathcal{B}_b|}$ instead of $\sum\limits_{d=1}^D \cdots$.
            \State \textbullet~Set $s:= s+1$ and the step size $\rho_s = (s+\tau)^{-\kappa}$. 
            \State \textbullet~Update $\bm \phi_{\theta d}^s := \widehat{\bm \phi}_{\theta d}$, $\bm \phi_{\xi d}^s := \widehat{\bm \phi}_{\xi d}$ for $d \in \mathcal{B}_b$. 
            \Comment{CAVI \emph{local} updates}
                \For{$\phi$ in $\{\bm \phi_\mu^\loc, \bm \phi_\mu^\sclsq, \bm \phi_\delta^\loc, \bm \phi_\delta^\sclsq, \bm \phi_\tau^\loc, \bm \phi_\tau^\sclsq\}$}
                \State \textbullet~Update $\phi^s := \rho_s \widehat{\phi} + (1-\rho_s) \phi^{s-1} $. 
                \Comment{CAVI \emph{global} updates}
                \EndFor
            \State \textbullet~Approximate $\ELBO{\bm \phi}$ by replacing $\sum\limits_{d=1}^D$ with $\frac{D}{|\mathcal{B}_b|} \sum\limits_{d \in \mathcal{B}_b} $ in (S.4.2). 
            %to track gradients of $\bm \phi_h^\loc, \bm \phi_h^\covm$.
            \State \textbullet~Track gradients of $\bm \phi_h^\loc, \bm \phi_h^\covm$ as by-product. 
            \State \textbullet~Update $\bm \phi_h^\loc, \bm \phi_h^\covm$ using Adam with learning rate $\alpha$. 
            \Comment{ADVI \emph{global} updates}
            \EndFor
        \State \textbullet~Evaluate $\ELBO{\bm \phi}$, $\mathsf{VAIC}$, $\mathsf{VBIC}$ using \eqref{eq:ELBO_general}, \eqref{eq:VAIC}, \eqref{eq:VBIC}.
        \Comment{optional}
		\EndFor
        \State \textbf{Return:} The last values $\bm \phi^{s}$.
	\end{algorithmic}
\end{algorithm}

%%%!!!!!!!!!!!!!!!!!!!!!!!!!!!!!!!!!!!!!!!!!!!!!!!!!!!!!!!!!!!!!!!!!!!!!!!%%%
%%%%%%%%%%%%               NEW SUBSECTION                   %%%%%%%%%%%%%%%%%
%%%!!!!!!!!!!!!!!!!!!!!!!!!!!!!!!!!!!!!!!!!!!!!!!!!!!!!!!!!!!!!!!!!!!!!!!!%%%
\subsection{Model evaluation} \label{subsec:variational_criteria}
Model criteria like $\mathsf{AIC}$ and $\mathsf{BIC}$ are usually used to evaluate and compare statistical models, helping to balance model fit and complexity.
Within a Bayesian context, also the \emph{deviance information criterion} \citep[$\mathsf{DIC}$,][]{Spiegelhalter_et_al:2002} is commonly used to compare different models. It resembles Akaike's $\mathsf{AIC}$ since it combines a goodness-of-fit measure with a penalisation for complexity $p_D$ which corresponds to the effective number of parameters. The $\mathsf{DIC}$ is obtained by:
%identifies models that best explain the observed data, but with the expectation that they are likely to minimise uncertainty about observations generated in the same way
$$
\mathsf{DIC} = -2 \log p(\mathbb{Y} | \widetilde{\bm \zeta}) + 2 p_D
\quad \text{with} \quad 
p_D = 2 \log p(\mathbb{Y} | \widetilde{\bm \zeta}) - 2 \E_{\bm \zeta | \mathbb{Y}} \left[\log p(\mathbb{Y} | \bm \zeta) \right],
$$
where $\widetilde{\bm \zeta}$ is a~Bayesian estimator of model parameters $\bm \zeta$, e.g., the posterior mean. 

A version of $\mathsf{DIC}$ for VI has been proposed by~\citet{McGrory_Titterington:2007_VIC} where the posterior is replaced with its variational approximation: 
\begin{equation} \label{eq:VAIC}
\mathsf{VAIC} = -2 \log p(\mathbb{Y} | \bm \zeta^\star) + 2 p_D^\star
\quad \text{with} \quad 
p_D^\star = 2 \log p(\mathbb{Y} | \bm \zeta^\star) - 2 \Eq \left[\log p(\mathbb{Y} | \bm \zeta) \right],
\end{equation}
where $\bm \zeta^\star = \Eq \bm \zeta$ is a~vector of variational means (either $\phi^\loc$ or $\phi^\shp / \phi^\rte$). 
Similar to~\citet{You+Ormerod+Muller:2014_VIC}, we will refer to this criterion as the \emph{variational Akaike information criterion} ($\mathsf{VAIC}$).

Motivated by the Laplace approximation of the $\mathsf{BIC}$, \citet{You+Ormerod+Muller:2014_VIC} propose in the context of Bayesian linear regression models the \emph{variational Bayesian information criterion} 
($\mathsf{VBIC}$) which is of the form 
\begin{equation} \label{eq:VBIC}
\mathsf{VBIC} = - 2 \ELBO{\bm \phi} + 2 \Eq \log p(\bm \zeta) 
%\overset{\eqref{eq:ELBO_general}}{=} 
%-2 \Eq \log p(\mathbb{Y} | \bm \zeta) + 2 \Eq \log q_{\bm \phi}(\bm \zeta),
.
\end{equation}
%which adds the negative \emph{reconstruction} and the \emph{entropy} term.  
%It was designed for the Bayesian linear regression model, hence, its use for other models is speculative.
%\textcolor{red}{However, they decided to approximate $\log p (\widehat{\bm \zeta}^\mathsf{ML})$ with $\Eq \log p (\bm \zeta)$, why not $\log p (\bm \zeta^\star)$?}
Equations \eqref{eq:VAIC} and \eqref{eq:VBIC}  provide the generic expressions of the $\mathsf{VAIC}$ and $\mathsf{VBIC}$ criteria; the formulations specific to our model are given in Section S.4.
In the following, we use the two criteria to compare and evaluate different model and VI specifications. 

%%%!!!!!!!!!!!!!!!!!!!!!!!!!!!!!!!!!!!!!!!!!!!!!!!!!!!!!!!!!!!!!!!!!!!!!!!%%%
%%%%%%%%%%%%               NEW SUBSECTION                   %%%%%%%%%%%%%%%%%
%%%!!!!!!!!!!!!!!!!!!!!!!!!!!!!!!!!!!!!!!!!!!!!!!!!!!!!!!!!!!!!!!!!!!!!!!!%%%
\subsection{Interpretation}\label{subsec:interpretation}

VI results in estimates for the variational parameters which imply a variational distribution that approximates the posterior distribution of the latent variables and parameters. We obtain point estimates for document intensities and term intensities based on the means induced by the variational distributions of these parameters. In the following, we discuss how to infer the time-varying prevalence of topics, the topical content as well as the consistency of topics over time based on these estimates. 

\subsubsection{Time-varying prevalence of topics} \label{subsec:topic_prevalence}
We determine the time-varying prevalence of topics to understand how the distribution of topics evolves over time. While our model does not include a parameter to directly capture the temporal evolution of topic prevalence, appropriate post-processing of the model parameters enables this analysis. 

% The effect of $h_{kv,t}$ is also accompanied by $\theta_{dk}$ that reflects individual documents from very long era. 
We determine a time period specific topic prevalence by averaging the estimated Poisson rates $\lambda_{dkv}$ over documents and terms. 
In particular, VI allows to approximate the posterior mean of $\lambda_{dkv}$ by
$$
\Eq \lambda_{dkv} 
= 
\Eq \left[ \theta_{dk} \cdot \exp\left\{ h_{kv,t} \right\} \right]
= 
\dfrac{\phi_{\theta dk}^\shp}{\phi_{\theta dk}^\rte} \cdot
\exp\left\{\phi_{hkv,t}^\loc + \frac{1}{2}\phi_{hkv,t}^\sclsq \right\}.
$$
%% $$
%% \Eq \lambda_{dkv} 
%% = 
%% \dfrac{\phi_{\theta dk}^\shp}{\phi_{\theta dk}^\rte} \cdot \beta_{kv,t}
%% \quad \text{where} \quad
%% \beta_{kv,t} 
%% := 
%% \Eq \exp\left\{ h_{kv,t} \right\}
%% = 
%% \exp\left\{\phi_{hkv,t}^\loc + \frac{1}{2}\phi_{hkv,t}^\sclsq \right\}.
%% $$
We sum over these approximate document-term specific rates for topic $k$ to obtain topic prevalences $\psi_{kt}$ for time period $t$:
\begin{align*}
\psi_{kt} 
&\propto
%\dfrac{1}{V \left|\mathcal{D}_{t}\right|} 
\sum\limits_{d \in \mathcal{D}_{t}} \sum\limits_{v=1}^V 
\dfrac{\phi_{\theta dk}^\shp}{\phi_{\theta dk}^\rte} 
\exp\left\{\phi_{hkv,t}^\loc + \frac{1}{2}\phi_{hkv,t}^\sclsq \right\} \\
&\propto
\left(
%\dfrac{1}{\left|\mathcal{D}_{t}\right|} 
\sum\limits_{d \in \mathcal{D}_{t}} \dfrac{\phi_{\theta dk}^\shp}{\phi_{\theta dk}^\rte} 
\right)
\left(
%\dfrac{1}{V}
\sum\limits_{v=1}^V 
\exp\left\{\phi_{hkv,t}^\loc + \frac{1}{2}\phi_{hkv,t}^\sclsq \right\}\right).
\end{align*}
\begin{comment}
$$
\psi_{kt} 
\propto
\sum\limits_{d \in \mathcal{D}_{t}} \sum\limits_{v=1}^V 
\dfrac{\phi_{\theta dk}^\shp}{\phi_{\theta dk}^\rte} 
\exp\left\{\phi_{hkv,t}^\loc + \frac{1}{2}\phi_{hkv,t}^\sclsq \right\} =
\left(
\sum\limits_{d \in \mathcal{D}_{t}} \dfrac{\phi_{\theta dk}^\shp}{\phi_{\theta dk}^\rte} 
\right)
\left(
\sum\limits_{v=1}^V 
\exp\left\{\phi_{hkv,t}^\loc + \frac{1}{2}\phi_{hkv,t}^\sclsq \right\}\right).
$$
\end{comment}
$\psi_{kt}$ are normalised so that they sum up to one over all topics for a given time period $t$ and thus correspond to time period specific topic proportions. 

%\textcolor{red}{Join this section with the following one?}
%\textcolor{red}{Paul: can we interpret the first phi ratio as the thetas? Honza: as aggregated thetas for certain topic over a time period.}
%For each time period~$t$ we compute the proportions $\psi_{kt} \;\big/ \sum\limits_{\kappa=1}^{K} \psi_{\kappa t}$ to depict the evolution of topic prevalence in a~spineplot.  

%%%!!!!!!!!!!!!!!!!!!!!!!!!!!!!!!!!!!!!!!!!!!!!!!!!!!!!!!!!!!!!!!!!!!!!!!!%%%
%%%%%%%%%%%%               NEW SUBSECTION                   %%%%%%%%%%%%%%%%%
%%%!!!!!!!!!!!!!!!!!!!!!!!!!!!!!!!!!!!!!!!!!!!!!!!!!!!!!!!!!!!!!!!!!!!!!!!%%%
\subsubsection{Topical content} \label{subsec:frex_measure}
The second key aspect in topic models is to assess the content of the topics. 
The content of a topic is usually characterised by the specific terms that are most representative of this topic. This allows to define the subject matter and themes that the topic covers. 

% One of the key features of our model is to determine a~set of the most used terms when discussing given topic~$k$ and how this set evolves though different time periods. 
The time period specific term effects for a topic are captured by $h_{kv,t}$. Thus an obvious choice for selecting the terms that are most representative of a topic is based on their exponentiated values such that they correspond to the intensities used to obtain the Poisson rates. 
However, this results in the identification of frequent terms which are not characteristic of a specific topic but are rather prevalent in many topics. To address this common problem when characterising topics, \citet{bischof_airoldi_2012_frequency_exclusivity} proposed the \emph{frequency-exclusivity} ($\mathsf{FREX}$) measure.  $\mathsf{FREX}$ reflects both the popularity of the term as well as its uniqueness across topics and has been used to characterise topics in a range of applications \citep[e.g.,][]{Roberts_etal_JASA:2016}. 

In our setting, the \emph{frequency} and \emph{exclusivity} measures are derived from the term intensities $\beta_{kv,t}$ obtained with
\begin{align*}
    \beta_{kv,t} &= 
\Eq \exp\left\{ h_{kv,t} \right\}
= 
\exp\left\{\phi_{hkv,t}^\loc + \frac{1}{2}\phi_{hkv,t}^\sclsq \right\}.
\end{align*}
In particular, the \emph{frequency} measure of term $v$ for topic~$k$ during time period~$t$ is determined by evaluating the  empirical distribution function constructed from the term intensities $\bm \beta_{k\bullet,t}$ (with the bullet representing all values associated with that index) at the corresponding value:
$$
    \mathsf{FR}_{kt}(v) = \mathsf{ECDF}_{\bm \beta_{k\bullet,t}} \left(\beta_{kv,t}\right) = \frac{1}{V} \sum\limits_{u=1}^V \indicator_{\left(\beta_{ku,t} \leq \beta_{kv,t}\right)}.
$$
%The closer to $1$, the more frequently used term is for topic~$k$ in time period~$t$. 
%If $\mathsf{FR}_{kt}(v) \approx 0$ then the word is used the least from all words, while words with $\mathsf{FR}_{kt}(v) \approx 1$ are the top used ones when discussing topic~$k$ in time period~$t$. 
Similarly, we define the \emph{exclusivity} measure of term $v$ for topic~$k$ during time period~$t$:
%as an empirical distribution function constructed from a~vector of comparisons $\overline{\bm \beta}_{k\bullet,t}$ evaluated in the corresponding values:
$$
    \mathsf{EX}_{kt}(v) = \mathsf{ECDF}_{\overline{\bm \beta}_{k\bullet,t}} \left(\overline{\beta}_{kv,t}\right) =  \frac{1}{V} \sum\limits_{u=1}^V \indicator_{\left(\overline{\beta}_{ku,t} \leq \overline{\beta}_{kv,t}\right)},
$$
where 
%$\overline{\beta}_{kv,t} = \beta_{kv,t} \;\big/ \sum\limits_{\kappa =1}^K \beta_{\kappa v,t}$ 
$\overline{\beta}_{kv,t} = \beta_{kv,t} / (\sum_{l=1}^K \beta_{lv,t})$ 
measures how prevalent the term is for topic~$k$ compared to all other topics. 
%High value of $\overline{\beta}_{kv,t} > 1/K$ confirms that term~$v$ is rather used for topic~$k$ compared to the others, thus, the word is more \emph{exclusive}. If all $\overline{\beta}_{kv,t} \approx 1/K$ then the word is prevalent in all the topics similarly. 

The $\mathsf{FREX}$ measure is defined as the reciprocal of the convex combination of the separate reciprocals of the \emph{frequency} and \emph{exclusivity} measure:
$$
    \mathsf{FREX}_{kt}(v) = \left(\dfrac{1-w}{\mathsf{FR}_{kt}(v)}  +  \dfrac{w}{\mathsf{EX}_{kt}(v)}\right)^{-1},
$$
where $w \in [0, 1]$ declares the weight on exclusivity. In the following we set $w=0.5$. Based on the $\mathsf{FREX}$ measure the top terms for a topic and time period can be easily extracted and inspected to characterise and manually label a topic.

%%%!!!!!!!!!!!!!!!!!!!!!!!!!!!!!!!!!!!!!!!!!!!!!!!!!!!!!!!!!!!!!!!!!!!!!!!%%%
%%%%%%%%%%%%               NEW SUBSECTION                   %%%%%%%%%%%%%%%%%
%%%!!!!!!!!!!!!!!!!!!!!!!!!!!!!!!!!!!!!!!!!!!!!!!!!!!!!!!!!!!!!!!!!!!!!!!!%%%
\subsubsection{Topic consistency} \label{subsec:topic_consistency}

In addition to characterising a topic based on the top-10 terms according to some measure of topical content, e.g., $\mathsf{FREX}_{kt}(v)$, a quantitative assessment of similarity of different topics at the same time period or the same topic across time periods helps to interpret the results obtained for the TPF model. 

Given two vectors of length $V$ characterising the topical content, a number of metrics were suggested to assess semantic congruence \citep{Bullinaria+Levy:2007}. We propose to compare the vectors not as $V$-dimensional vectors but as $V$-dimensional probability distributions. Since term prevalence is solely based on the parameter $h_{kv,t}$, we suggest to use the variational approximation of its posterior to assess congruence. This implies that we need to compare $V$ pairs of independent univariate normal distributions. 
We opt for the symmetrised KL divergence as dissimilarity measure (which is not a metric as it does not fulfil the triangle inequality) for the following two reasons: (1) the independence of individual distribution pairs yields a~sum of dissimilarities across the whole vocabulary and (2) KL divergence is invariant to transformations, i.e., the results do not depend on the choice of the scale. In particular the later aspect differentiates this choice from other popular dissimilarity measures used in this context, such as cosine dissimilarity. 

In general, the symmetrised KL divergence between two (log-)normal distributions is 
$$
d_{KL}\left( \norm{\mu_x}{\sigma_x^2}, \norm{\mu_y}{\sigma_y^2} \right) 
    = 
    \frac{1}{4}
    \dfrac{
        \left(\sigma_x^2 - \sigma_y^2\right)^2 + \left(\sigma_x^2 + \sigma_y^2\right) \left(\mu_x - \mu_y\right)^2
    }{
        \sigma_x^2\sigma_y^2
    }.
$$
The dissimilarity of topical content ($\mathsf{DTC}$) between two topic-time period combinations $(k_1, t_1)$, $(k_2, t_2)$ is given by:
$$
%d_{KL}
\mathsf{DTC}
\left( 
k_1, t_1
%\bm \phi_{h, k_1, t_1} 
%\bm \beta_{k_1 \bullet, t_1}
|
k_2, t_2
%\bm \phi_{h, k_2, t_2} 
%\bm \beta_{k_2 \bullet, t_2} 
\right) 
    =
    \sum\limits_{v=1}^V
    d_{KL}\left( \norm{\phi_{hk_1 v, t_1}^\loc}{\phi_{hk_1 v, t_1}^\sclsq}, \norm{\phi_{hk_2 v, t_2}^\loc}{\phi_{hk_2 v, t_2}^\sclsq} \right) .
$$
When comparing the content between two topics across all time periods, we determine $\mathsf{DTC}$ based on two multivariate normal distributions:
$$
\mathsf{DTC}
\left( 
k_1 | k_2
\right) 
    =
    \sum\limits_{v=1}^V
    \frac{1}{T} d_{KL}\left( \knorm{T}{\bm \phi_{hk_1 v}^\loc}{\bm \phi_{hk_1 v}^\covm}, \norm{\bm \phi_{hk_2 v}^\loc}{\bm \phi_{hk_2 v}^\covm} \right),
$$
where the symmetrised KL divergence
\begin{multline*}
d_{KL}\left( \knorm{T}{\bm\mu_x}{\bm\Sigma_x}, \knorm{T}{\bm\mu_y}{\bm\Sigma_y} \right) 
    = 
    \frac{1}{4}
    \left[
        \T{\left(\bm \mu_y - \bm \mu_x\right)} \bm \Sigma_y^{-1} \left(\bm \mu_y - \bm \mu_x\right)
        \right. \\ \left.
        + 
        \T{\left(\bm \mu_x - \bm \mu_y\right)} \bm \Sigma_x^{-1} \left(\bm \mu_x - \bm \mu_y\right)
        + \Tr{\bm\Sigma_y^{-1} \bm\Sigma_x} + \Tr{\bm\Sigma_x^{-1} \bm\Sigma_y} - 2T
    \right]
\end{multline*}
reduces to the sum of $d_{KL}$ values determined based on univariate normal distributions only if the covariance matrices are diagonal, with $\Tr{()}$ denoting the trace of a matrix.

%Thus, $\mathsf{DTC}$ measures can be easily derived from the point estimates of the variational parameters. 

%\paragraph{Implementation}
%\footnote{TPF is implemented within the Tensorflow environment and is publicly available at Github~\citep{Github_TPF}.}

%% file: section04_final.tex
%%%!!!!!!!!!!!!!!!!!!!!!!!!!!!!!!!!!!!!!!!!!!!!!!!!!!!!!!!!!!!!!!!!!!!!!!!%%%
%%%%%%%%%%%%%%%%%%%%%%%%%%%%%%%%%%%%%%%%%%%%%%%%%%%%%%%%%%%%%%%%%%%%%%%%%%%%%
%%%%%%%%%%%%               NEW SECTION                      %%%%%%%%%%%%%%%%%
%%%%%%%%%%%%%%%%%%%%%%%%%%%%%%%%%%%%%%%%%%%%%%%%%%%%%%%%%%%%%%%%%%%%%%%%%%%%%
%%%!!!!!!!!!!!!!!!!!!!!!!!!!!!!!!!!!!!!!!!!!!!!!!!!!!!!!!!!!!!!!!!!!!!!!!!%%%
\section{U.S. Senate speeches (1981--2016)} \label{sec:senate_speech_data}
We apply TPF to a dataset provided by 
\cite{Gentzkow+Shapiro+Taddy:2018}. This dataset contains all speeches in the U.S.\ Senate during the Congress sessions 97--114 (1981--2016). We aim to conduct an analysis at the speech level, where topic-specific term distributions evolve over time. Our TPF model is well-suited for this purpose, whereas the DPF model, which relies on repeated user-item matrices across time periods, is not applicable in this setting.

% We are interested in an analysis on speech level where the term distributions of the topics change over time. For such an analysis, our TPF model is suitable, whereas the DPF model, which requires the availability of repeated user-item matrices over time for analysis, cannot be employed.
Following \cite{Hofmarcher+Vavra+Adhikari:2025}, we conducted several pre-processing steps. We restricted our analysis to speeches given by Senators. % only which results in 309 different senators. 
We removed punctuation and numbers, changed the text to lower-case and eliminated stop words. For tokenisation, we used bigrams.
%Further, we removed speakers who gave less than 24 speeches in a particular session as well as bigrams which were used by less than 10 speakers in a particular session\textcolor{red}{Paul: I can't remember but I think we discussed this once :-)}. 
The time periods correspond to the single sessions, yielding $T=18$ time periods, each spanning two years. Each time period starts on January 1 of an odd year and ends on December 31 of the following even year. For these time periods the composition of the Senate is rather stable within but changes between due to election results leading to members leaving and entering the Senate. They are also short enough to reflect smooth transitions in vocabulary. 
We obtained a DTM $\mathbb{Y}$ consisting of $D = 732\,110$ documents and a vocabulary with $V = 12\,791$ unique bigrams. According to \citet{Gentzkow_etal_2019} and \citet{Hofmarcher+Vavra+Adhikari:2025}, assuming $K = 25$ latent topics is an appropriate choice for this dataset. 
%Section~S.6 
Section~\ref{supsec:processing_senate_data} 
provides more details on how the documents were transformed to obtain the DTM $\mathbb{Y}$.

%%%!!!!!!!!!!!!!!!!!!!!!!!!!!!!!!!!!!!!!!!!!!!!!!!!!!!!!!!!!!!!!!!!!!!!!!!%%%
%%%%%%%%%%%%               NEW SUBSECTION                   %%%%%%%%%%%%%%%%%
%%%!!!!!!!!!!!!!!!!!!!!!!!!!!!!!!!!!!!!!!!!!!!!!!!!!!!!!!!!!!!!!!!!!!!!!!!%%%
\subsection{Comparing different model settings} \label{subsec:model_selection} 
We estimate TPF under four settings. We consider two different model specifications: the auto-regressive coefficient $\delta_{kv}$ is either kept constant ($\delta_{kv} = 1$, representing a random walk) or allowed to vary ($\delta_{kv} \in \mathbb{R}$, resulting in $2KV$ more variational parameters to estimate). In addition, we use two versions for VI: the posterior distribution of $\bm{h}_{kv}$ is approximated either with independent variational distributions ($\bm{\phi}_{hkv}^\mathrm{cov}$ diagonal, involving $TKV$ variational parameters) or with a multivariate normal distribution ($\bm{\phi}_{hkv}^\mathrm{cov}$ general, involving $\frac{1}{2} T(T+1)KV$ variational parameters). %as outlined in Section~\ref{subsec:variational_inference}. 
This results in the following four settings: 
(A) $\delta_{kv} \in \mathbb{R}$, $\bm{\phi}_{hkv}^\mathrm{cov}$ general;
(B) $\delta_{kv} \in \mathbb{R}$, $\bm{\phi}_{hkv}^\mathrm{cov}$ diagonal;
(C) $\delta_{kv} = 1$, $\bm{\phi}_{hkv}^\mathrm{cov}$ general;
(D) $\delta_{kv} = 1$, $\bm{\phi}_{hkv}^\mathrm{cov}$ diagonal.

\begin{figure}[t!]
    \centering
    % \begin{subfigure}{0.485\textwidth}
    %     \includegraphics[width=\textwidth]{models_comparison_VAIC.png}
    %     \caption{$\mathsf{VAIC}$. }
    %     \label{fig:models_comparison_VAIC}
    % \end{subfigure}
    % ~
    % \begin{subfigure}{0.485\textwidth}
    %     \includegraphics[width=\textwidth]{models_comparison_VBIC.png}
    %     \caption{$\mathsf{VBIC}$. }
    %     \label{fig:models_comparison_VBIC}
    % \end{subfigure}
    % ~
    \begin{subfigure}{0.485\textwidth}
        \includegraphics[width=\textwidth]{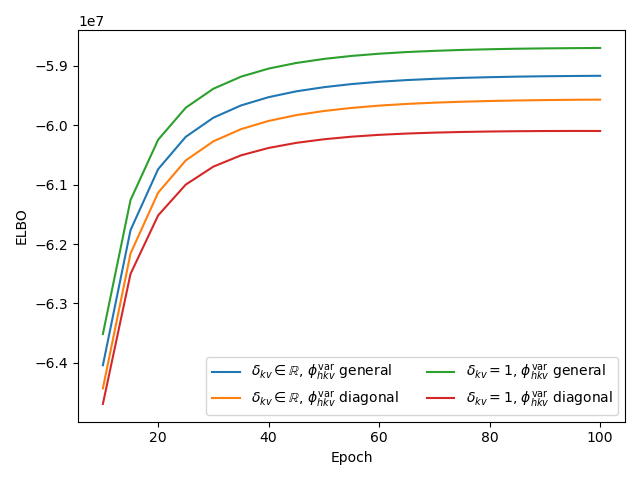}
        \caption{$\mathsf{ELBO}$. }
        \label{fig:models_comparison_ELBO}
    \end{subfigure}
    ~
    \begin{subfigure}{0.485\textwidth}
        \includegraphics[width=\textwidth]{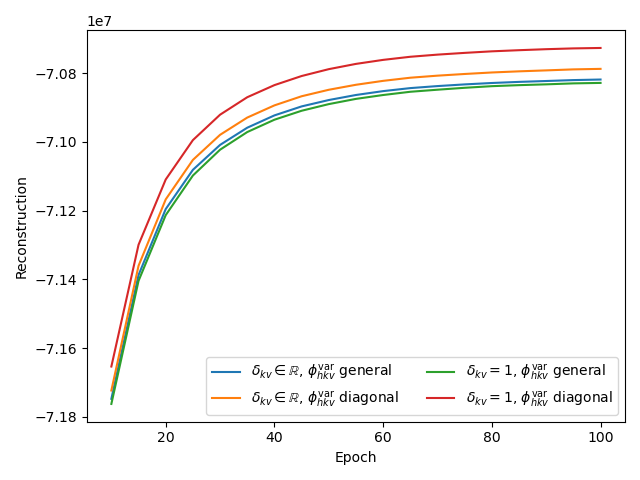}
        \caption{Reconstruction. }
        \label{fig:models_comparison_recpnstruction}
    \end{subfigure}
    % ~
    % \begin{subfigure}{0.485\textwidth}
    %     \includegraphics[width=\textwidth]{models_comparison_log_prior_plus_entropy.png}
    %     \caption{Log-prior and entropy. }
    %     \label{fig:models_comparison_log_prior_plus_entropy}
    % \end{subfigure}
    % ~
    % \begin{subfigure}{0.485\textwidth}
    %     \includegraphics[width=\textwidth]{models_comparison_effective_number_of_parameters.png}
    %     \caption{Effective number of parameters $p_D$. }
    %     \label{fig:models_comparison_sec_effective_number_of_parameters}
    % \end{subfigure}
    %~
    %\begin{subfigure}{0.485\textwidth}
    %    \includegraphics[width=\textwidth]{models_comparison_sec_ELBO.png}
    %    \caption{Seconds per $\mathsf{ELBO}$ evaluation. }
    %    \label{fig:models_comparison_sec_ELBO}
    %\end{subfigure}
    %~
    %\begin{subfigure}{0.485\textwidth}
    %    \includegraphics[width=\textwidth]{models_comparison_sec_ELBO.png}
    %    \caption{Seconds per $\mathsf{ELBO}$ evaluation. }
    %    \label{fig:models_comparison_sec_ELBO}
    %\end{subfigure}
    \caption{Evolution of $\mathsf{ELBO}$ and reconstruction values over epochs under different settings. }
    \label{fig:models_comparison}
\end{figure}

We iterated Algorithm~\ref{alg:TVPF_CAVI_and_SVI} for $E = \epochs$ epochs for each  of the four settings. Figure~\ref{fig:models_comparison} provides the change in $\mathsf{ELBO}$ as well as reconstruction values, i.e., the sum of the first three terms in~\eqref{eq:ELBO_general}, over epochs for all four settings. The $\mathsf{ELBO}$ curve shows the model's improvement in capturing the data distribution including log-prior and entropy, while the reconstruction curve reflects the model's performance in reconstructing the input data from its latent representation only. Both curves indicate that a reasonable level of convergence has been reached for all four settings after $\epochs$~epochs. 

\begin{comment}
\begin{table}
    \centering
    \resizebox{\textwidth}{!}{
        \input{table_ELBO}
    }
    \caption{Comparison of 4 models after $E=\epochs$ epochs.}
    \label{tab:table_ELBO}
\end{table}
\end{comment}

\begin{table}
    \centering
    \resizebox{0.99\textwidth}{!}{
        \input{models_VIC}
    }
    \caption{Comparison of the four model settings after $E=\epochs$ epochs
    with respect to goodness-of-fit criteria ($\mathsf{ELBO}$, reconstruction, $\mathsf{VAIC}$, $\mathsf{VBIC}$)
    and average computation time per epoch (``sec/epoch'' denotes the average time per epoch to update the variational parameters and the approximate $\mathsf{ELBO}$ values; ``sec/$\mathsf{ELBO}$'' denotes the average time per epoch to determine exact $\mathsf{ELBO}$ values and the information criteria). The value of the best performing model on the specific criterion is shown in bold font.}
    \label{tab:models_VIC}
\end{table}

All four combinations are compared using the variational information criteria suggested in Section~\ref{subsec:variational_criteria}. In addition, Table~\ref{tab:models_VIC} provides the final $\mathsf{ELBO}$ and reconstruction values after $\epochs$ epochs as well as insights into the run-times. Run-time is assessed in two ways: (1) the average number of seconds the VI algorithm needed per epoch to update the variational parameters and determine the approximate $\mathsf{ELBO}$ values (in column ``sec/epoch'') and (2) the average number of seconds needed to determine the exact 
$\mathsf{ELBO}$ values as well as the $\mathsf{VAIC}$ and $\mathsf{VBIC}$ values per epoch (indicated as optional in Algorithm~\ref{alg:TVPF_CAVI_and_SVI}, in column ``sec/$\mathsf{ELBO}$'').

Settings (B) and (D) with diagonal covariance matrix $\bm \phi_{hkv}^\covm$ reach lower $\mathsf{ELBO}$ values compared to the general specification.
%which is due to different values for the log-prior and entropy (i.e., stemming from different constants). 
However, they result in a~better fit with respect to the reconstruction term and also $\mathsf{VAIC}$ and $\mathsf{VBIC}$ suggest to prefer these settings. The diagonal specification also results in lower computational costs as the average run-times increase substantially when the independence assumption within the variational family is relaxed. 
Clearly, dealing with non-diagonal elements of $\bm \phi_{hkv}^\covm$ considerably slows down the algorithm. However, the inference  results are rather comparable. Selected results for the four settings are shown in 
%Section~S.7 
Section~\ref{supsec:additional_outputs} 
to indicate that similar insights are gained. Overall, these results suggest that the strongest differences are obtained when inspecting the results based on the $\mathsf{DTC}$ between pairs of topics across time points and when comparing models fitted using independent univariate instead of a multivariate distribution.

% Moreover, looking at Figure~\ref{fig:models_comparison_sec_effective_number_of_parameters} we clearly see that the effective number of parameters $p_D$ has dramatic change depending on the prior choice for $\delta_{kv}$ while being comparable for the choice of $\bm \phi_{hkv}^\covm$. On average the estimated mean for $\delta_{kv}$ parameters was $0.541$~{\footnotesize(0.044)} with average scale parameter of $0.142$~{\footnotesize(0.006)}, which resembles the chosen prior of $\norm{0.5}{0.2^2}$.  

%\textcolor{red}{Either that or the convergence is  quicker with when less parameters is present, therefore, with the same amount of epochs it reaches lower values faster. I shall increase the number of epochs to converge properly.} 

Overall, the simplest setting (D) with fixed $\delta=1$ and diagonal variance matrix $\bm \phi_{hkv}^\covm$ reaches the lowest $\mathsf{VAIC}$ and $\mathsf{VBIC}$ values. The average run-times for the VI algorithm as well as the full evaluation of $\mathsf{ELBO}$ and variational criteria are also the lowest for this setting. The next section thus presents the results of setting (D), the random walk model with a simplified diagonal structure for the covariance matrix $\bm \phi_{hkv}^\covm$.

%Nevertheless, we present the results of the proposed TPF model with autoregressive parametrization and its full-covariance approximation of the posterior.

%%%!!!!!!!!!!!!!!!!!!!!!!!!!!!!!!!!!!!!!!!!!!!!!!!!!!!!!!!!!!!!!!!!!!!!!!!%%%
%%%%%%%%%%%%               NEW SUBSECTION                   %%%%%%%%%%%%%%%%%
%%%!!!!!!!!!!!!!!!!!!!!!!!!!!!!!!!!!!!!!!!!!!!!!!!!!!!!!!!!!!!!!!!!!!!!!!!%%%
\subsection{Inspecting results} \label{subsec:results}

In the following, we assess the change of prevalence of each topic and the similarity of topical content over time for one topic and between pairs of topics across time periods. For ease of interpretation we manually assign labels to the $K=25$ topics. We inspect the most prevalent terms according to the $\textsf{FLEX}$ measure for each topic-time period combination (e.g., such as shown in Table~\ref{tab:vocabulary_evolution_frex_50_12} for topic~12) and infer the most prominent terms to characterise a topic. For the resulting labels see Table~\ref{tab:topic_labels}. 

\begin{table}
    \centering
    \resizebox{\textwidth}{!}{
        \input{topic_labels}
    }
    \caption{Labels assigned to the $K=25$ topics.}
    \label{tab:topic_labels}
\end{table}

\begin{figure}[t!]
    \centering
    \includegraphics[scale=0.42]{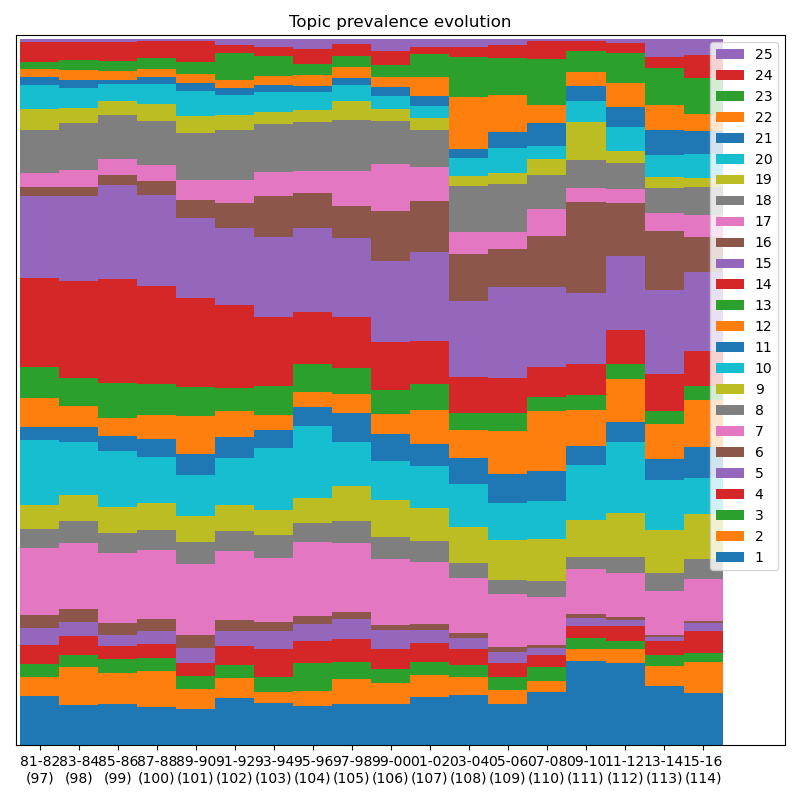}
    \caption{Evolution of topic prevalences $\psi_{kt}$ over time. }
    \label{fig:spineplot_avg_rates_topics}
\end{figure}

Figure~\ref{fig:spineplot_avg_rates_topics} provides a stacked bar plot showing the prevalence of topics over time. For each time period the relative proportion of a topic prevalence $\psi_{kt}$ is determined. Clearly, topics vary in general prevalence over time. Topics~14 and 15 seem to be the most prevalent ones at the beginning. However, while topic~15 remains highly prevalent, the prevalence of topic~14 decreases over time. Inspecting the topical content of these topics indicates that topic~14 is dominated by terms like \emph{United States}, \emph{trade}, and \emph{foreign relations}.
Topic~15 is characterised at the beginning by common political rhetoric and formal expressions often used in the U.S.\ Senate, e.g., phrases like \emph{distinguished Republican} and \emph{on the other side of the aisle} are typical of respectful address and bipartisan discourse within the legislative context. Over time, however, the focus of topic~15 shifts from these general political expressions to more specific discussions about immigration. 
Figure~\ref{fig:spineplot_avg_rates_topics} also indicates that topic~12 (climate change) and 16 (affordable health care) show increasing prevalence over time.
Notably, topic~1, which addresses taxes and business, is the most prevalent topic during the financial crisis. Overall these results indicate that the TPF model captures that the prevalence of topics is time-varying. 
%Smooth transitions across time periods are obtained despite these time specific topic prevalences only being implicitly derived from the document intensities where the time information is not explicitly accounted for. 
In particular, the rather smooth changes in prevalence are remarkable given that the model does not explicitly impose any smoothness in this respect.

%[width=\textwidth]

\begin{figure}[t!]
    \centering
    \begin{subfigure}{0.49\textwidth}
        \includegraphics[width=\textwidth]{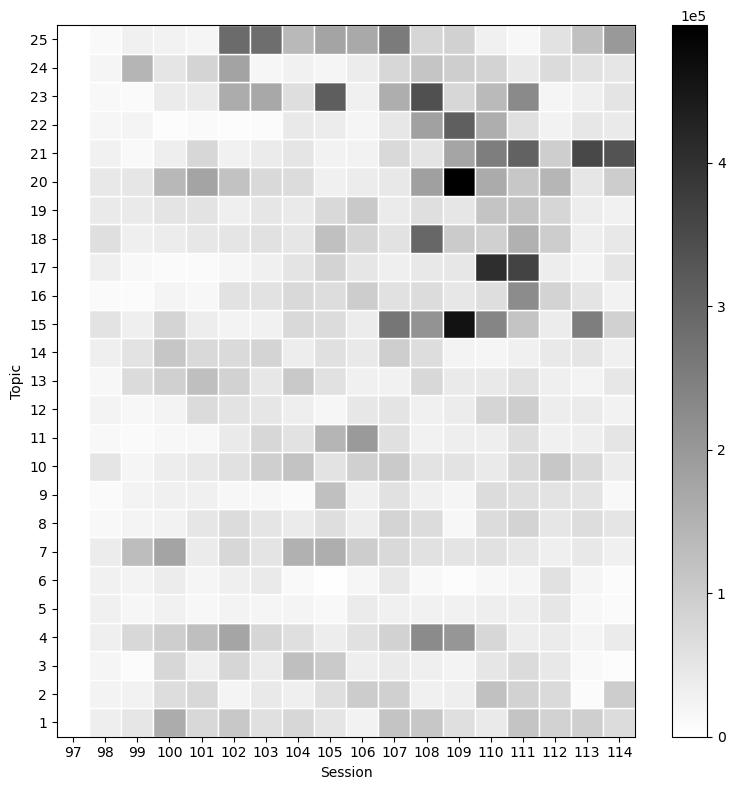}
        \caption{Consecutive sessions -- $\mathsf{DTC}(k,t-1|k,t)$. }
        \label{fig:KL_dissimilarity_in_time}
    \end{subfigure}
    ~
    \begin{subfigure}{0.48\textwidth}
        \includegraphics[width=\textwidth]{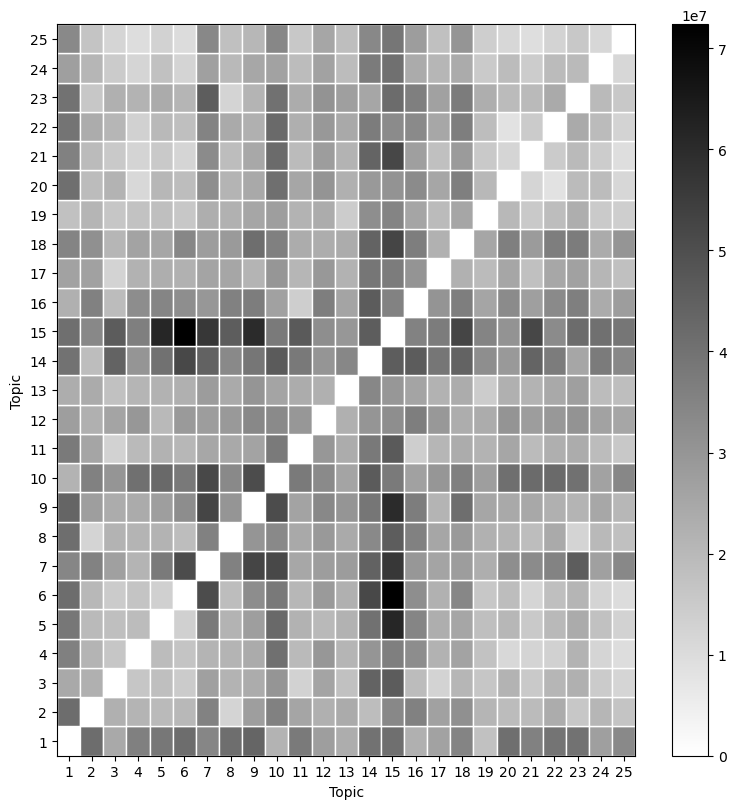}
        \caption{Across topics -- $\mathsf{DTC}(k_1|k_2)$. }
        \label{fig:KL_dissimilarity_topics_multivariate}
    \end{subfigure}
    %\caption{$\mathsf{DTC}$ across consecutive sessions or averaged within sessions for U.S. Senate speeches. }
    \caption{The dissimilarity of topical content measured by $\mathsf{DTC}$. }
    \label{fig:KL_dissimilarity}
\end{figure}

%After inspecting the topic prevalences, we focus on
To inspect topic consistency, we quantify the dissimilarity of topics for consecutive sessions as well as the dissimilarity of pairs of topics across time periods based on $\mathsf{DTC}$.
%yielded similar results as applying the cosine similarity to $\bm \phi_{hkv}^\loc$. 
Figure~\ref{fig:KL_dissimilarity} summarises the results in two heatmaps. 
Figure~\ref{fig:KL_dissimilarity_in_time} provides the heatmap for the comparison across consecutive sessions with the time periods on the $x$-axis and the topics on the $y$-axis and darker colours indicate a stronger discrepancy between the topical content of the current and the previous time period. In particular, the topical content of topic~17 (education) changes considerably in session~110 and then changes back afterwards. Reasons might only be speculated, but this could be due to the implementation of the heavily discussed America COMPETES Act in 2007. For topic~20, a change in topical content is also evident following the appointment of two new judges -- Samuel Alito and John Roberts -- to the Supreme Court. Here we do not observe a change back which coincides with both judges keeping their mandate for the next sessions. In general, the strongest changes in topical content are observed for topic~15 which 
transforms from a topic that captures rather stylistic phrases in speech into a topic about immigration. We clearly see that the main change occurs in session 109. This observation is also supported by inspecting the evolution of the top-10 terms
%(see Table~S2 in Section~S.7). 
(see Table~\ref{tab:vocabulary_evolution_frex_50_15} in Section~\ref{supsec:additional_outputs}). 

Figure~\ref{fig:KL_dissimilarity_topics_multivariate} provides the heatmap for the dissimilarity of pairs of topics across time periods with topics on both axes. Note that the range covered by the dissimilarities between topics, however, is two magnitudes larger than for the dissimilarity within a topic across consecutive sessions. This is indicative of the consistency of the topics over time as well as the distinctiveness of pairs of different topics for the same time period. Table~\ref{tab:topic_labels} suggests that topics 11 and 16 are of similar content as both are about health care, as well as topics 20 and 22 which are both about courts. This
similarity between these pairs of topics is also reflected in Figure~\ref{fig:KL_dissimilarity_topics_multivariate} where the corresponding cells are light grey shaded. 

To indicate how TPF captures differences over time in topic prevalence and topical content, we focus in the following on topic~12 which is about climate change. Obviously this topic increases in popularity over time. Its topical content evolves slowly without any dramatic changes as supported by the light grey shades indicating only slight dissimilarity in Figure~\ref{fig:KL_dissimilarity_in_time}. Table~\ref{tab:vocabulary_evolution_frex_50_12} provides the evolution of the top-10 terms over time determined based on the $\mathsf{FREX}$ measure with equal weight on frequency and exclusivity. In the 80's the topic was mainly driven by \emph{acid rain}, then \emph{clean air act} took over a more dominant position. Around 2000 \emph{foreign oil} was heavily discussed and from 2011 onwards, the \emph{keystone xl} pipeline project also began to receive significant attention. As one would expect, topic~5 about natural resources, national parks, water conservation funds, etc.\ is the closest topic in terms of $\mathsf{DTC}$, see Figure~\ref{fig:KL_dissimilarity_topics_multivariate}.

\begin{table}[t!]
    \centering
    \resizebox{\textwidth}{!}{
        \input{\bestmodel_vocabulary_evolution_frex_50_12}
    }
    \caption{Evolution of the top-10 terms selected based on $\mathsf{FREX}$ for topic 12 -- Climate change.}
    \label{tab:vocabulary_evolution_frex_50_12}
\end{table}

\begin{figure}
    \centering
     \begin{subfigure}{\textwidth}
        \includegraphics[width=\textwidth]{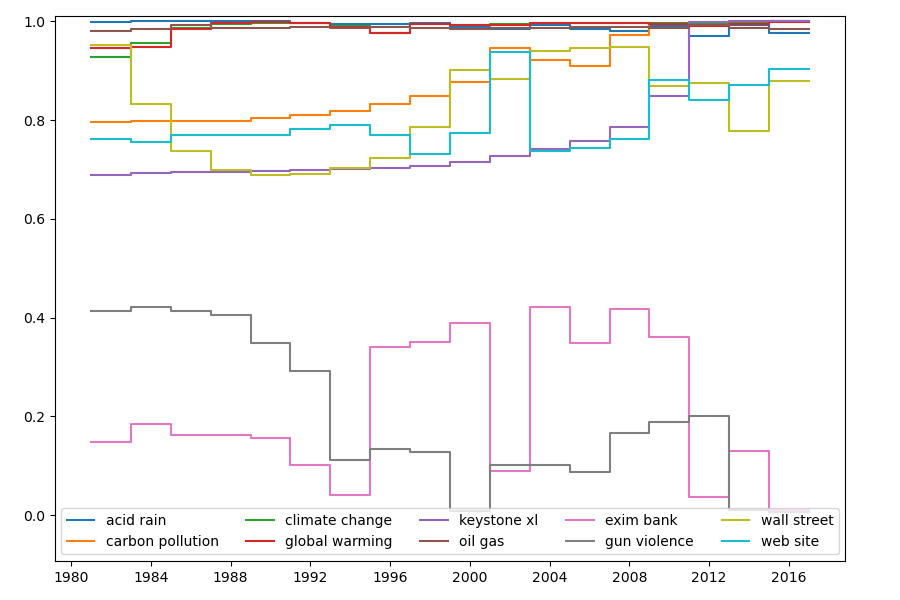}
        \caption{$\mathsf{FREX}$ measure. }
        \label{fig:words_climate_change_frex}
    \end{subfigure}
    ~
    \begin{subfigure}{\textwidth}
        \includegraphics[width=\textwidth]{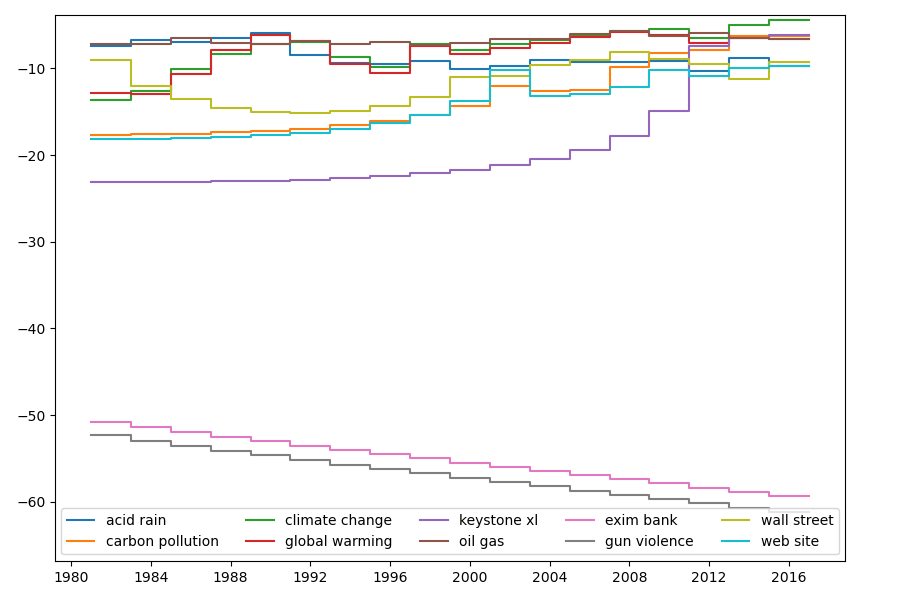}
        \caption{$\AR{1}$ sequence $\bm h_{kv}$ estimated by $\bm \phi_{hkv}^\loc$. }
        \label{fig:words_climate_change_ar}
    \end{subfigure}
   \caption{Evolution of selected 10 terms for topic 12 -- Climate change. }
    \label{fig:words_climate_change}
\end{figure}

To provide more details on how the model captures the change of the term intensities over time, we inspect the evolution of $\bm \phi_{hkv}^\loc$ and $\mathsf{FREX}$ over time for topic~12 (climate change) based on six terms which are important when discussing climate change and include for comparison four terms which have no obvious relation with climate change.  Figure~\ref{fig:words_climate_change_frex} shows that the relevant terms \emph{acid rain}, \emph{climate change}, \emph{global warming} and \emph{oil gas} are at the top with $\mathsf{FREX}$ values close to 1. The terms \emph{carbon pollution} and \emph{keystone xl} which are only relevant during the last sessions have lower values at the beginning but also reach values close to 1 at the end of the observation period. Still these two terms have $\mathsf{FREX}$ values considerably higher than zero for the first sessions, presumably because of the fitted random walk process which enforces a smooth transition to the high values in later sessions. This smooth step-by-step transition 
for terms essentially absent in earlier sessions is also clearly visible in Figure~\ref{fig:words_climate_change_ar}. The term \emph{web site} should not be present in the 80's and early 90's. However, its later partial importance for this topic forces $\bm h_{12\,\text{web site}}$ to already have higher values in these early time periods than completely insignificant terms to avoid huge steps in the sequence (and thus reduce its variability). This is also reflected in the $\mathsf{FREX}$ measure values which are around 0.8 even though this term could not have been used in that time period.

%\textcolor{red}{Should we then start the AR sequence for each bigram from the session it first appears? But would be tricky to implement.}

\begin{comment}
\begin{enumerate}
    \item 4 model comparison via VIC - choose the best (hopefully AR MVnormal) + how different are they in results and in time, convergence, point to plots (all 4 models) in the appendix
    \item outputs for the selected model, climate change topic, acid rain, climate change 
\end{enumerate}

list of outputs:
\begin{itemize}
    \item topic prevalence in time (unsorted for comparability with other models)
    \item chosen topic evolution of term usage using $\mathsf{FREX}$ measure
    \item[$\pm$] evolution of chosen word (\emph{web site}, \emph{terrorist attack} \emph{soviet union}, \ldots)
    \item topic dissimilarity between consecutive time periods 
    \item[$\pm$] dissimilarity among topics (averaged, min, max over time)
    \item[$\pm$] comparison to DPF outputs
    \item[$\pm$] Democrats vs Republicans
\end{itemize}
\end{comment}

%% file: models_VIC.tex
\begin{tabular}{lll|rrrrrr}
\toprule
\multicolumn{1}{c}{Setting}&\multicolumn{1}{c}{$\delta_{kv}$} & \multicolumn{1}{c}{$\bm \phi_{hkv}^\covm$} & \multicolumn{1}{c}{$\mathsf{ELBO}$} & \multicolumn{1}{c}{Reconstr.} & \multicolumn{1}{c}{$\mathsf{VAIC}$} & \multicolumn{1}{c}{$\mathsf{VBIC}$} & \multicolumn{1}{c}{sec/epoch} & \multicolumn{1}{c}{sec/$\mathsf{ELBO}$}\\
\midrule
(A)&$\in \R$ & general & $-$59\,168\,608 & $-$70\,818\,840 & 144\,875\,424 & 94\,996\,600 & 930.16 & 303.45\\
(B)&$\in \R$ & diagonal & $-$59\,569\,440 & $-$70\,787\,832 & 144\,816\,640 & 94\,815\,824 & 605.40 & 173.30\\
(C)&$= 1$ & general & \contour{black}{$-$58\,700\,208} & $-$70\,828\,688 & 144\,858\,736 & 100\,482\,816 & 914.34 & 302.52\\
(D)&$= 1$ & diagonal & $-$60\,096\,988 & \contour{black}{$-$70\,727\,016} & \contour{black}{144\,654\,016} & \contour{black}{89\,400\,552} & 599.50 & 172.36\\
\bottomrule
\end{tabular}

%% file: topic_labels.tex
\begin{tabular}{llllll}
\toprule
1 & Taxes, business & 9 & Commemoration & 17 & Education \\
2 & World conflicts & 10 & Budget & 18 & Environment preservation \\
3 & Children, families & 11 & Health, diseases & 19 & Finance \\
4 & Law enforcement & 12 & Climate change & 20 & Supreme court \\
5 & Natural environment & 13 & Agriculture & 21 & Civil rights \\
6 & Senate coordination & 14 & United States & 22 & Circuit courts \\
7 & Federal government & 15 & Rhetoric $\rightarrow$ Immigration & 23 & War \\
8 & Armed services & 16 & Affordable health care & 24 & Social security \\
& & & & 25 & Crime \\
\bottomrule
\end{tabular}

%% file: RW_normal_vocabulary_evolution_frex_50_12.tex
\begin{tabular}{|l|l|l|l|l|l|}
\toprule
\multicolumn{1}{c}{97: 1981--1982} & \multicolumn{1}{c}{98: 1983--1984} & \multicolumn{1}{c}{99: 1985--1986} & \multicolumn{1}{c}{100: 1987--1988} & \multicolumn{1}{c}{101: 1989--1990} & \multicolumn{1}{c}{102: 1991--1992} \\
\midrule
natural gas & natural gas & natural gas & nuclear waste & air act & natural gas\\
nuclear waste & acid rain & acid rain & acid rain & natural gas & energy policy\\
energy policy & energy security & energy policy & natural gas & acid rain & air act\\
strategic petroleum & air act & gas industry & clean coal & carbon dioxide & energy efficiency\\
air act & energy policy & energy security & energy security & energy policy & energy strategy\\
acid rain & sulfur dioxide & clean coal & air act & sulfur dioxide & carbon dioxide\\
foreign oil & foreign oil & foreign oil & energy policy & clean coal & energy security\\
energy security & gas industry & domestic oil & greenhouse effect & air pollutants & foreign oil\\
energy sources & energy sources & air act & carbon dioxide & dioxide emissions & greenhouse gases\\
alternative energy & price natural & energy independence & sulfur dioxide & foreign oil & greenhouse gas\\
\midrule
\multicolumn{1}{c}{103: 1993--1994} & \multicolumn{1}{c}{104: 1995--1996} & \multicolumn{1}{c}{105: 1997--1998} & \multicolumn{1}{c}{106: 1999--2000} & \multicolumn{1}{c}{107: 2001--2002} & \multicolumn{1}{c}{108: 2003--2004} \\
\midrule
air act & nuclear waste & foreign oil & energy policy & natural gas & energy policy\\
natural gas & natural gas & greenhouse gas & foreign oil & energy policy & foreign oil\\
energy policy & foreign oil & gas emissions & energy security & foreign oil & fuel cell\\
energy efficiency & energy policy & greenhouse gases & greenhouse gas & renewable portfolio & greenhouse gas\\
gas industry & energy security & carbon dioxide & clean coal & energy sources & comprehensive energy\\
foreign oil & energy efficiency & energy policy & energy efficiency & energy efficiency & clean coal\\
energy security & gas industry & energy efficiency & domestic oil & portfolio standard & gas emissions\\
energy sources & domestic oil & energy security & energy sources & energy supply & energy efficiency\\
domestic oil & clean air & oil percent & gas emissions & greenhouse gas & energy sources\\
wind energy & air act & reduce greenhouse & domestic energy & billion barrels & wind energy\\
\midrule
\multicolumn{1}{c}{109: 2005--2006} & \multicolumn{1}{c}{110: 2007--2008} & \multicolumn{1}{c}{111: 2009--2010} & \multicolumn{1}{c}{112: 2011--2012} & \multicolumn{1}{c}{113: 2013--2014} & \multicolumn{1}{c}{114: 2015--2016} \\
\midrule
energy policy & foreign oil & greenhouse gas & energy policy & energy efficiency & fossil fuel\\
foreign oil & energy policy & energy efficiency & foreign oil & carbon pollution & keystone xl\\
energy independence & greenhouse gas & gas emissions & domestic energy & keystone xl & energy efficiency\\
greenhouse gas & energy independence & foreign oil & energy production & xl pipeline & carbon pollution\\
energy efficiency & energy efficiency & greenhouse gases & wind energy & energy policy & energy policy\\
gas emissions & gas emissions & energy policy & carbon dioxide & greenhouse gas & xl pipeline\\
oil natural & energy sources & energy independence & keystone xl & carbon dioxide & energy security\\
energy sources & billion barrels & carbon dioxide & greenhouse gas & fossil fuels & fossil fuels\\
greenhouse gases & carbon dioxide & energy sources & energy efficiency & energy independence & greenhouse gas\\
clean coal & greenhouse gases & hydraulic fracturing & energy independence & energy savings & carbon dioxide\\
\midrule
\end{tabular}

%% file: section05_final.tex
%%%!!!!!!!!!!!!!!!!!!!!!!!!!!!!!!!!!!!!!!!!!!!!!!!!!!!!!!!!!!!!!!!!!!!!!!!%%%
%%%%%%%%%%%%%%%%%%%%%%%%%%%%%%%%%%%%%%%%%%%%%%%%%%%%%%%%%%%%%%%%%%%%%%%%%%%%%
%%%%%%%%%%%%               NEW SECTION                      %%%%%%%%%%%%%%%%%
%%%%%%%%%%%%%%%%%%%%%%%%%%%%%%%%%%%%%%%%%%%%%%%%%%%%%%%%%%%%%%%%%%%%%%%%%%%%%
%%%!!!!!!!!!!!!!!!!!!!!!!!!!!!!!!!!!!!!!!!!!!!!!!!!!!!!!!!!!!!!!!!!!!!!!!!%%%
\section{Conclusion} \label{sec:conclusion}

Our TPF model extends the PF model for high-dimensional sparse matrices of count data taking the time information of the observations into account. The time-varying term intensities of the latent topics are modelled using a non-centred auto-regressive formulation with the auto-regressive parameter either being estimated or potentially fixed to one to induce a random walk. The careful choice of priors enables the estimation using VI based on mean-field inference where CAVI and ADVI update steps are combined with batching of the observations. 

We applied the model to a dataset covering decades of intercourse in the U.S.\ Senate to investigate the change in topic prevalence as well as the consistency and variation of topical content over time. We compared results using different model specification settings where we contrast the use of the random walk with the more flexible auto-regressive specification and investigated the use of a multivariate variational family to relax the assumption of independence for the temporal sequence. 

Results justify the general preference of the random walk specification and do not provide evidence for the superiority of the more complicated approaches involving also the multivariate VI approach, which was the only case we compared to the completely independent specification. To obtain additional insights into the benefits of taking dependence into account and using a block structure, one might employ a multivariate normal variational family with tridiagonal precision matrix $\left(\bm\phi_{hkv}^\covm\right)^{-1}$ that should be flexible enough to capture the essence of the $\AR{1}$ sequence. This could represent a suitable compromise between faster evaluation and a better posterior approximation. 

This work offers several avenues for further development and extension. One potential direction is adopting a more flexible model specification. Our current framework enforces a degree of smoothness in temporal changes, which may limit its ability to capture rapidly emerging terms. To address this, model extensions—such as incorporating latent change point models—could be explored. Additionally, a time-varying prior structure on the document intensities $\theta_{dk}$
could be introduced to explicitly model their temporal evolution.

In the model evaluations and comparisons, we focused on the use of variational in-sample criteria. Alternatively, within a Bayesian context, one could also consider out-of-sample criteria based on the predictive posterior. Such an approach would require inserting missing values to create an out-of-sample data. In the context of topic models where documents are available over time, the insertion of missing values is not straightforward, but different ways are possible. One could either omit a set of time points by omitting all documents from these time points, single documents from all time points or even single cells in the DTM. This ambiguity suggests that future work could focus on a detailed analysis of using out-of-sample criteria for model evaluation in the context of the TPF model. 

In addition, future work could also rigorously investigate identifiability issues specific to the TPF model building on results provided by~\citet{ShikunOleaNesbit2024RobustMLAlgforTextAnalysis} for LDA models where they suggest a robust prior structure to ensure identifiability.

%% file: appendix_final.tex
\appendix
% \section*{Appendices}
%%%!!!!!!!!!!!!!!!!!!!!!!!!!!!!!!!!!!!!!!!!!!!!!!!!!!!!!!!!!!!!!!!!!!!!!!!%%%
%%%%%%%%%%%%%%%%%%%%%%%%%%%%%%%%%%%%%%%%%%%%%%%%%%%%%%%%%%%%%%%%%%%%%%%%%%%%%
%%%%%%%%%%%%               NEW SECTION                      %%%%%%%%%%%%%%%%%
%%%%%%%%%%%%%%%%%%%%%%%%%%%%%%%%%%%%%%%%%%%%%%%%%%%%%%%%%%%%%%%%%%%%%%%%%%%%%
%%%!!!!!!!!!!!!!!!!!!!!!!!!!!!!!!!!!!!!!!!!!!!!!!!!!!!!!!!!!!!!!!!!!!!!!!!%%%
%%%!!!!!!!!!!!!!!!!!!!!!!!!!!!!!!!!!!!!!!!!!!!!!!!!!!!!!!!!!!!!!!!!!!!!!!!%%%
%%%%%%%%%%%%%%%%%%%%%%%%%%%%%%%%%%%%%%%%%%%%%%%%%%%%%%%%%%%%%%%%%%%%%%%%%%%%%
%%%%%%%%%%%%               NEW SECTION                      %%%%%%%%%%%%%%%%%
%%%%%%%%%%%%%%%%%%%%%%%%%%%%%%%%%%%%%%%%%%%%%%%%%%%%%%%%%%%%%%%%%%%%%%%%%%%%%
%%%!!!!!!!!!!!!!!!!!!!!!!!!!!!!!!!!!!!!!!!!!!!!!!!!!!!!!!!!!!!!!!!!!!!!!!!%%%
\section{CAVI Updates} \label{appsec:CAVI}

In this section, we provide details on all CAVI updates. The updated variational parameters are denoted by $\widehat{\bm \phi}_\bullet$, where $\bullet$ is a placeholder for the parameter name and necessary indices depending on context. The only variational parameters, where no CAVI updates are available, are the variational parameters $\bm \phi_{hkv}^\loc$ and $\bm \phi_{hkv}^\covm$ for $\bm h_{kv}$. This is due to the inclusion of the corresponding parameters in  the Poisson rates after exponentiation.

\citet{BleiVI:2017} provide a general approach to find CAVI updates, which we follow. In addition, some of our updates are inspired by~\citet{gopalanhofmanblei2015scalableHPoisF} who introduce auxiliary counts $\widetilde{y}_{dkv}$ corresponding to latent variables and summing up to $y_{dv}$ over topics. The multinomial distribution is used as variational distribution for these counts with the variational parameters (proportions) $\phi_{dkv}^y$ summing up to one over topics.   
%In general, the most straightforward way to obtain CAVI updates is to differentiate~\eqref{eq:ELBO_general} with respect to the current block of variational parameters and find the optimal update by solving the system of estimating equations constructed by zeroing the gradients.
More details can be found in 
%Section~S.2. 
Section~\ref{supsec:auxiliary_variables}.
We update them as follows:
\begin{equation} %\label{eq:cavi_aux_prop}
\widehat{\phi}_{dkv}^y \propto
\exp \left\{
\psi(\phi_{\theta dk}^\shp) - \log\phi_{\theta dk}^\rte
+ \phi_{hkv,t_d}^\loc
\right\}
\end{equation}
where $\psi()$ denotes the digamma function.
Updates of the variational parameters for $\bm \theta$ and related auxiliary variables resemble the ones by~\citet{gopalanhofmanblei2015scalableHPoisF}:
\begin{align*}
    \widehat{\phi}_{\theta dk}^\shp &= a^\theta + \sum\limits_{v=1}^V y_{dv} \phi_{dkv}^y
	&&\text{and}& \quad
	\widehat{\phi}_{\theta dk}^\rte &= \dfrac{\phi_{\xi d}^\shp}{\phi_{\xi d}^\rte}
	+ \sum\limits_{v=1}^V \exp\left\{ \phi_{hkv,t_d}^\loc + \frac{1}{2}\phi_{hkv,t_d}^\sclsq\right\},
    \\
	\widehat{\phi}_{\xi d}^\shp &= a^{\xi} + K a^\theta 
	&&\text{and}& \quad
	\widehat{\phi}_{\xi d}^\rte &= b^{\xi} + \sum\limits_{k=1}^K \dfrac{\phi_{\theta dk}^\shp}{\phi_{\theta dk}^\rte}.
\end{align*}
Under a batching mechanism, we update only the (local) parameters for documents $d$ in the current batch. 

We continue with the updates for precision parameters $\tau_{kv}$. The variational shape parameters have a constant update that can be set from the very beginning:
$
\widehat{\phi}_{\tau kv }^\shp = a^{\tau} + \frac{T}{2}.
$
However, the variational rate parameters have to be updated with every iteration of the algorithm:
\begin{multline*}
    \widehat{\phi}_{\tau kv }^\rte = b^{\tau} + \frac{1}{2} 
        %\left[
        %    \Eq \T{\left(\bm h_{kv} - \ones \mu_{kv}\right)} \Delta_T\left(\delta_{kv}\right) \left(\bm h_{kv} - \ones \mu_{kv}\right)
        %\right].
        \left[
            \left(1+ \Eq \left(\delta_{kv}\right)^2\right) \sum \limits_{t=1}^{T-1}  \Eq \left(h_{kv,t} - \mu_{kv}\right)^2
            + \Eq \left(h_{kv,T} - \mu_{kv}\right)^2
            - \right. \\ \left.-   
            2\phi_{\delta kv }^\loc \sum\limits_{t=2}^{T} \Eq \left(h_{kv,t}^{} - \mu_{kv}\right) \left(h_{kv,t-1}^{} - \mu_{kv}\right)
        \right],
\end{multline*}
where 
$\Eq \left(\delta_{kv}\right)^2 = 
   \left(\phi_{\delta kv}^\loc\right)^2 + \phi_{\delta kv}^\sclsq$.
The updates for the auto-regressive coefficients $\delta_{kv}$ are obtainable as they appear only in combinable quadratic forms. In particular,
\begin{align*}
\widehat{\phi}_{\delta kv }^\sclsq &= 
%\dfrac{1}{
\left[
    \dfrac{1}{\left(\sigma^{\delta}\right)^2} 
    + \dfrac{\phi_{\tau kv }^\shp}{\phi_{\tau kv }^\rte}
    \sum\limits_{t=1}^{T-1} \Eq \left(h_{kv,t} - \mu_{kv} \right)^2
    %\Tr^0 \left( \Eq \left(\bm h_{kv} - \ones \mu_{kv}\right) \T{\left(\bm h_{kv} - \ones \mu_{kv}\right)} \right)
\right]^{-1}
    %}
    ,
\\
\widehat{\phi}_{\delta kv }^\loc &= 
    \widehat{\phi}_{\delta kv }^\sclsq
    \cdot
    \left[
        \dfrac{\mu^{\delta}}{\left(\sigma^{\delta}\right)^2} 
        + \dfrac{\phi_{\tau kv }^\shp}{\phi_{\tau kv }^\rte}
        \sum\limits_{t=2}^T \Eq \left(h_{kv,t} - \mu_{kv} \right) \left(h_{kv,t-1} - \mu_{kv} \right)
        %\Tr^1 \left( \Eq \left(\bm h_{kv} - \ones \mu_{kv}\right) \T{\left(\bm h_{kv} - \ones \mu_{kv}\right)} \right)
    \right],
\end{align*}
\begin{comment}
where analogously to Lemma~\ref{lemma:E_quadratic}
$$
\Eq \left(\bm h_{kv} - \ones \mu_{kv}\right) \T{\left(\bm h_{kv} - \ones \mu_{kv} \right)}
= 
\bm \phi_{hkv}^\covm + \left(\bm \phi_{hkv}^\loc - \ones \phi_{\mu kv}^\loc \right) \T{\left(\bm \phi_{hkv}^\loc - \ones \phi_{\mu kv}^\loc \right)} + \phi_{\mu kv}^\sclsq \ones \T{\ones}.
$$
Note that these updates would hold even if truncated normal distribution had been used for both prior and variational distribution. 
\end{comment}
where (stripped of indices) we abbreviate
\begin{align*}
\Eq \left(h_t - \mu\right)^2 &= \phi_{h,t}^\sclsq + (\phi_{h,t}^\loc - \phi_{\mu}^\loc)^2 + \phi_{\mu}^\sclsq, 
\\
\Eq \left(h_t - \mu\right)\left(h_{t-1} - \mu\right) &= \phi_{h,t,t-1}^\covm + (\phi_{h,t}^\loc - \phi_{\mu}^\loc)(\phi_{h,t-1}^\loc - \phi_{\mu}^\loc) + \phi_{\mu}^\sclsq.
\end{align*}

The updates for the auto-regressive means $\mu_{kv}$ have the following form:
\begin{align*}
\widehat{\phi}_{\mu kv }^\sclsq &= 
\left[
%\dfrac{1}{
    \dfrac{1}{\left(\sigma^{\mu}\right)^2} 
    + \dfrac{\phi_{\tau kv }^\shp}{\phi_{\tau kv }^\rte}
    %\; \T{\ones} \Eq \Delta_T\left(\delta_{kv}\right) \ones
    \left[ 
        1 + (T-1)(1 - 2\phi_{\delta kv }^\loc + \Eq \left(\delta_{kv} \right)^2)
    \right]
\right]^{-1}
    %}
    ,
\\
\widehat{\phi}_{\mu kv }^\loc &= 
    \widehat{\phi}_{\mu kv }^\sclsq
    \cdot
    \left(
        \dfrac{\mu^{\mu}}{\left(\sigma^{\mu}\right)^2} 
        + \dfrac{\phi_{\tau kv }^\shp}{\phi_{\tau kv }^\rte}
        %\; \T{\ones} \Eq \Delta_T\left(\delta_{kv}\right) \bm \phi_{hkv}^\loc
        \; \widetilde{\mu}
    \right),  
\end{align*}
where
$$
\widetilde{\mu} = 
\left(1+\Eq \left(\delta_{kv} \right)^2 \right) 
\sum\limits_{t=1}^{T-1}\phi_{hkv,t}^\loc 
+ \phi_{hkv,T}^\loc 
- \phi_{\delta kv }^\loc \sum\limits_{t=2}^{T} \left(\phi_{hkv,t-1}^\loc + \phi_{hkv,t}^\loc\right).
$$

%% file: supplement01_TPF_summary.tex
%%%!!!!!!!!!!!!!!!!!!!!!!!!!!!!!!!!!!!!!!!!!!!!!!!!!!!!!!!!!!!!!!!!!!!!!!!%%%
%%%%%%%%%%%%%%%%%%%%%%%%%%%%%%%%%%%%%%%%%%%%%%%%%%%%%%%%%%%%%%%%%%%%%%%%%%%%%
%%%%%%%%%%%%               NEW SECTION                      %%%%%%%%%%%%%%%%%
%%%%%%%%%%%%%%%%%%%%%%%%%%%%%%%%%%%%%%%%%%%%%%%%%%%%%%%%%%%%%%%%%%%%%%%%%%%%%
%%%!!!!!!!!!!!!!!!!!!!!!!!!!!!!!!!!!!!!!!!!!!!!!!!!!!!!!!!!!!!!!!!!!!!!!!!%%%
%\section{TPF Model Specification} \label{supsec:tpf}
\section{TPF model specification} \label{supsec:tpf}
\subsection{Notation} \label{supsubsec:tpf_summary}

\begin{tabular}{rlrl}
    $d$ & index of a document & $D$ & number of documents \\
    $v$ & index of a term (word) & $V$ & number of terms (vocabulary size) \\
    $k$ & index of a topic & $K$ & number of topics \\
    $t$ & index of a time period & $T$ & number of time periods \\
    $t_d$ & time period of document~$d$ & $\mathcal{D}_t$ & document indices in time period $t$ \\
    $a$ & index of an author & $A$ & number of authors \\
    $y_{dv}$ & term~$v$ counts in document $d$ & $\mathbb{Y}$ & DTM \\
    $\widetilde{y}_{dkv}$ & latent $dv$ counts for topic $k$& $\widetilde{\mathbb{Y}}$ & latent topic-specific DTMs \\
    $\knorm{T}{\bm \mu}{\bm \Sigma}$ & \multicolumn{3}{l}{$T$-variate normal distribution with mean $\bm \mu$ and covariance matrix $\bm \Sigma$} \\
    $\gammadist{a}{b}$ & \multicolumn{3}{l}{gamma distribution with shape~$a$ and rate~$b$} \\
    $\Pois{\lambda}$ & \multicolumn{3}{l}{Poisson distribution with rate $\lambda$}\\
    $\Mult{K}{T}{\bm \pi}$&\multicolumn{3}{l}{$K$-dimensional multinomial distribution with number of trials $T$}\\
    &\multicolumn{3}{l}{and success probabilities $\bm \pi$}
\end{tabular}

\subsection{Hierarchical Structure} \label{supsubsec:tpf_hierarchy}

The TPF model assumes the following data generating process for the document-term matrix $\mathbb{Y}$:
\begin{equation} \label{supeq:tvpf_rates}
y_{dv} \sim \Pois{\lambda_{dv}}
\qquad \text{with rate} \qquad 
\lambda_{dv} 
= \sum \limits_{k=1}^K \lambda_{dkv}
= \sum \limits_{k=1}^K \theta_{dk} \exp 
\left\{ 
    h_{kv,t_d}
\right\}.
\end{equation}
We assume the following hierarchical prior for the document intensities $\theta_{dk}$:
\begin{equation} \label{supeq:tvpf_prior_theta}
    \theta_{dk} \sim \gammadist{a^\theta}{\xi_d}
    \qquad \text{and} \qquad
    \xi_d \sim \gammadist{a^{\xi}}{b^{\xi}}
\end{equation}
with document-specific rates $\xi_d$ \citep{gopalanhofmanblei2015scalableHPoisF}. 

We use a~shifted auto-regressive prior for $h_{kv,t}$, parameterised in the following way:
\begin{equation}
    \left. h_{t}- \mu \;\right|\; h_{t-1} \sim \norm{\delta
    \left(h_{t-1} - \mu\right)}{\tau^{-1}}
    \quad \text{and} \quad 
    h_{1} - \mu \sim \norm{0}{\tau^{-1}},
\end{equation}
where we drop for clarity all other indices except time. The sequence is centred around the mean $\mu$. The centred sequence then follows an $\AR{1}$ process where $\delta$ is the auto-regressive coefficient. The precision $\tau$ controls the variability of the innovations added to the centred sequence including the first innovation from an artificially introduced observation $h_0 - \mu = 0$. 

We write this prior in vectorised form using:
\begin{equation} \label{supeq:tvpf_prior_ar_kv}
    \bm h \sim \sAR{T}{\mu}{\delta}{\tau} 
    \Longleftrightarrow
    \bm h \sim \knorm{T}{\mu \ones}{
    \tau^{-1}
    \Delta_T(\delta)^{-1}
    },
\end{equation}
where the precision matrix $\Delta_T(\delta)$ is a tridiagonal matrix:
\begin{equation} \label{supeq:Delta_matrix}
    \Delta_T(\delta)
    %=
    %\mathbb{I}_T
    %+ \delta^2 \mathbb{I}_{T,0}
    %- \delta \left(\mathbb{I}_{T,-1} + \mathbb{I}_{T,1}\right)
    =
    \begin{pmatrix}
    1+\delta^2 & - \delta & 0 & \ldots & 0 & 0\\
    -\delta & 1+\delta^2 & - \delta & \ldots & 0 & 0\\
    0 & -\delta & 1+\delta^2 & \ldots & 0 & 0\\
    \vdots & \vdots & \vdots & \vdots & \vdots & \vdots \\
    0 & 0 & 0 & \ldots & 1+\delta^2 & -\delta \\
    0 & 0 & 0 & \ldots & -\delta & \textcolor{red}{1} 
    \end{pmatrix}.
\end{equation}
Including the indices, we assume $\bm h_{kv} \sim \sAR{T}{\mu_{kv}}{\delta_{kv}}{\tau_{kv}}$.

For the parameters of the auto-regressive sequence, we assume
\begin{equation} \label{supeq:mu_delta_tau_prior}
\mu_{kv} \sim \norm{\mu^{\mu}}{\left(\sigma^{\mu}\right)^2}
, \quad 
\delta_{kv} \sim \norm{\mu^{\delta}}{\left(\sigma^{\delta}\right)^2}
, \quad
\tau_{kv} \sim \gammadist{a^{\tau}}{b^{\tau}}.
\end{equation}
Moreover, in the simulation study and the empirical application, we also consider a deterministic distribution for $\delta_{kv}$ ($\delta_{kv} = 1$, the random walk), and we include in the simulation study a truncated normal distribution where the support is restricted to the interval $[-1, 1]$ (to ensure stationarity).

%% file: supplement02_auxiliary_counts.tex
\section{Latent variables} \label{supsec:auxiliary_variables}

Under \eqref{supeq:tvpf_rates}, the contribution $\Eq \log p(\mathbb{Y} | \bm \zeta)$ to $\ELBO{\bm \phi}$ is given by
$$
\Eq \log p(\mathbb{Y} | \bm \zeta) 
=
\sum\limits_{d=1}^D  \sum\limits_{v=1}^V
    \left[
        y_{dv} \Eq\log \left(\sum\limits_{k=1}^K \lambda_{dkv}\right)
        - \sum\limits_{k=1}^K \Eq \lambda_{dkv} 
        - \log \left(y_{dv}!\right)
    \right],
$$
which is difficult to evaluate due to $\Eq\log \sum\limits_{k=1}^K$. 

We follow \citet{gopalanhofmanblei2015scalableHPoisF} to overcome this issue and introduce latent variables by decomposing the word counts $y_{dv}$ into auxiliary individual contributions to each of the topics $\widetilde{y}_{dkv}$. This trick eliminates the sum over the topics in the Poisson rates $\lambda_{dv} = \sum\limits_{k=1}^K \lambda_{dkv}$ and leads to updates in closed form. 
In particular, this corresponds to assume the following data generating process for the latent variables:
\begin{equation} \label{supeq:Pois_auxiliary}
    \widetilde{y}_{dkv} | \bm \zeta \sim \Pois{\lambda_{dkv}}
    \quad \text{with} \quad 
    y_{dv} = \sum\limits_{k=1}^K \widetilde{y}_{dkv}.
\end{equation}
%This implies that the distribution $\mathbb{Y} | \bm \zeta$ is decomposed into a deterministic component $\mathbb{Y} | \widetilde{\mathbb{Y}}$ and the additional component $\widetilde{\mathbb{Y}} | \bm \zeta$.  
$\widetilde{\mathbb{Y}} = (\widetilde{y}_{dkv})_{d,k,v=1}^{D,K,V}$ represents new latent variables, for which a variational family is set up. For $y_{dv} > 0$, the multinomial distribution $\Mult{K}{y_{dv}}{\bm \phi_{dv}^y}$ is a suitable option for $\widetilde{\bm y}_{dv}$ where the variational parameters $0 < \phi_{dkv}^y \leq 1$ satisfy $\phi_{d1v}^y + \cdots + \phi_{dKv}^y = 1$. Under $y_{dv} = 0$, we immediately obtain $\widetilde{y}_{dkv} = 0$.

Auxiliary counts contribute to $\ELBO{\bm \phi}$ with
$$
\Eq \log p(\mathbb{Y} | \widetilde{\mathbb{Y}}) + \Eq \log p(\widetilde{\mathbb{Y}} | \bm \zeta) - \Eq \log q_{\bm \phi}(\widetilde{\mathbb{Y}}).
$$
Notice that $\mathbb{Y} | \widetilde{\mathbb{Y}}$ is deterministic, hence, its contribution to $\mathsf{ELBO}(\bm \phi)$ is zero. The other two combine into
\begin{align} 
    \Eq &\log p(\widetilde{\mathbb{Y}} | \bm \zeta) - \Eq \log q_{\bm \phi}(\widetilde{\mathbb{Y}})
    =\nonumber \\
    &\sum\limits_{d=1}^D  \sum\limits_{v=1}^V
    \left[
        y_{dv} \sum\limits_{k=1}^K \phi_{dkv}^y \left(\Eq\log \lambda_{dkv} - \log \phi_{dkv}^y \right)
        - \sum\limits_{k=1}^K \Eq \lambda_{dkv} 
        - \log \left(y_{dv}!\right)
    \right] \label{supeq:reconstruction_tilde}
    \\ 
    &\overset{(A.1)}{=} 
    \sum\limits_{d=1}^D  \sum\limits_{v=1}^V
    \left[
        y_{dv} 
        \left[
            c_{dv} +
            \log \left(\sum\limits_{k=1}^K \exp \left\{ \Eq \log \lambda_{dkv} - c_{dv} \right\}\right)
        \right]
        - \right.\nonumber\\        &\left.\qquad\qquad\qquad\qquad\qquad\qquad\qquad\qquad\qquad\qquad\qquad
        \sum\limits_{k=1}^K \Eq \lambda_{dkv} 
        - \log \left(y_{dv}!\right)
    \right], \nonumber
\end{align}
where we inserted the CAVI updates $\widehat{\bm \phi}^y$ for $\bm \phi^y$ (A.1) to simplify the computation and a~constant $c_{dv}$ to ensure numerical stability. In particular, we use $c_{dv} = \max\limits_{k=1, \ldots, K} \Eq \log \lambda_{dkv}$. Inserting the moments of the gamma and the log-normal distribution, we obtain
\begin{align*}
\Eq \log \lambda_{dkv} 
&= 
\psi(\phi_{\theta dk}^\shp) - \log(\phi_{\theta dk}^\rte) + \phi_{hkv, t_d}^\loc,
\\
\Eq \lambda_{dkv} 
&= 
\dfrac{\phi_{\theta dk}^\shp}{\phi_{\theta dk}^\rte} 
\exp\left\{ \phi_{hkv,t_d}^\loc + \frac{1}{2}\phi_{hkv,t_d}^\sclsq\right\}.
\end{align*}

%% file: supplement03_Eq_quadratic.tex
%%%!!!!!!!!!!!!!!!!!!!!!!!!!!!!!!!!!!!!!!!!!!!!!!!!!!!!!!!!!!!!!!!!!!!!!!!%%%
%%%%%%%%%%%%%%%%%%%%%%%%%%%%%%%%%%%%%%%%%%%%%%%%%%%%%%%%%%%%%%%%%%%%%%%%%%%%%
%%%%%%%%%%%%               NEW SECTION                      %%%%%%%%%%%%%%%%%
%%%%%%%%%%%%%%%%%%%%%%%%%%%%%%%%%%%%%%%%%%%%%%%%%%%%%%%%%%%%%%%%%%%%%%%%%%%%%
%%%!!!!!!!!!!!!!!!!!!!!!!!!!!!!!!!!!!!!!!!!!!!!!!!!!!!!!!!!!!!!!!!!!!!!!!!%%%
%\section{Variational Mean of a~Quadratic Form} \label{supsec:Eq_quadratic_form}
\section{Variational mean of a~quadratic form} \label{supsec:Eq_quadratic_form}

Variational means of gamma and log-normal families are trivial to obtain. However, to evaluate the ELBO efficiently or to compute CAVI updates we also need the variational means of quadratic forms which may not be obvious to obtain at first sight.
For scalar random variables, the result is trivial using Lemma~\ref{lemma:E_square}. 
For a~quadratic form involving random vectors and a random matrix, the result is obtained using Lemma~\ref{lemma:E_quadratic}. 
\begin{lemma} \label{lemma:E_square}
    Let the scalar random variables $X$ and $Y$ be independent and $\E X^2$, $\E Y^2 < \infty$, then $\E \left(X - Y\right)^2 = \var X + \left(\E X - \E Y\right)^2 + \var Y$.
\end{lemma}
\begin{lemma} \label{lemma:E_quadratic}
    Let the random vectors $\bm X$ and $\bm Y$ have finite second moments and let the random matrix $\Delta$ have mean $\E \Delta$. Then, under the independence of $\bm X$, $\bm Y$ and $\Delta$, the quadratic form $\E \T{(\bm X - \bm Y)} \Delta (\bm X - \bm Y)$ can be evaluated as
    \begin{equation*}
        \Tr \left(\E \Delta \; \cov \bm X\right)
        + 
        \T{\left(\E \bm X - \E \bm Y\right)} \E \Delta \; \left(\E \bm X - \E \bm Y\right)
        +
        \Tr \left(\E \Delta \; \cov \bm Y \right).
    \end{equation*}
\end{lemma}
The proofs of both lemmas are straightforward to derive using some basic statistics and probability results involving matrices. 

Using Lemma~\ref{lemma:E_quadratic} we derive the contribution of $\bm h$ to the \emph{log-prior} of $\ELBO{\bm \phi}$ under the multivariate normal variational family. Focusing on a single pair of $k$ and $v$ we obtain from~\eqref{supeq:tvpf_prior_ar_kv}: 
\begin{multline} \label{supeq:E_quadratic_h_mu}
    \Eq \T{\left(\bm h - \mu\ones\right)} \Delta_T\left(\delta\right) \left(\bm h - \mu\ones\right)
=
    \Tr \left(\Eq \Delta_T(\delta) \; \bm \phi_h^\covm\right)
    + \\
    \T{\left(\bm \phi_h^\loc - \ones \phi_\mu^\loc\right)} \Eq \Delta_T(\delta) \; \left(\bm \phi_h^\loc - \ones \phi_\mu^\loc\right)
    +
    \phi_\mu^\sclsq \T{\ones} \Eq \Delta_T(\delta) \ones,
\end{multline}
where we drop indices $k,v$ for ease of notation. 

Matrix $\Delta_T(\delta)$ is tridiagonal~\eqref{supeq:Delta_matrix} and allows for the following decomposition:
$$
\Delta_T\left(\delta\right) 
=
\mathbb{I}_T
+ \delta^2 \mathbb{I}_{T,0}
- \delta \left(\mathbb{I}_{T,-1} + \mathbb{I}_{T,1}\right),
$$
where $\mathbb{I}_T$ is the identity matrix, $\mathbb{I}_{T,0} = \diagzero(1, \ldots, 1) = \diag(1, \ldots, 1, 0)$ is the identity matrix where the last element is zero, $\mathbb{I}_{T,-1} = \subdiag(1, \ldots, 1)$ is the subdiagonal and $\mathbb{I}_{T,1} = \superdiag(1, \ldots, 1)$ is the superdiagonal.
Then, $\Eq \Delta_T\left(\delta\right)$ is also a tridiagonal matrix with analogous decomposition:
$$
\Eq \Delta_T\left(\delta\right) 
=
\mathbb{I}_T
+ \left[ \left(\phi_\delta^\loc\right)^2 + \phi_\delta^\sclsq\right] \mathbb{I}_{T,0}
- \phi_\delta^\loc \left(\mathbb{I}_{T,-1} + \mathbb{I}_{T,1}\right),
$$
%$(1 + \Eq \delta^2, \ldots, 1+\Eq \delta^2, 1)$ on the main diagonal and $-\phi_\delta^\loc$ all around it. 
where by Lemma~\ref{lemma:E_square}, $\Eq \delta^2 = \left(\phi_\delta^\loc\right)^2 + \phi_\delta^\sclsq$. 
If we denote by $\Tr^k (A) := \Tr(\mathbb{I}_{T,k} A)$, $k \in \{-1, 0, 1\}$, i.e., either the sum of the subdiagonal elements, the diagonal elements except the last element or the superdiagonal elements of the $T\times T$ matrix $A$, then we can abbreviate this as:
$$
\Tr \left( \Eq \Delta_T\left(\delta\right) A \right) 
= 
\Tr (A) 
+ \left[ \left(\phi_\delta^\loc\right)^2 + \phi_\delta^\sclsq\right] \Tr^0 (A)
- \phi_\delta^\loc \left(\Tr^{-1}(A) + \Tr^1(A)\right).
$$
For a~symmetric matrix $A$, e.g. $\bm\phi_h^\covm$, $\ones \T{\ones}$, $\ldots$, this reduces to 
$$
\Tr \left( \Eq \Delta_T\left(\delta\right) A \right) 
= 
\Tr (A) 
+ \left[ \left(\phi_\delta^\loc\right)^2 + \phi_\delta^\sclsq\right] \Tr^0 (A)
- 2\phi_\delta^\loc \Tr^1(A),
$$
which helps us evaluate~\eqref{supeq:E_quadratic_h_mu} more efficiently.

The results hold even under the random walk setting ($\delta = 1$), where the variational family is deterministic with $\phi_\delta^\loc = 1$ and $\phi_\delta^\sclsq = 0$.

%% file: supplement04_ELBO.tex
%%%!!!!!!!!!!!!!!!!!!!!!!!!!!!!!!!!!!!!!!!!!!!!!!!!!!!!!!!!!!!!!!!!!!!!!!!%%%
%%%%%%%%%%%%%%%%%%%%%%%%%%%%%%%%%%%%%%%%%%%%%%%%%%%%%%%%%%%%%%%%%%%%%%%%%%%%%
%%%%%%%%%%%%               NEW SECTION                      %%%%%%%%%%%%%%%%%
%%%%%%%%%%%%%%%%%%%%%%%%%%%%%%%%%%%%%%%%%%%%%%%%%%%%%%%%%%%%%%%%%%%%%%%%%%%%%
%%%!!!!!!!!!!!!!!!!!!!!!!!!!!!!!!!!!!!!!!!!!!!!!!!!!!!!!!!!!!!!!!!!!!!!!!!%%%
\section{Evaluating the ELBO} \label{supsec:ELBO}

Inference is based on $\ELBO{\bm \phi}$ 
%\begin{equation} \label{supeq:ELBO_general}
%\mathsf{ELBO}(\bm \phi) = \Eq \left[ \log p(\mathbb{Y} | \bm \zeta) + \log p (\bm \zeta) - \log q_{\bm \phi} (\bm \zeta) \right],
%\end{equation}
where the components can be decomposed into:
\begin{multline} \label{supeq:ELBO_coloured}
    \ELBO{\bm \phi} 
    = 
    \underbrace{{\color{red} 
    \Eq \log p(\mathbb{Y} | \widetilde{\mathbb{Y}}) + 
    \Eq \log p(\widetilde{\mathbb{Y}} | \bm \zeta) - 
    \Eq \log q_{\bm \phi}(\widetilde{\mathbb{Y}})}}_{\text{reconstruction}}
    + \\ +
    \underbrace{{\color{blue} \Eq \log p (\bm \zeta)}}_{\text{log-prior}}
    + 
    \underbrace{{\color{olive} (-\Eq \log q_{\bm \phi} (\bm \zeta))}}_{\text{entropy}}.
\end{multline}
The sum of the first three terms is referred to as \emph{reconstruction} (red). The remaining terms are called \emph{log-prior} (blue) and \emph{entropy} (olive). 
%In Section~\ref{supsec:auxiliary_variables}, we introduced auxiliary variables $\widetilde{\mathbb{Y}}$ that make the reconstruction term technically zero. However, we decided to call by \emph{reconstruction} also the contribution of auxiliary counts and its entropy, equation~\eqref{supeq:reconstruction_tilde}.

For evaluating $\ELBO{\bm \phi}$ the entropies of the variational families are required. In the following we exemplify them for each parametric family for one set of parameters:
\begin{itemize}
    \item $\gammadist{\phi_{\theta dk}^\shp}{\phi_{\theta dk}^\rte}$:
	\begin{align*}
	-\Eq \log q_{\bm \phi} (\theta_{dk}) &=\left(1-\phi_{\theta dk}^\shp\right)
	    %\left(
	    \psi(\phi_{\theta dk}^\shp) - \log \phi_{\theta dk}^\rte 
	    %\right)
	    +
	    \phi_{\theta dk}^\shp + \log \Gamma \left(\phi_{\theta dk}^\shp\right) ;
	\end{align*}
	\item $\norm{\phi_{\delta kv}^\loc}{\phi_{\delta kv}^\sclsq}$:
	\begin{align*}
	-\Eq \log q_{\bm \phi} (\delta_{kv}) &=    \log \phi_{\delta kv}^\scl + \frac{1}{2}\log(2\pi e);
	\end{align*}
	\item $\knorm{T}{\bm\phi_{h kv}^\loc}{\bm\phi_{h kv}^\covm}$:
	\begin{align*}
	    -\Eq \log q_{\bm \phi} (\bm h_{kv}) &=\frac{1}{2} \log \left| \bm \phi_{h kv}^\covm \right| + \frac{T}{2}\log(2\pi e),
	\end{align*}
\end{itemize}
where $\psi(\cdot)$ is the digamma function.
\begin{comment}
	\item $\Mult{K}{y_{dv}}{\bm \phi_{dv}^y}$:
	\begin{multline*}
		-\Eq \log q_{\bm \phi} (\widetilde{\bm y}_{dv}) =
		-\log(y_{dv}!) - y_{dv} \sum_{k=1}^K \phi_{dkv}^y \log(\phi_{dkv}^y) +\\
		\sum_{k=1}^K \sum_{x_k=0}^{y_{dv}}
		\binom{y_{dv}}{x_k} \left(\phi_{dkv}^y\right)^{x_k}
		\left(1-\phi_{dkv}^y\right)^{y_{dv}-x_k}\log(x_k!).
	\end{multline*}
\end{comment}

We now present the full evaluation of the $\ELBO{\bm \phi}$. To easily see where some of the CAVI updates come from, we arrange all contributions in the following way:
\begin{align} 
\mathsf{EL}&\mathsf{BO}\left(\bm \phi\right) = 
    {\color{red} \mathsf{reconstruction}} 
    + 
    {\color{blue} \mathsf{log\mbox{-}prior}} 
    + 
    {\color{olive} \mathsf{entropy}}
    = 
    {\color{red}
        - \sum\limits_{d=1}^D \sum\limits_{k=1}^K \sum\limits_{v=1}^V 
        y_{dv} \phi_{dkv}^y \log \phi_{dkv}^y
    }
    + \nonumber\\ &+
    \sum\limits_{d=1}^D \sum\limits_{k=1}^K
    \left\{
        {\color{violet} \left(\psi(\phi_{\theta dk}^\shp) - \log\phi_{\theta dk}^\rte\right) }
        \left[
            {\color{blue}a^\theta} 
            + {\color{red}\sum\limits_{v=1}^V y_{dv} \phi_{dkv}^y}
            - {\color{olive} \phi_{\theta dk}^\shp}
        \right]
        %+ {\color{olive}\phi_{\theta dk}^\shp - \phi_{\theta dk}^\shp \log\left(\phi_{\theta dk}^\rte\right) + \log \Gamma \left(\phi_{\theta dk}^\shp\right)}
        -
        {\color{blue}
            \dfrac{\phi_{\xi d}^\shp}{\phi_{\xi d}^\rte}
            \dfrac{\phi_{\theta dk}^\shp}{\phi_{\theta dk}^\rte}
        }
    \right\}
    + \nonumber\\ &+
    \sum\limits_{d=1}^D \sum\limits_{k=1}^K
    \left[
        {\color{olive}
        %\left(\psi(\phi_{\theta dk}^\shp) - \log\phi_{\theta dk}^\rte\right) +
        \phi_{\theta dk}^\shp + \log \Gamma \left(\phi_{\theta dk}^\shp\right) - \phi_{\theta dk}^\shp \log\left(\phi_{\theta dk}^\rte\right)}
    \right]
    + \nonumber\\ &+
    \sum\limits_{d=1}^D 
    \left\{
        {\color{blue} \left(\psi(\phi_{\xi d}^\shp) - \log\phi_{\xi d}^\rte\right)}
        \left[
            {\color{blue} a^{\xi} + a^\theta K}
            - {\color{olive} \phi_{\xi d}^\shp}
        \right]
        %+ 
        %{\color{olive}
        %    \phi_{\xi d}^\shp - \phi_{\xi d}^\shp \log\left(\phi_{\xi d}^\rte\right) + \log \Gamma \left(\phi_{\xi d}^\shp\right)
        %}
        -
        {\color{blue} b^{\xi}\dfrac{\phi_{\xi d}^\shp}{\phi_{\xi d}^\rte} }
    \right\}
    + \nonumber\\ &+
    \sum\limits_{d=1}^D \left[
    {\color{olive}
        % \left(\psi(\phi_{\xi d}^\shp) - \log\phi_{\xi d}^\rte\right) +
        \phi_{\xi d}^\shp + \log \Gamma \left(\phi_{\xi d}^\shp\right) - \phi_{\xi d}^\shp \log\left(\phi_{\xi d}^\rte\right)
    }
    \right]
    + \nonumber\\ &+
    \sum\limits_{k=1}^K \sum\limits_{v=1}^V \left\{
    {\color{blue} \left(\psi(\phi_{\tau kv}^\shp) - \log\phi_{\tau kv}^\rte\right)}
    \left[
        {\color{blue}a^{\tau} + \frac{T}{2}}
        - {\color{olive} \phi_{\tau kv}^\shp}
    \right]
    - 
    {\color{blue}b^{\tau} \dfrac{\phi_{\tau kv}^\shp}{\phi_{\tau kv}^\rte}}
    %+ 
    %{\color{olive}
    %    \phi_{\tau kv }^\shp - \phi_{\tau kv }^\shp \log\left(\phi_{\tau kv }^\rte\right) + \log \Gamma \left(\phi_{\tau kv }^\shp\right)
    %}
    \right\}
    + \nonumber\\ &+
    \sum\limits_{k=1}^K \sum\limits_{v=1}^V \left[
    {\color{olive}
        % \left(\psi(\phi_{\tau kv}^\shp) - \log\phi_{\tau kv}^\rte\right) +
        \phi_{\tau kv }^\shp + \log \Gamma \left(\phi_{\tau kv }^\shp\right) - \phi_{\tau kv }^\shp \log\left(\phi_{\tau kv }^\rte\right)
    }
    \right]
    - \nonumber\\ &-
    {\color{blue}
        \frac{1}{2\left(\sigma^{\mu}\right)^2}
        \sum\limits_{k=1}^K \sum\limits_{v=1}^V
        \left[
            \left(\phi_{\mu kv }^\loc - \mu^{\mu} \right)^2 + \phi_{\mu kv }^\sclsq
        \right]
    }
    +
    {\color{olive}
        \sum\limits_{k=1}^K \sum\limits_{v=1}^V \log \phi_{\mu kv }^\scl
    }
    - \nonumber\\ &-
    {\color{blue}
        \frac{1}{2\left(\sigma^{\delta}\right)^2}
        \sum\limits_{k=1}^K \sum\limits_{v=1}^V
        \left[
            \left(\phi_{\delta kv }^\loc - \mu^{\delta} \right)^2 + \phi_{\delta kv }^\sclsq
        \right]
    }
    +
    {\color{olive}
        \sum\limits_{k=1}^K \sum\limits_{v=1}^V \log \phi_{\delta kv }^\scl
    }
    + \nonumber\\ &+
    {\color{red}
        \sum\limits_{d=1}^D \sum\limits_{k=1}^K \sum\limits_{v=1}^V
        \left[ y_{dv} \phi_{dkv}^y \phi_{hkv,t_d}^\loc
            %-  y_{dv} \phi_{dkv}^y \log \phi_{dkv}^y
            - 
            \dfrac{\phi_{\theta dk}^\shp}{\phi_{\theta dk}^\rte} 
            \exp\left\{
                \phi_{hkv,t_d}^\loc + \frac{1}{2} \phi_{hkv,t_d}^\sclsq
            \right\}
        \right]
    }
    - \nonumber\\ &-
    {\color{blue} 
        \frac{1}{2} \sum\limits_{k=1}^K \sum\limits_{v=1}^V
        \dfrac{\phi_{\tau kv }^\shp}{\phi_{\tau kv }^\rte}
        \left[
            \Eq \T{\left(\bm h_{kv} - \ones \mu_{kv}\right)} \Delta_T\left(\delta_{kv}\right) \left(\bm h_{kv} - \ones \mu_{kv}\right)
        \right]
    }
    + \nonumber\\ &+
    {\color{olive}
        \frac{1}{2} \sum\limits_{k=1}^K \sum\limits_{v=1}^V 
        \log \left| \bm \phi_{hkv}^\covm \right|
    } + c,\label{supeq:ELBO_detail}
\end{align}
where $c$ is an additive constant. 

Note that only the last three rows (where~\eqref{supeq:E_quadratic_h_mu} is applied) depend on variational parameters $\bm \phi_h^\loc$ and $\bm \phi_h^\covm$. Since only gradients with respect to these parameters are desired, other terms could be omitted from evaluation to speed up computation. However, the rest (including constant $c$) is still needed for $\ELBO{\bm \phi}$ and variational criteria.

The variational criteria are, in general, defined as 
\begin{align*}
    \mathsf{VAIC} 
    &= 
    -2 \log p(\mathbb{Y} | \bm \zeta^\star) + 2 p_D^\star
    \quad \text{with} \quad 
    p_D^\star = 2 \log p(\mathbb{Y} | \bm \zeta^\star) - 2 \Eq \left[\log p(\mathbb{Y} | \bm \zeta) \right]
    , \\
    \mathsf{VBIC} 
    &= 
    - 2 \ELBO{\bm \phi} + 2 \Eq \log p(\bm \zeta) 
    ,
\end{align*}
where $\bm \zeta^\star = \Eq \bm \zeta$ is a~vector of variational means. Determining $\log p (\mathbb{Y} | \bm \zeta^\star)$ is straightforward for the Poisson distribution~\eqref{supeq:tvpf_rates}. Without the auxiliary latent counts $\widetilde{\mathbb{Y}}$, evaluating $\Eq \left[\log p(\mathbb{Y} | \bm \zeta) \right]$ is complicated and we thus replace it with the reconstruction as suggested in Section~\ref{supsec:auxiliary_variables}. Rearranging the terms, we obtain:
\begin{align*}
    \mathsf{VAIC} 
    &= 
    2 \log p(\mathbb{Y} | \bm \zeta^\star) - 4 {\color{red} \mathsf{reconstruction}}
    , \\
    \mathsf{VBIC} 
    &= 
    - 2 {\color{red} \mathsf{reconstruction}} - 2 {\color{olive} \mathsf{entropy}} 
    .
\end{align*}

%% file: supplement06_simulation.tex
%%%!!!!!!!!!!!!!!!!!!!!!!!!!!!!!!!!!!!!!!!!!!!!!!!!!!!!!!!!!!!!!!!!!!!!!!!%%%
%%%%%%%%%%%%%%%%%%%%%%%%%%%%%%%%%%%%%%%%%%%%%%%%%%%%%%%%%%%%%%%%%%%%%%%%%%%%%
%%%%%%%%%%%%               NEW SECTION                      %%%%%%%%%%%%%%%%%
%%%%%%%%%%%%%%%%%%%%%%%%%%%%%%%%%%%%%%%%%%%%%%%%%%%%%%%%%%%%%%%%%%%%%%%%%%%%%
%%%!!!!!!!!!!!!!!!!!!!!!!!!!!!!!!!!!!!!!!!!!!!!!!!!!!!!!!!!!!!!!!!!!!!!!!!%%%
%\section{Simulation Study} \label{supsec:simulation}
\section{Simulation study} \label{supsec:simulation}

To compare different TPF settings as well as TPF to DPF, we designed the following simulation study. For each time period $t \in \{1, \ldots, T\}$, where $T=10$ or $T=20$, we generated $A=1\,000$ documents ($D = AT$) so that  the data format is suitable for fitting both models, TPF and DPF. We create a synthetic vocabulary of size $V=500$ and $K=6$ latent topics. The counts are generated via TPF~\eqref{supeq:tvpf_rates}, where $\theta_{dk}$ are specific to each document and are not tied to the document with the same index in the next time period (unlike DPF). 

For simplicity, we generate $\theta_{dk} \sim \mathsf{Unif}\{0.8, 0.9, 1.0, 1.1, 1.2\}$ to avoid extremely low or high values. By contrast, $h_{kv,t}$ are generated with wide variability. We fix $\tau_{kv} = \tau = 10$ and $\delta_{kv} = \delta$ for all topic-word combinations $kv$. The study focuses on the performance regarding the identification of the true value of $\delta$, where we consider three scenarios: $\delta = 0$ (random noise), $\delta = 0.5$ (stationary auto-regressive sequence) and $\delta = 1$ (random walk). We define the auto-regressive sequence $h_{kv,t} = \mu_{kv} + \widetilde{h}_{v,t}$  where $\mu_{kv}$ has a certain structure and the evolution in time $\widetilde{h}_{v,t} \sim \sAR{T}{0}{\delta}{\tau}$ is the same across topics. In this way, we can create an interesting link between a word and the topic it affiliates with and how the word use evolves over time. 

\begin{figure}[t]
    \centering
    \includegraphics[width=0.8\textwidth]{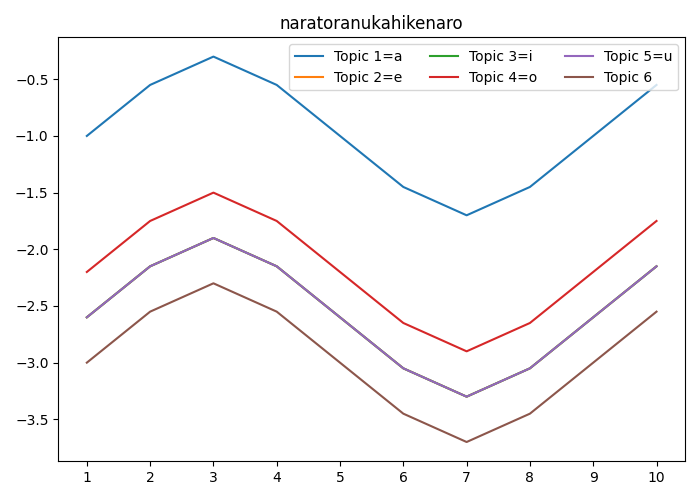}
    \caption{Illustrative example of sampled $h_{kv,t}$ for $v=\mathtt{naratoranukahikenaro}$. }
    \label{fig:naratoranukahikenaro}
\end{figure}

First, each word~$v$ is assigned $T$ randomly generated vowels $\{\mathtt{a}, \mathtt{e}, \mathtt{i}, \mathtt{o}, \mathtt{u}\}$. We sample them from a distribution where one random vowel has probability $\frac{5}{9}$ and all others have the probability $\frac{1}{9}$. This ensures that the frequency of one of the vowels in most cases dominates the other vowels. We then align each topic with a certain vowel: $k=1 \leftrightarrow \mathtt{a}, \ldots, k=5 \leftrightarrow \mathtt{u}$. The last topic, $k=6$, is a~redundant topic not aligned with any vowel. Then, we define
$$
\mu_{kv} = -3 + 4 \frac{\{\#\; \text{$k$-th vowel}\; \text{in word}\; v\}}{T}. 
$$
This implies that the mean of the sequence $h_{kv,t}$ is high for a~word that contains the $k$-th vowel frequently and that the same word is less relevant for other topics. The last topic is the least prevalent one. Next, we generate $\widetilde{h}_{v,t} \sim \sAR{T}{0}{\delta}{\tau}$ and use these values to determine $T$ consonants for this word. We create an equidistant grid of 21 ($=$ number of consonants) intervals. The middle interval around zero corresponds to consonant $\mathtt{n}$, while the intervals to the left for negative values correspond to consonants earlier in the alphabet and the intervals to the right to later consonants in the alphabet. Depending on where the values of $\widetilde{h}_{v,t}$ land in the grid we assign the corresponding consonant to this word resulting in a set of $T$ consonants. A word is then created by pasting one generated vowel behind one generated consonant, thus obtaining a~word of $2T$ letters. Its composition with respect to vowels and consonants indicates the topic it relates to and how it evolves over time. For example, the word $\mathtt{naratoranukahikenaro}$ is highly relevant for topic~1 and follows a~sinusoid curve, see Figure~\ref{fig:naratoranukahikenaro}.

Based on the rates $\lambda_{dv} = \sum\limits_{k=1}^{6} \theta_{dk} \exp\{ h_{kv,t}\}$, we generate $10$ replicates of document-term matrices $\mathbb{Y}$. Varying $\delta \in \{0, 0.5, 1\}$ and $T \in \{10, 20\}$ results in 60 different DTMs. For each DTM, we estimate seven different models: 
\begin{enumerate}[label = (\Alph*)]
    \item TPF with $\delta_{kv} \in \R$ (normal) and general $\bm \phi_{hkv}^\covm$,
    \item TPF with $\delta_{kv} \in \R$ (normal) and diagonal $\bm \phi_{hkv}^\covm$,
    \item TPF with $\delta_{kv} = 1$ (deterministic) and general $\bm \phi_{hkv}^\covm$,
    \item TPF with $\delta_{kv} = 1$ (deterministic) and diagonal $\bm \phi_{hkv}^\covm$,
    \item TPF with $\delta_{kv} \in [-1, 1]$ (truncated normal) and general $\bm \phi_{hkv}^\covm$,
    \item TPF with $\delta_{kv} \in [-1, 1]$ (truncated normal) and diagonal $\bm \phi_{hkv}^\covm$,
    \item DPF.
\end{enumerate}
Different from the empirical application, we flattened the prior distribution for $\delta$ to $\norm{0.5}{1}$ when fitting the TPF model. 

To estimate the DPF, we slightly adjusted the implementation of TPF and kept some of its specifications. For example, the prior specification of auto-regression is the same than for TPF (i.e., the mean is in the prior and not in the formula for the rates). In particular, the version of DPF we work with is defined by
\begin{align*}
    y_{av,t} &\sim \Pois{\lambda_{av,t}}
\qquad \text{with rate} \qquad 
\lambda_{av,t} 
= \sum \limits_{k=1}^K \lambda_{akv,t}
= \sum \limits_{k=1}^K \exp 
\left\{ 
    g_{ak,t} + h_{kv,t}
\right\}, \\
\bm g_{ak} &\sim \sAR{T}{\mu_{ak}^g}{\delta_{ak}^g}{\tau_{ak}^g}, \\
\bm h_{kv} &\sim \sAR{T}{\mu_{kv}^h}{\delta_{kv}^h}{\tau_{kv}^h}
\end{align*}
and priors given by~\eqref{supeq:mu_delta_tau_prior}. To resemble the original DPF formulation, we assume $\delta_{ak}^g, \delta_{kv}^h = 1$ and a diagonal structure for $\bm \phi_{gak}^\covm$ and $\bm \phi_{hkv}^\covm$.
Direct CAVI updates are also applied where possible. Hence, our implementation of DPF is not identical to the implementation proposed by~\citet{charlin2015DPF} but it closely corresponds it while being implemented within the Tensorflow environment. The code is also available at our Github repository
(\url{https://github.com/vavrajan/TPF}).
%~\citep{Github_TPF}. 

\begin{comment}
\begin{table}
    \centering
    \resizebox{0.99\textwidth}{!}{
        \input{model_comparison.tex}
    }
    \caption{Comparison of six settings in terms of VIC and computation time.}
    \label{tab:model_comparison}
\end{table}
\end{comment}

\begin{table}
    \centering
    \resizebox{0.97\textwidth}{!}{
        \input{model_comparison_tall.tex}
    }
    \caption{Comparison of seven model specifications under six simulation settings averaged over 10 replications in terms of variational information criteria ($\mathsf{VAIC}$, $\mathsf{VBIC}$; with values divided by 1000) and computation time (``sec/epoch'' denotes the average time per epoch to update the variational parameters and the approximate $\mathsf{ELBO}$ values; ``sec/$\mathsf{ELBO}$'' denotes the average time per epoch to determine exact $\mathsf{ELBO}$ values and the information criteria). The best value among seven models is highlighted by bold font. }
    \label{tab:model_comparison_tall}
\end{table}

\begin{figure}[t!]
    \centering
    \includegraphics[width=\textwidth]{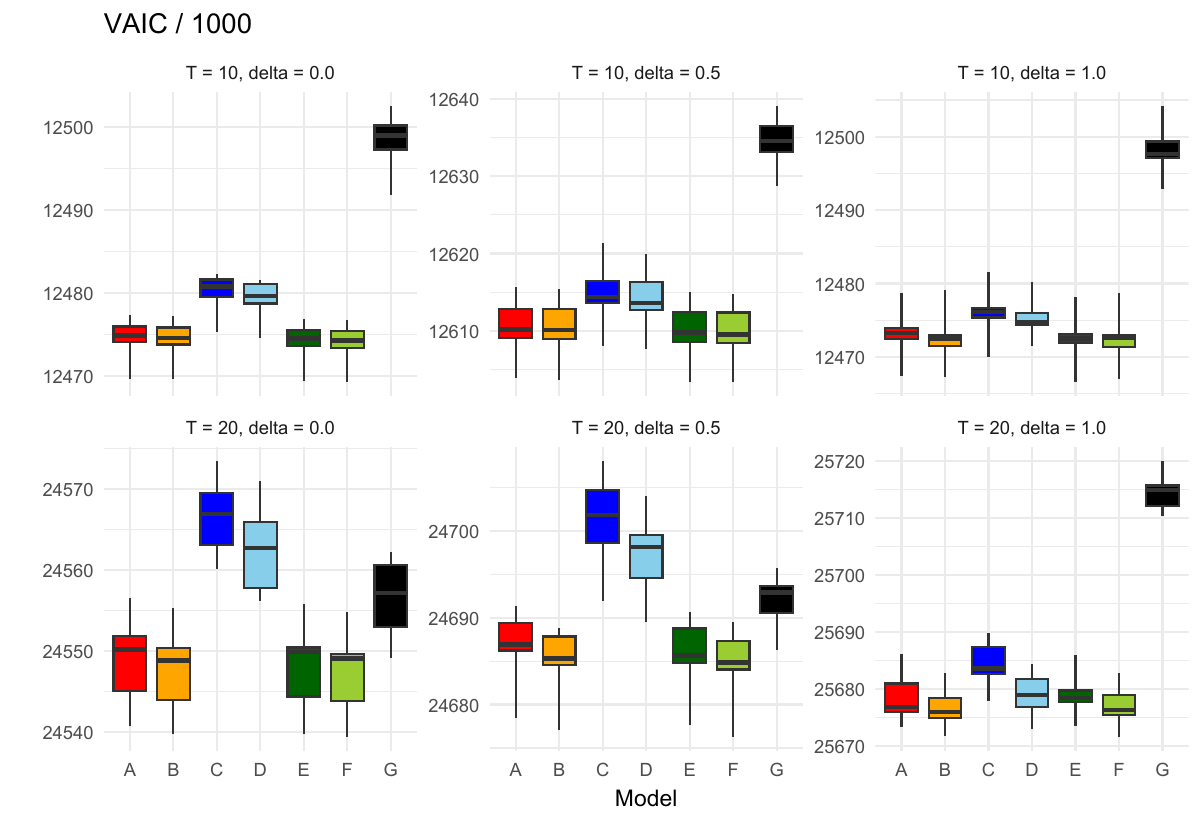}
    \caption{Boxplots of $\mathsf{VAIC}$ values over 10 replications for estimated models under six simulation settings. }
    \label{fig:boxplot_VAIC}
\end{figure}

Table~\ref{tab:model_comparison_tall} summarises the results based on variational criteria and the computational costs (averaged over 10 replications) of the seven estimated models under the six different model settings (combination of $T$ and the true value of $\delta$). According to  $\mathsf{VBIC}$, DPF performs best in all settings. However, these results might be taken rather cautiously given that $\mathsf{VBIC}$ was developed and proposed for Bayesian linear regression. We thus focus in the following on $\mathsf{VAIC}$. In this case the DPF model performs the worst among the seven considered models in almost all simulation settings, see Figure~\ref{fig:boxplot_VAIC}. The comparison of different TPF models shows that specifying a diagonal covariance matrix $\bm \phi_{hkv}^\covm$  (light colour) is  in terms of $\mathsf{VAIC}$ only negligibly better than using a general covariance matrix (dark colour). Assuming different assumptions on $\delta$ induces much higher differences in $\mathsf{VAIC}$ values. Specifying $\delta = 1$ if the true value $\delta = 0$ yields higher $\mathsf{VAIC}$ values, i.e., using the random-walk parameterisation when no temporal dependence is present considerably reduces the model fit. Even if the true value $\delta = 1$ coincides with the model specification in models C and D, these models still have slightly higher $\mathsf{VAIC}$ values but the difference to other models is smaller than if $\delta < 1$. The shorter time sequence ($T=10$) leads to very similar $\mathsf{VAIC}$ values for the other models (A, B, E, F), with a negligible preference for models restricting $\delta$ to $[-1, 1]$. For the longer sequence ($T=20$), the models where $\delta$ is restricted to $[-1, 1]$  perform slightly worse than the models where $\delta \in \R$ in case the true value $\delta = 1$. For this setting the non-stationarity leads to wilder sequences $h_{kv,t}$ than for the shorter sequence. 

Regarding the computational costs, the TPF models with diagonal $\bm \phi_{hkv}^\covm$ have the lowest run-times for one epoch as well as $\mathsf{ELBO}$ and variational criteria evaluation. In addition the models that restrict $\delta \in [-1, 1]$ are in general a bit slower than models without any restriction or fixing $\delta$ to $1$. Despite also using a diagonal structure, DPF does not exhibit the same efficiency. According to the run-times, DPF is more comparable to the TPF models with general covariance $\bm \phi_{hkv}^\covm$. We conjecture that this might be the case because of the additional time-varying component $g_{ak,t}$ (which is estimated with a stochastic gradient approach) despite the equivalent dimension ($AKT = DK$) of $\theta_{dk}$. 

%% file: model_comparison_tall.tex
\begin{tabular}{cc|cccc|rrrr}
\toprule
\multirow{2}{*}{$T$} & \multirow{2}{*}{$\delta$} & \multicolumn{4}{c|}{Model} & \multirow{2}{*}{$\mathsf{VAIC}/1000$} & \multirow{2}{*}{$\mathsf{VBIC}/1000$} & \multirow{2}{*}{sec/epoch} & \multirow{2}{*}{sec/$\mathsf{ELBO}$} \\
 & & & & $\delta$ & $\bm\phi_{hkv}^\covm$ &  &  &  & \\
\midrule
\multirow{7}{*}{10} & \multirow{7}{*}{0.0} & A & TPF & $\R$ & general & $12\,474.6$ & $12\,798.6$ & $2.74$ & $2.51$ \\
 & & B & TPF & $\R$ & diagonal & $12\,474.5$ & $12\,798.8$ & $1.99$ & \contour{black}{$1.52$} \\
 & & C & TPF & $1$ & general & $12\,480.0$ & $12\,779.0$ & $2.71$ & $2.53$ \\
 & & D & TPF & $1$ & diagonal & $12\,479.2$ & $12\,779.0$ & \contour{black}{$1.98$} & $1.53$ \\
 & & E & TPF & $[-1, 1]$ & general & $12\,474.2$ & $12\,802.4$ & $2.93$ & $3.30$ \\
 & & F & TPF & $[-1, 1]$ & diagonal & \contour{black}{$12\,474.1$} & $12\,802.2$ & $2.17$ & $2.29$ \\
 & & G & DPF & $1$ & diagonal & $12\,498.5$ & \contour{black}{$12\,427.4$} & $4.06$ & $2.81$ \\
\midrule
\multirow{7}{*}{10} & \multirow{7}{*}{0.5} & A & TPF & $\R$ & general & $12\,610.5$ & $12\,931.8$ & $2.75$ & $2.52$ \\
 & & B & TPF & $\R$ & diagonal & $12\,610.4$ & $12\,932.0$ & $1.99$ & \contour{black}{$1.53$} \\
 & & C & TPF & $1$ & general & $12\,615.0$ & $12\,921.1$ & $2.71$ & $2.80$ \\
 & & D & TPF & $1$ & diagonal & $12\,614.2$ & $12\,920.8$ & \contour{black}{$1.98$} & $1.54$ \\
 & & E & TPF & $[-1, 1]$ & general & $12\,610.0$ & $12\,935.6$ & $2.93$ & $3.30$ \\
 & & F & TPF & $[-1, 1]$ & diagonal & \contour{black}{$12\,609.9$} & $12\,935.9$ & $2.17$ & $2.29$ \\
 & & G & DPF & $1$ & diagonal & $12\,634.8$ & \contour{black}{$12\,565.0$} & $4.04$ & $2.81$ \\
\midrule
\multirow{7}{*}{10} & \multirow{7}{*}{1.0} & A & TPF & $\R$ & general & $12\,473.1$ & $12\,769.4$ & $2.75$ & $2.52$ \\
 & & B & TPF & $\R$ & diagonal & $12\,472.5$ & $12\,769.6$ & $1.99$ & \contour{black}{$1.53$} \\
 & & C & TPF & $1$ & general & $12\,476.0$ & $12\,765.5$ & $2.71$ & $2.53$ \\
 & & D & TPF & $1$ & diagonal & $12\,475.2$ & $12\,765.9$ & \contour{black}{$1.97$} & $1.54$ \\
 & & E & TPF & $[-1, 1]$ & general & $12\,472.5$ & $12\,772.6$ & $2.93$ & $3.29$ \\
 & & F & TPF & $[-1, 1]$ & diagonal & \contour{black}{$12\,472.4$} & $12\,773.0$ & $2.17$ & $2.29$ \\
 & & G & DPF & $1$ & diagonal & $12\,498.1$ & \contour{black}{$12\,432.4$} & $4.07$ & $2.81$ \\
\midrule
\multirow{7}{*}{20} & \multirow{7}{*}{0.0} & A & TPF & $\R$ & general & $24\,548.9$ & $25\,198.3$ & $11.27$ & $6.75$ \\
 & & B & TPF & $\R$ & diagonal & $24\,547.6$ & $25\,197.9$ & $6.07$ & \contour{black}{$3.79$} \\
 & & C & TPF & $1$ & general & $24\,566.8$ & $25\,129.6$ & $11.06$ & $6.79$ \\
 & & D & TPF & $1$ & diagonal & $24\,562.5$ & $25\,128.6$ & \contour{black}{$6.04$} & $3.83$ \\
 & & E & TPF & $[-1, 1]$ & general & $24\,548.2$ & $25\,200.0$ & $11.63$ & $8.32$ \\
 & & F & TPF & $[-1, 1]$ & diagonal & \contour{black}{$24\,547.4$} & $25\,200.6$ & $6.48$ & $5.34$ \\
 & & G & DPF & $1$ & diagonal & $24\,556.5$ & \contour{black}{$24\,407.8$} & $11.69$ & $7.37$ \\
\midrule
\multirow{7}{*}{20} & \multirow{7}{*}{0.5} & A & TPF & $\R$ & general & $24\,686.6$ & $25\,315.2$ & $11.27$ & $6.78$ \\
 & & B & TPF & $\R$ & diagonal & $24\,684.8$ & $25\,315.2$ & $6.10$ & \contour{black}{$3.82$} \\
 & & C & TPF & $1$ & general & $24\,701.2$ & $25\,281.6$ & $11.04$ & $6.78$ \\
 & & D & TPF & $1$ & diagonal & $24\,697.0$ & $25\,280.6$ & \contour{black}{$6.02$} & $3.84$ \\
 & & E & TPF & $[-1, 1]$ & general & $24\,685.5$ & $25\,318.1$ & $11.62$ & $8.32$ \\
 & & F & TPF & $[-1, 1]$ & diagonal & \contour{black}{$24\,684.4$} & $25\,318.7$ & $6.52$ & $5.43$ \\
 & & G & DPF & $1$ & diagonal & $24\,692.1$ & \contour{black}{$24\,545.0$} & $11.69$ & $7.37$ \\
\midrule
\multirow{7}{*}{20} & \multirow{7}{*}{1.0} & A & TPF & $\R$ & general & $25\,678.7$ & $26\,212.0$ & $11.27$ & $6.76$ \\
 & & B & TPF & $\R$ & diagonal & \contour{black}{$25\,676.8$} & $26\,215.0$ & $6.10$ & \contour{black}{$3.81$} \\
 & & C & TPF & $1$ & general & $25\,684.2$ & $26\,205.7$ & $11.05$ & $6.80$ \\
 & & D & TPF & $1$ & diagonal & $25\,679.2$ & $26\,204.4$ & \contour{black}{$6.03$} & $3.98$ \\
 & & E & TPF & $[-1, 1]$ & general & $25\,679.2$ & $26\,218.4$ & $11.61$ & $8.32$ \\
 & & F & TPF & $[-1, 1]$ & diagonal & $25\,677.3$ & $26\,221.0$ & $6.44$ & $5.35$ \\
 & & G & DPF & $1$ & diagonal & $25\,714.4$ & \contour{black}{$25\,603.0$} & $11.80$ & $7.40$ \\
\bottomrule
\end{tabular}

%% file: supplement07_senate_preprocessing.tex
%%%!!!!!!!!!!!!!!!!!!!!!!!!!!!!!!!!!!!!!!!!!!!!!!!!!!!!!!!!!!!!!!!!!!!!!!!%%%
%%%%%%%%%%%%%%%%%%%%%%%%%%%%%%%%%%%%%%%%%%%%%%%%%%%%%%%%%%%%%%%%%%%%%%%%%%%%%
%%%%%%%%%%%%               NEW SECTION                      %%%%%%%%%%%%%%%%%
%%%%%%%%%%%%%%%%%%%%%%%%%%%%%%%%%%%%%%%%%%%%%%%%%%%%%%%%%%%%%%%%%%%%%%%%%%%%%
%%%!!!!!!!!!!!!!!!!!!!!!!!!!!!!!!!!!!!!!!!!!!!!!!!!!!!!!!!!!!!!!!!!!!!!!!!%%%
%\section{Pre-processing the U.S. Senate Speeches} \label{supsec:processing_senate_data}
\section{Pre-processing the U.S. Senate speeches} \label{supsec:processing_senate_data}

The results can differ substantially depending on how the speeches are converted to the DTM $\mathbb{Y}$. %By trial and error we have converged to the following procedure. 
The main function to transform speech data to counts is the \texttt{CountVectorizer} function from the \textbf{scikit-learn} module \citep{scikit-learn}. We decided to limit its use to bigrams only. The combination of two words carries the ideological thought behind their use much better than just a~single word, while the number of higher $n$-grams grows quickly but are only used very rarely. The list of stop words was taken from \citet{Vafa_etal_2020}.

We processed the speeches of each session separately with \texttt{CountVectorizer} to create session specific vocabularies. These vocabularies were combined into the final vocabulary of size $V = 12\,791$. Each included bigram satisfied the condition on minimum and maximum frequency set to 0.001 and 0.3 on session level for at least one session. Finally, we called \texttt{CountVectorizer} on the set of all documents with this combined vocabulary and eliminated the empty rows. The final $\mathbb{Y}$ is a~sparse $732\,110 \times 12\,791$ matrix with $11\,111\,756$ stored elements ($0.119\%$).

This procedure was pursued because calling \texttt{CountVectorizer} on the complete set of speeches from all 18 Congress session resulted in a final vocabulary of size $V = 6\,468$ which contained only the most general bigrams (\emph{united states}, \emph{republican democrat}) that appear consistently across all time-periods. Bigrams that were specific for only a~brief era (\emph{acid rain}, \emph{hurricane katrina}) were not included because they did not meet the condition on appearing in at least $0.05\%$ of documents across the whole corpus.

%% file: supplement08_senate_outputs.tex
%%%!!!!!!!!!!!!!!!!!!!!!!!!!!!!!!!!!!!!!!!!!!!!!!!!!!!!!!!!!!!!!!!!!!!!!!!%%%
%%%%%%%%%%%%%%%%%%%%%%%%%%%%%%%%%%%%%%%%%%%%%%%%%%%%%%%%%%%%%%%%%%%%%%%%%%%%%
%%%%%%%%%%%%               NEW SECTION                      %%%%%%%%%%%%%%%%%
%%%%%%%%%%%%%%%%%%%%%%%%%%%%%%%%%%%%%%%%%%%%%%%%%%%%%%%%%%%%%%%%%%%%%%%%%%%%%
%%%!!!!!!!!!!!!!!!!!!!!!!!!!!!!!!!!!!!!!!!!!!!!!!!!!!!!!!!!!!!!!!!!!!!!!!!%%%
%\section{Additional Empirical Results} \label{supsec:additional_outputs}
\section{Additional empirical results} \label{supsec:additional_outputs}
\renewcommand\thesubfigure{\Alph{subfigure}}

\begin{figure}
    \centering
    \begin{subfigure}[t]{0.48\textwidth}
        \includegraphics[width=\textwidth]{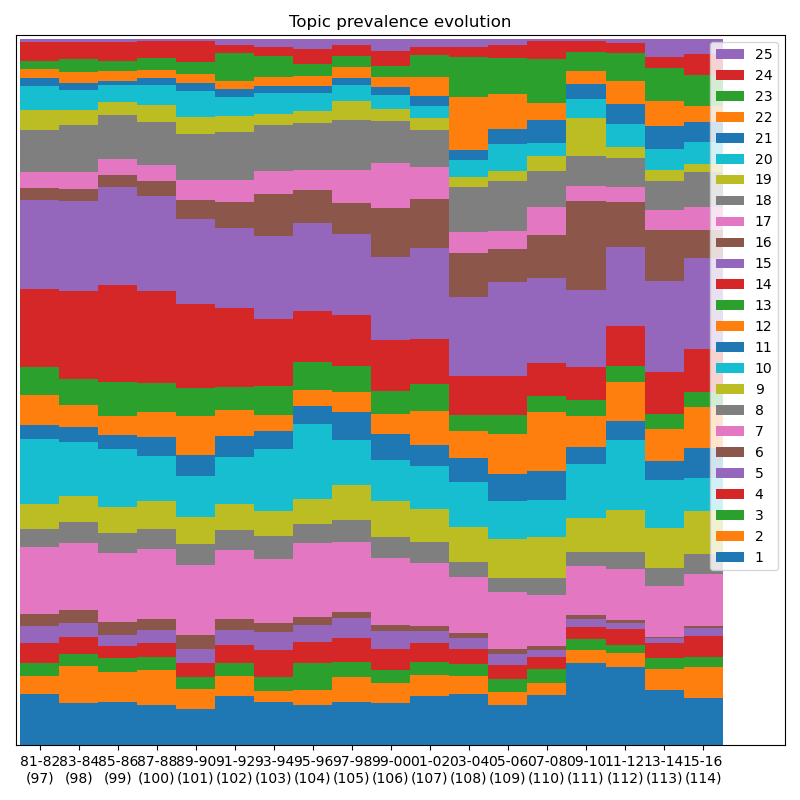}
        \caption{$\delta_{kv} \in \R$, $\bm \phi_{hkv}^\covm$ general. }
        \label{fig:AR_MVnormal_spineplot_avg_rates_topics}
    \end{subfigure}
    ~
    \begin{subfigure}[t]{0.48\textwidth}
        \includegraphics[width=\textwidth]{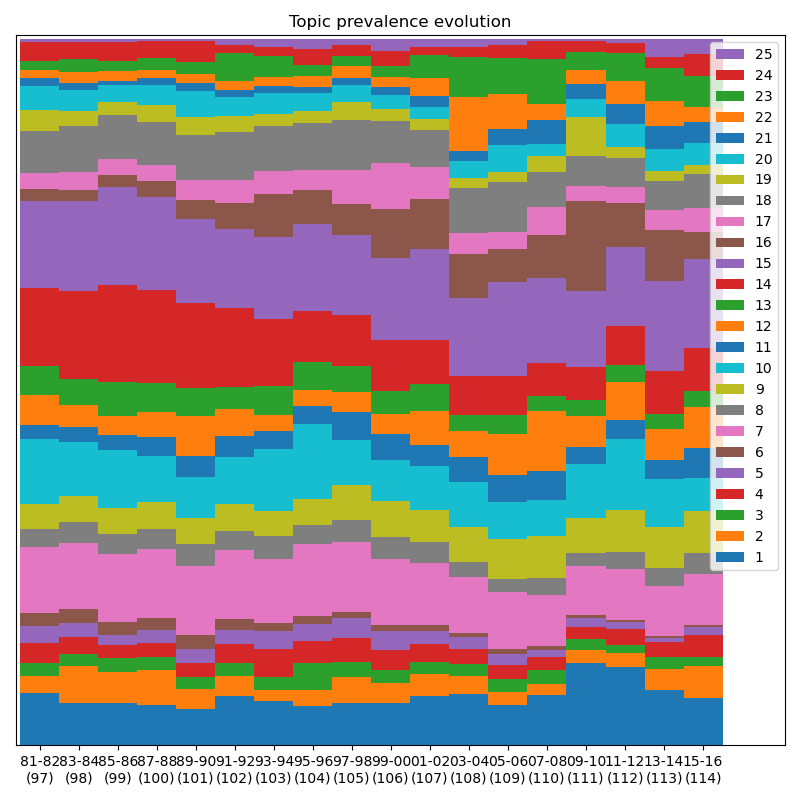}
        \caption{$\delta_{kv} \in \R$, $\bm \phi_{hkv}^\covm$ diagonal. }
        \label{fig:AR_normal_spineplot_avg_rates_topics}
    \end{subfigure}
    ~
    \begin{subfigure}[b]{0.48\textwidth}
        \includegraphics[width=\textwidth]{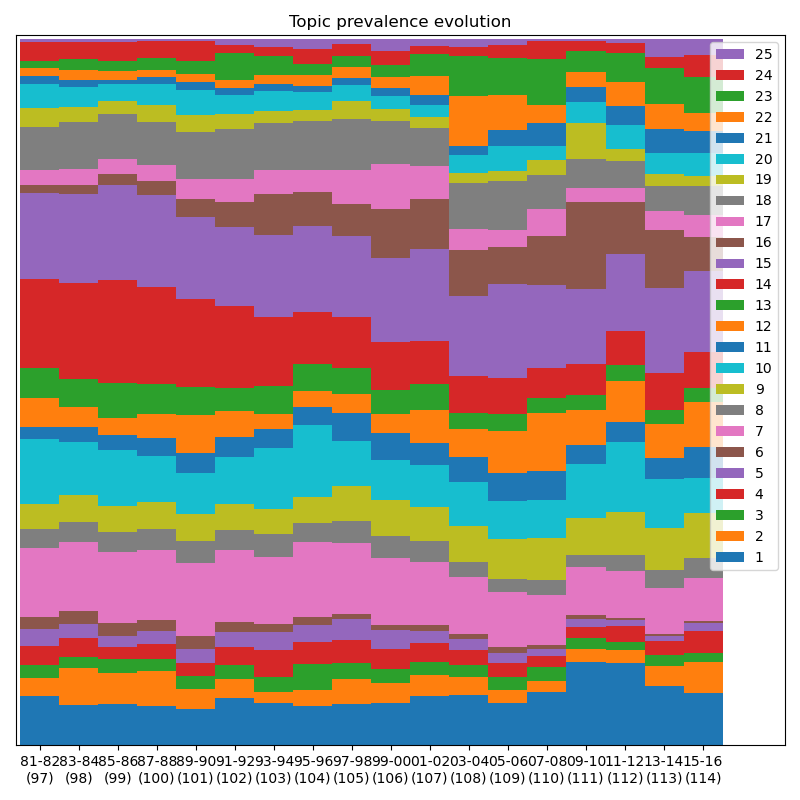}
        \caption{$\delta_{kv} = 1$, $\bm \phi_{hkv}^\covm$ general. }
        \label{fig:RW_MVnormal_spineplot_avg_rates_topics}
    \end{subfigure}
    ~
    \begin{subfigure}[b]{0.48\textwidth}
        \includegraphics[width=\textwidth]{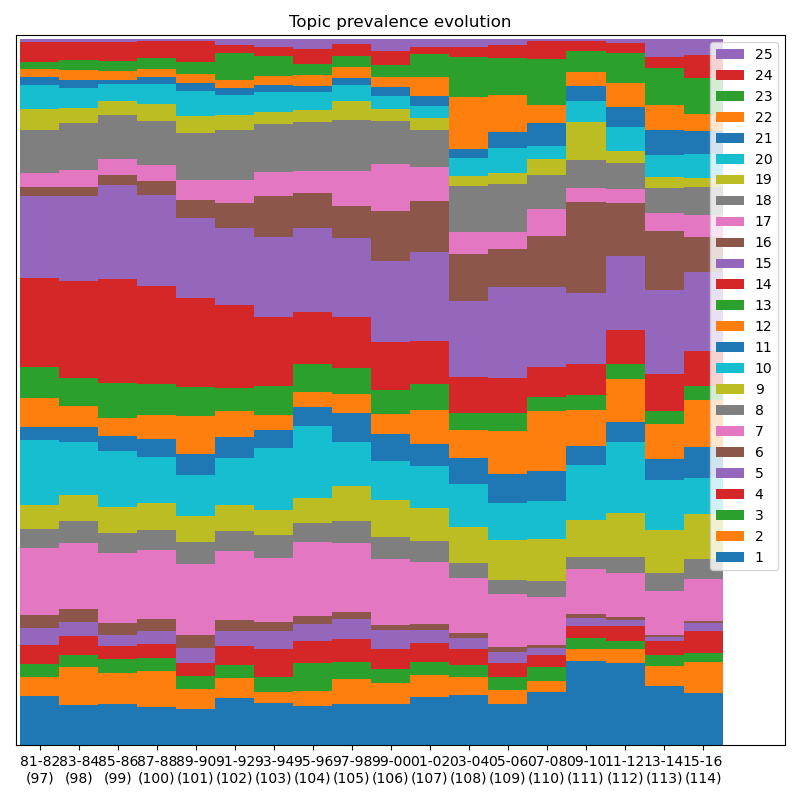}
        \caption{$\delta_{kv} = 1$, $\bm \phi_{hkv}^\covm$ diagonal. }
        \label{fig:RW_normal_spineplot_avg_rates_topics}
    \end{subfigure}
    \caption{Evolution of topic prevalence $\psi_{kt}$ over time.} 
    \label{fig:all_models_spineplot_avg_rates_topics}
\end{figure}

\begin{figure}
    \centering
    \begin{subfigure}[t]{0.48\textwidth}
        \includegraphics[width=\textwidth]{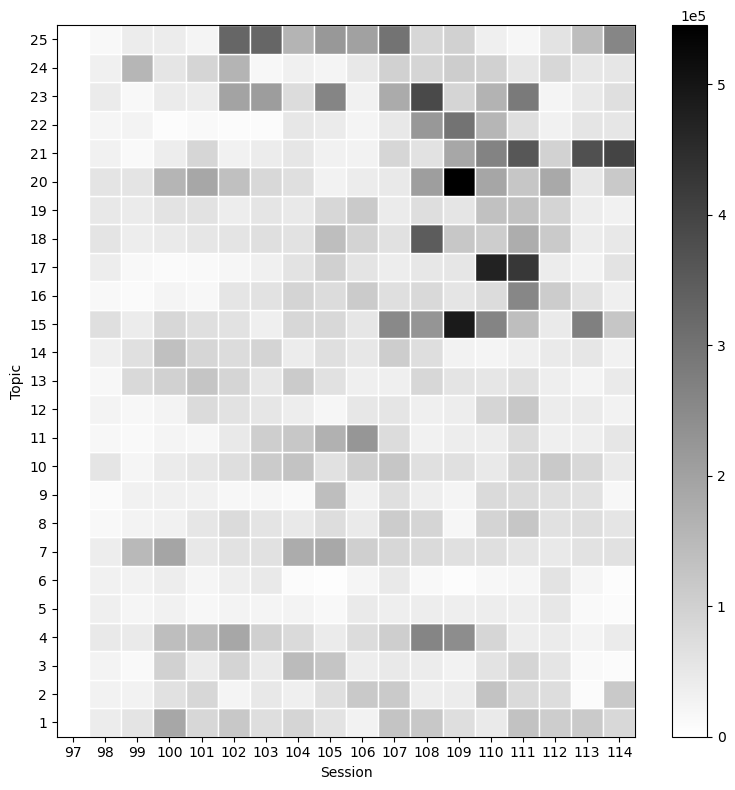}
        \caption{$\delta_{kv} \in \R$, $\bm \phi_{hkv}^\covm$ general. }
        \label{fig:AR_MVnormal_KL_dissimilarity_in_time}
    \end{subfigure}
    ~
    \begin{subfigure}[t]{0.48\textwidth}
        \includegraphics[width=\textwidth]{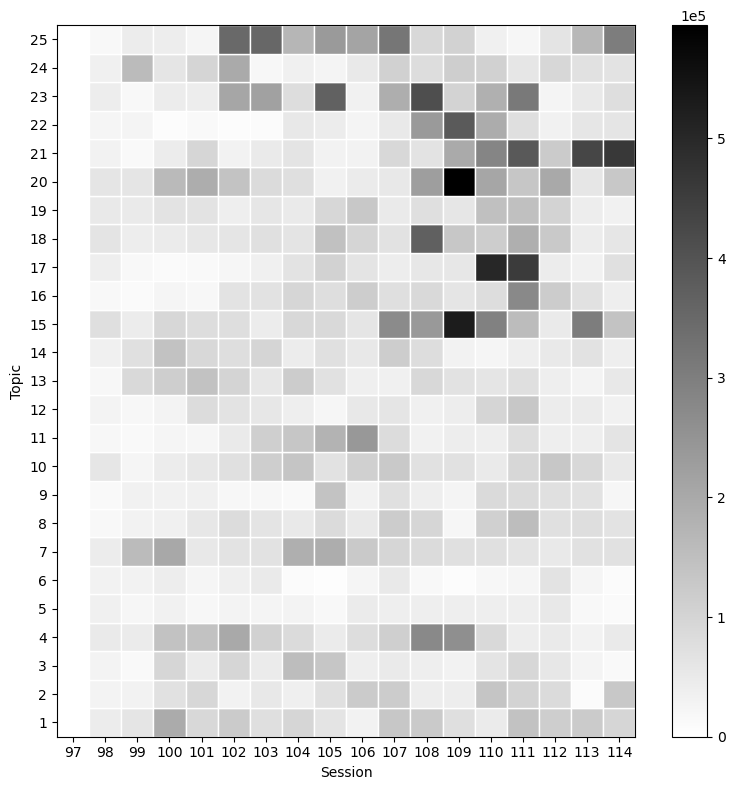}
        \caption{$\delta_{kv} \in \R$, $\bm \phi_{hkv}^\covm$ diagonal. }
        \label{fig:AR_normal_KL_dissimilarity_in_time}
    \end{subfigure}
    ~
    \begin{subfigure}[b]{0.48\textwidth}
        \includegraphics[width=\textwidth]{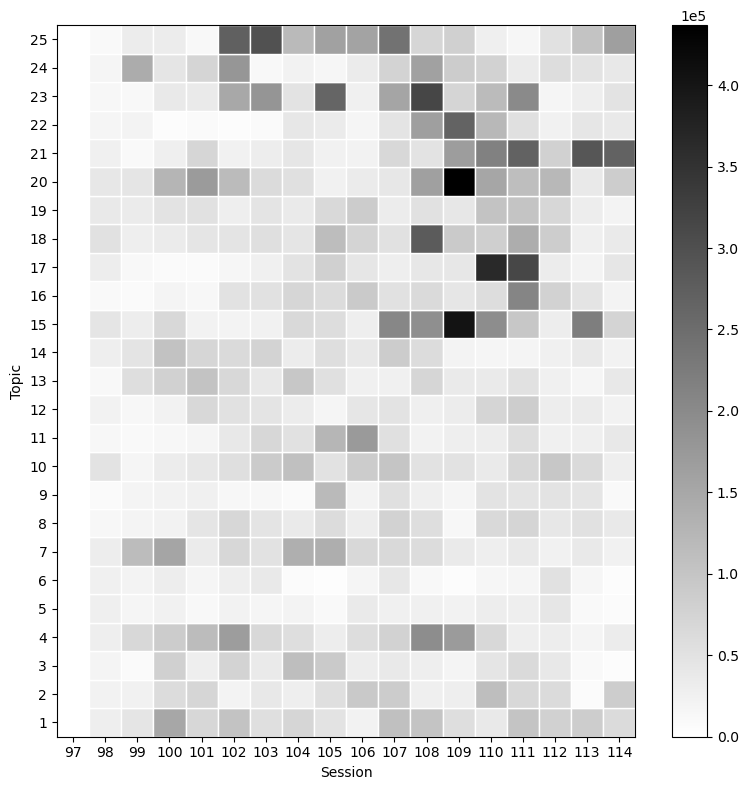}
        \caption{$\delta_{kv} = 1$, $\bm \phi_{hkv}^\covm$ general. }
        \label{fig:RW_MVnormal_KL_dissimilarity_in_time}
    \end{subfigure}
    ~
    \begin{subfigure}[b]{0.48\textwidth}
        \includegraphics[width=\textwidth]{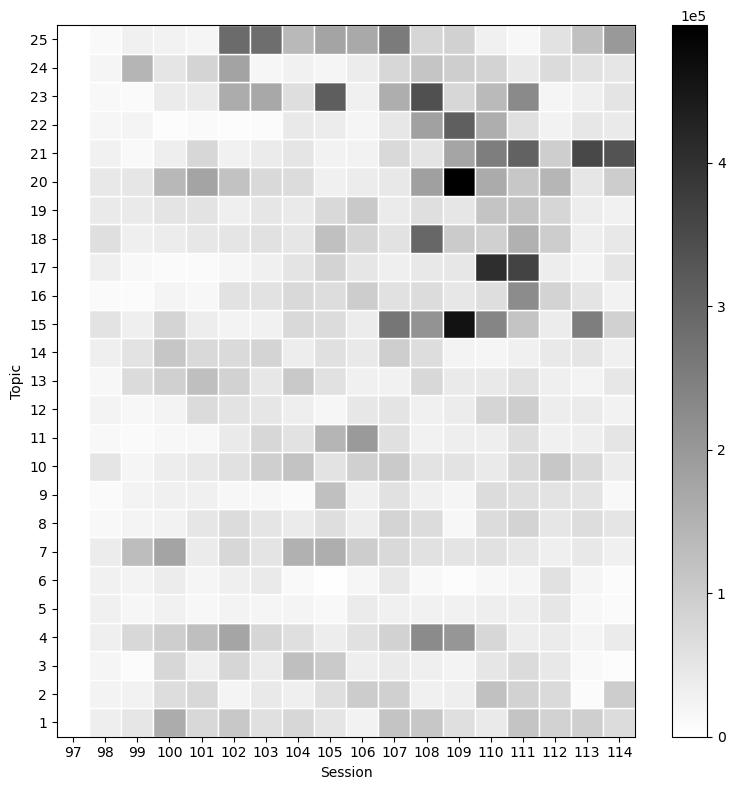}
        \caption{$\delta_{kv} = 1$, $\bm \phi_{hkv}^\covm$ diagonal. }
        \label{fig:RW_normal_KL_dissimilarity_in_time}
    \end{subfigure}
    \caption{The dissimilarity of topical content across consecutive sessions measured by $\mathsf{DTC}(k,t-1|k,t)$.} 
    \label{fig:all_models_KL_dissimilarity_in_time}
\end{figure}

\begin{figure}
    \centering
    \begin{subfigure}[t]{0.48\textwidth}
        \includegraphics[width=\textwidth]{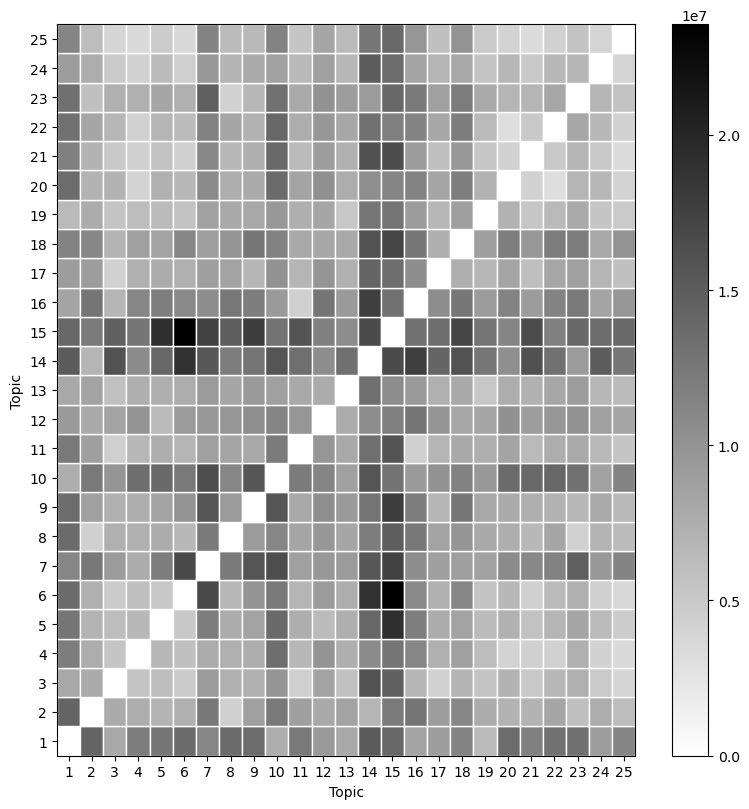}
        \caption{$\delta_{kv} \in \R$, $\bm \phi_{hkv}^\covm$ general. }
        \label{fig:AR_MVnormal_KL_dissimilarity_topics_multivariate}
    \end{subfigure}
    ~
    \begin{subfigure}[t]{0.48\textwidth}
        \includegraphics[width=\textwidth]{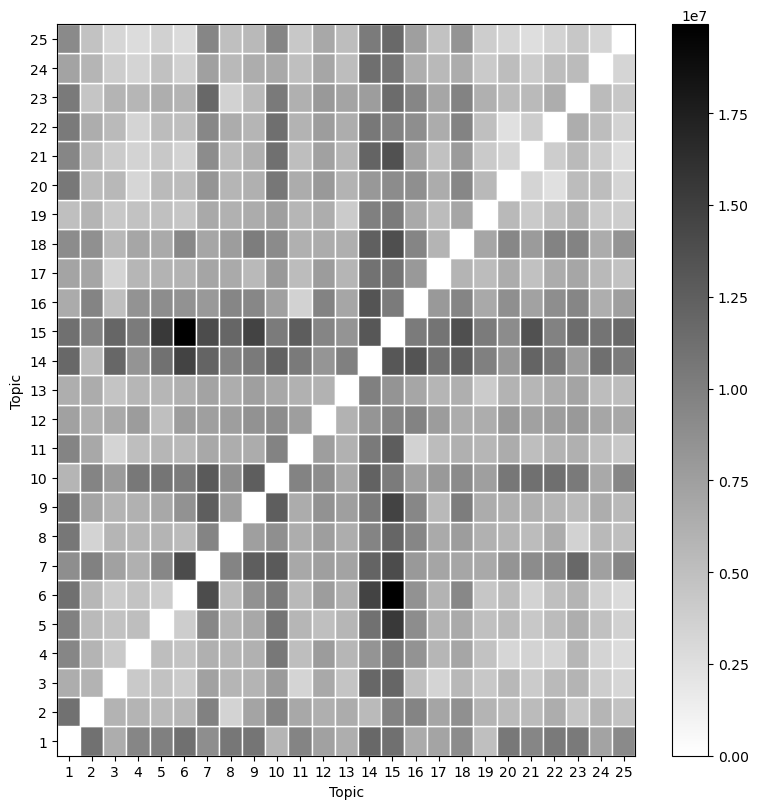}
        \caption{$\delta_{kv} \in \R$, $\bm \phi_{hkv}^\covm$ diagonal. }
        \label{fig:AR_normal_KL_dissimilarity_topics_multivariate}
    \end{subfigure}
    ~
    \begin{subfigure}[b]{0.48\textwidth}
        \includegraphics[width=\textwidth]{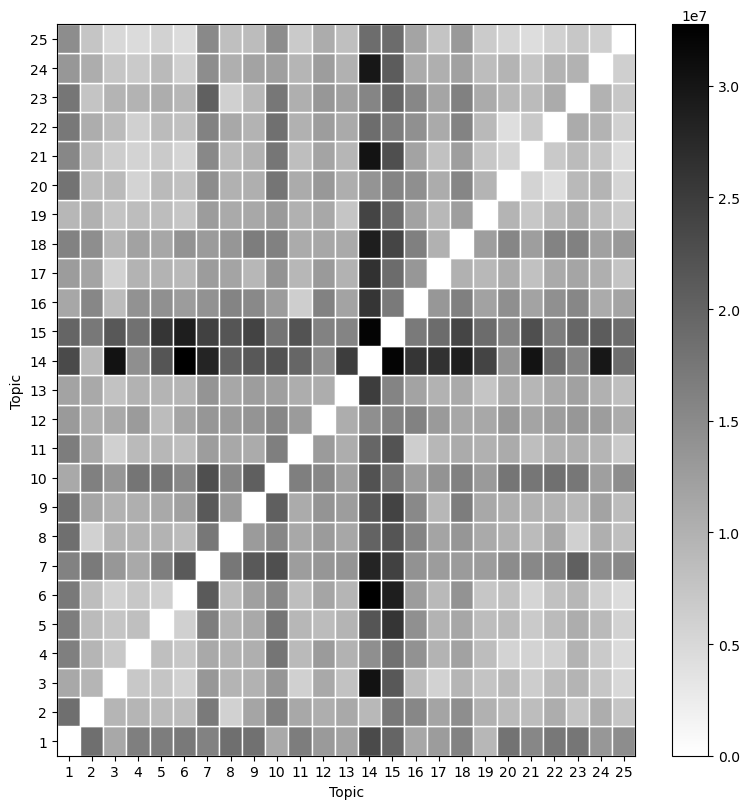}
        \caption{$\delta_{kv} = 1$, $\bm \phi_{hkv}^\covm$ general. }
        \label{fig:RW_MVnormal_KL_dissimilarity_topics_multivariate}
    \end{subfigure}
    ~
    \begin{subfigure}[b]{0.48\textwidth}
        \includegraphics[width=\textwidth]{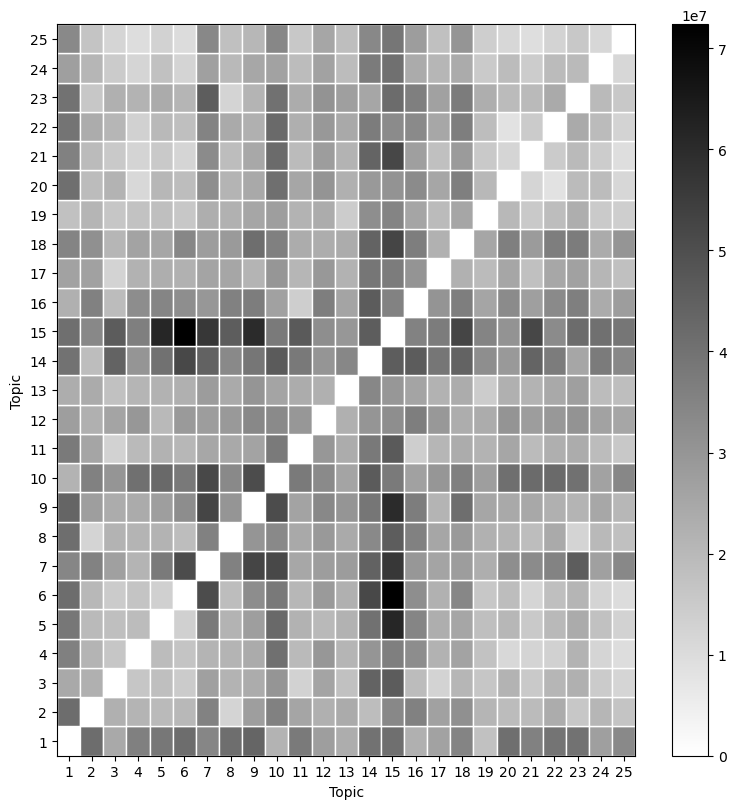}
        \caption{$\delta_{kv} = 1$, $\bm \phi_{hkv}^\covm$ diagonal. }
        \label{fig:RW_normal_KL_dissimilarity_topics_multivariate}
    \end{subfigure}
    \caption{The dissimilarity of topical content across topics measured by $\mathsf{DTC}(k_1|k_2)$.} 
    \label{fig:all_models_KL_dissimilarity_topics_multivariate}
\end{figure}

\begin{figure}
    \centering
    \begin{subfigure}[t]{0.48\textwidth}
        \includegraphics[width=\textwidth]{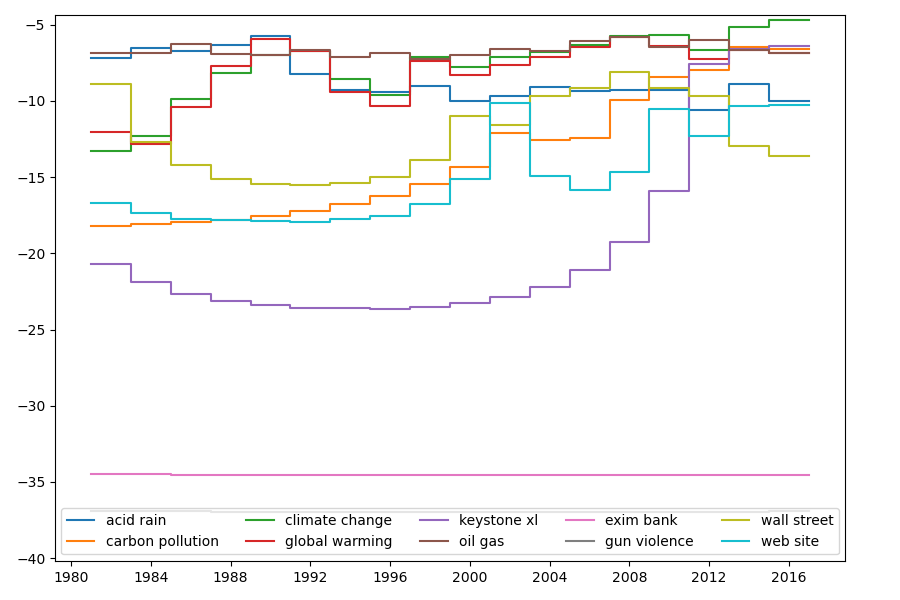}
        \caption{$\delta_{kv} \in \R$, $\bm \phi_{hkv}^\covm$ general. }
        \label{fig:AR_MVnormal_evolution_words_for_climate_change_topic_ar}
    \end{subfigure}
    ~
    \begin{subfigure}[t]{0.48\textwidth}
        \includegraphics[width=\textwidth]{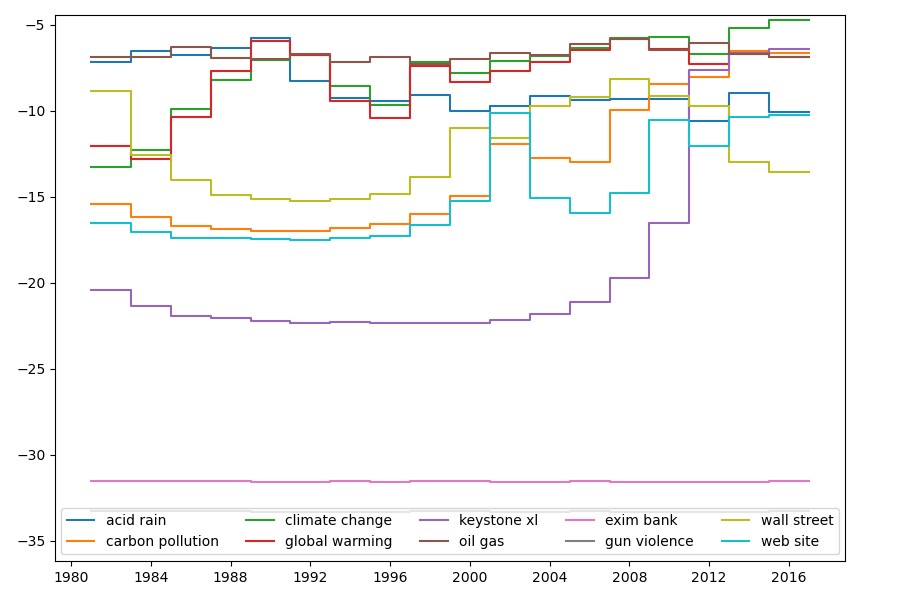}
        \caption{$\delta_{kv} \in \R$, $\bm \phi_{hkv}^\covm$ diagonal. }
        \label{fig:AR_normal_evolution_words_for_climate_change_topic_ar}
    \end{subfigure}
    ~
    \begin{subfigure}[b]{0.48\textwidth}
        \includegraphics[width=\textwidth]{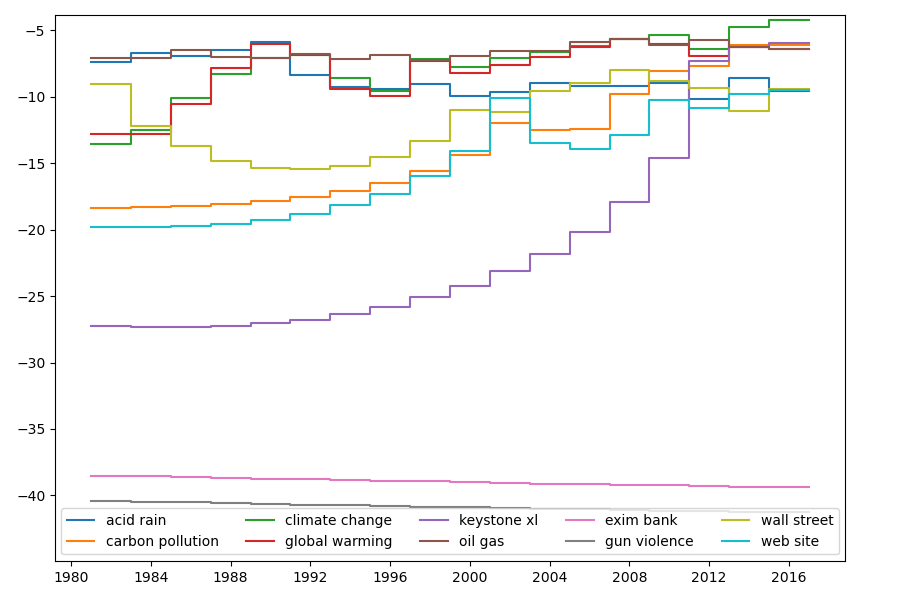}
        \caption{$\delta_{kv} = 1$, $\bm \phi_{hkv}^\covm$ general. }
        \label{fig:RW_MVnormal_evolution_words_for_climate_change_topic_ar}
    \end{subfigure}
    ~
    \begin{subfigure}[b]{0.48\textwidth}
        \includegraphics[width=\textwidth]{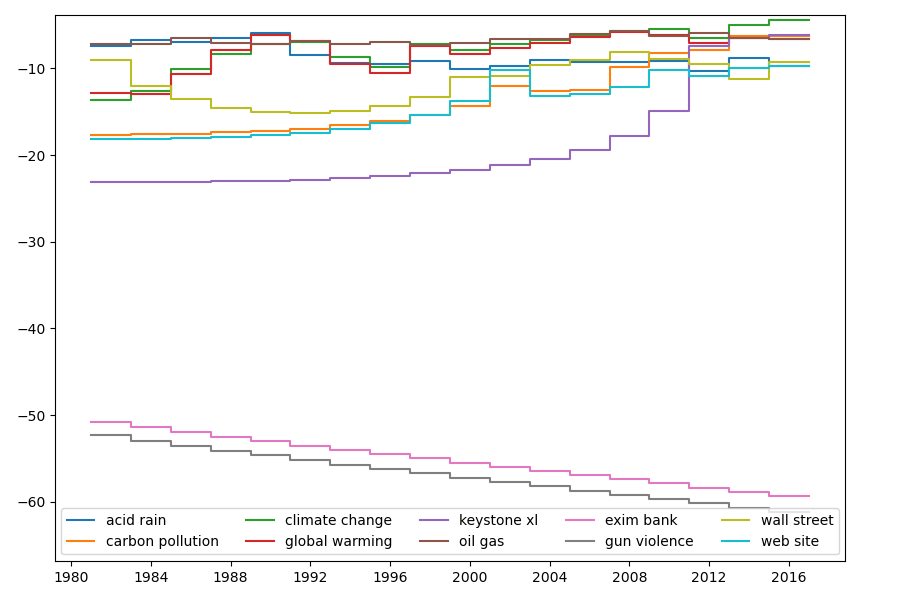}
        \caption{$\delta_{kv} = 1$, $\bm \phi_{hkv}^\covm$ diagonal. }
        \label{fig:RW_normal_evolution_words_for_climate_change_topic_ar}
    \end{subfigure}
    \caption{Evolution of $h_{kv}$ for selected 10 terms for topic 12 -- Climate change.} 
    \label{fig:all_models_evolution_words_for_climate_change_topic_ar}
\end{figure}

\begin{figure}
    \centering
    \begin{subfigure}[t]{0.48\textwidth}
        \includegraphics[width=\textwidth]{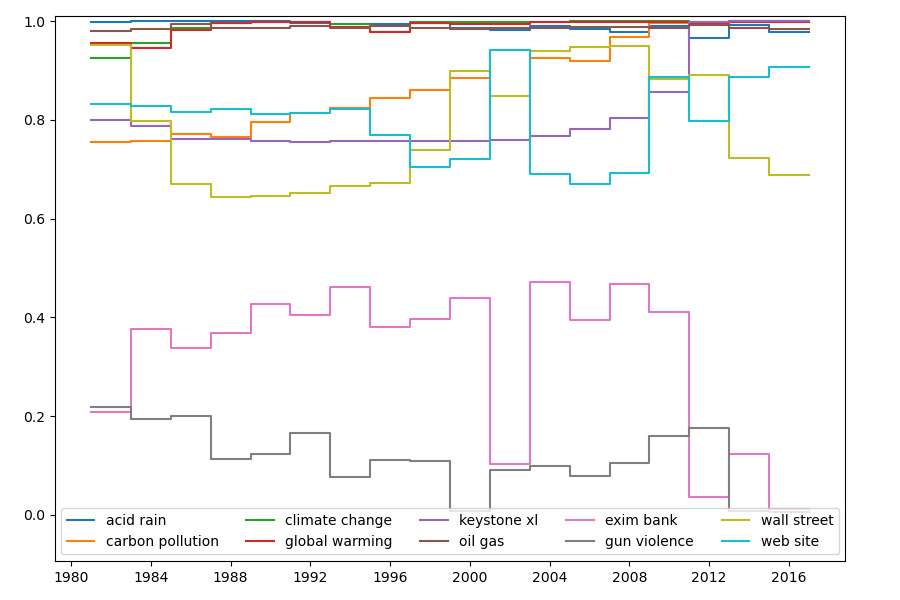}
        \caption{$\delta_{kv} \in \R$, $\bm \phi_{hkv}^\covm$ general. }
        \label{fig:AR_MVnormal_evolution_words_for_climate_change_topic_ef_w_50}
    \end{subfigure}
    ~
    \begin{subfigure}[t]{0.48\textwidth}
        \includegraphics[width=\textwidth]{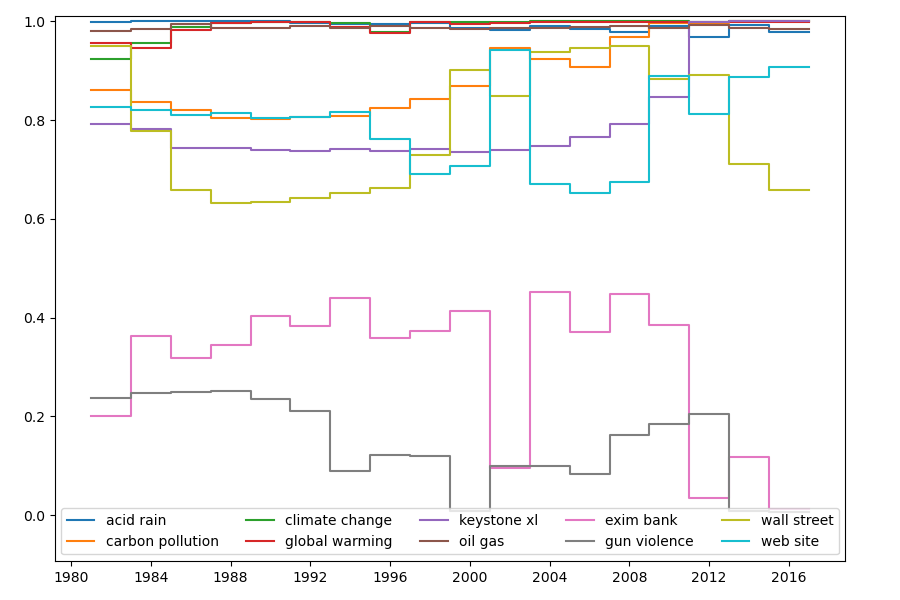}
        \caption{$\delta_{kv} \in \R$, $\bm \phi_{hkv}^\covm$ diagonal. }
        \label{fig:AR_normal_evolution_words_for_climate_change_topic_ef_w_50}
    \end{subfigure}
    ~
    \begin{subfigure}[b]{0.48\textwidth}
        \includegraphics[width=\textwidth]{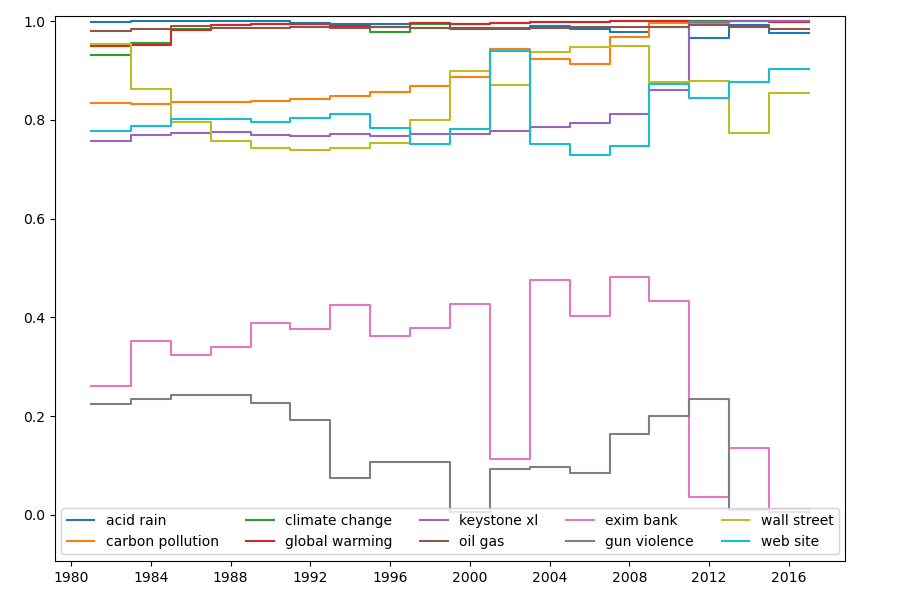}
        \caption{$\delta_{kv} = 1$, $\bm \phi_{hkv}^\covm$ general. }
        \label{fig:RW_MVnormal_evolution_words_for_climate_change_topic_ef_w_50}
    \end{subfigure}
    ~
    \begin{subfigure}[b]{0.48\textwidth}
        \includegraphics[width=\textwidth]{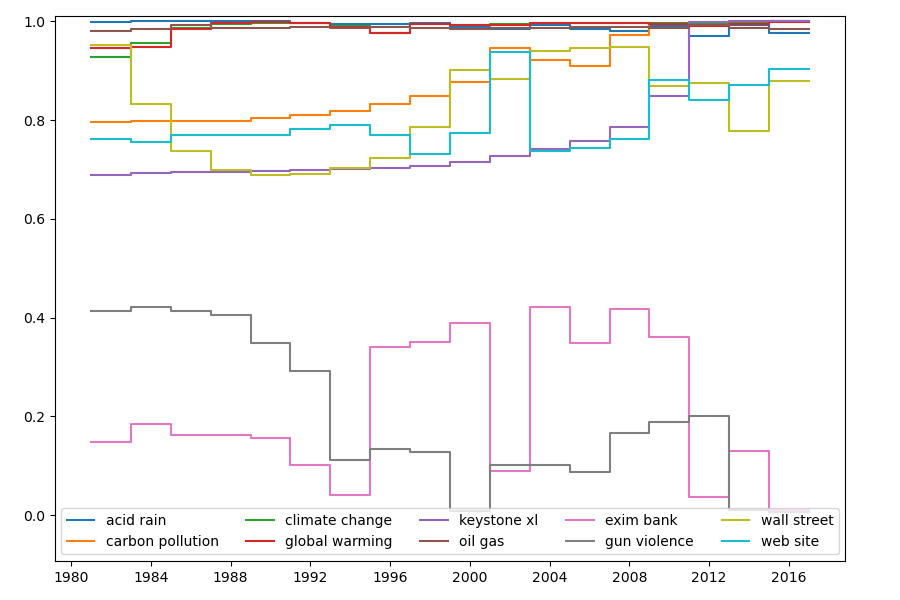}
        \caption{$\delta_{kv} = 1$, $\bm \phi_{hkv}^\covm$ diagonal. }
        \label{fig:RW_normal_evolution_words_for_climate_change_topic_ef_w_50}
    \end{subfigure}
    \caption{Evolution of $\mathsf{FREX}$ for selected 10 terms for topic 12 -- Climate change.} 
    \label{fig:all_models_evolution_words_for_climate_change_topic_ef_w_50}
\end{figure}

Figures 
\ref{fig:all_models_spineplot_avg_rates_topics}, \ref{fig:all_models_KL_dissimilarity_in_time}, \ref{fig:all_models_KL_dissimilarity_topics_multivariate}, \ref{fig:all_models_evolution_words_for_climate_change_topic_ar}, \ref{fig:all_models_evolution_words_for_climate_change_topic_ef_w_50}
(analogous to the figures in the main paper)
show that the differences between the estimated models (A -- D) are negligible. The only remarkable difference can be seen in Figure~\ref{fig:all_models_KL_dissimilarity_topics_multivariate} where the dissimilarity in topical content $\mathsf{DTC}$ depends on non-diagonal elements of $\bm \phi_{hkv}^\covm$. Models (B) and (D) with diagonal $\bm \phi_{hkv}^\covm$ seem to have a much lower amount of highly dissimilar topics than models (A) and (C) with general $\bm \phi_{hkv}^\covm$. But even in this case the overall pattern still seems to rather be the same.  
In addition,  Table~\ref{tab:vocabulary_evolution_frex_50_15} shows the evolution of the top-10 words based on the $\mathsf{FREX}$ measure for topic~15, which shifts from general political expressions to more specific discussions about immigration, based on model (D).

\begin{sidewaystable}[h!]
    \centering
    \resizebox{\textwidth}{!}{
        \input{\bestmodel_vocabulary_evolution_frex_50_15}
    }
    \caption{Evolution of top-10 terms selected based on $\mathsf{FREX}$ for topic 15 -- Rhetoric $\rightarrow$ Immigration.}
    \label{tab:vocabulary_evolution_frex_50_15}
\end{sidewaystable}

%% file: RW_normal_vocabulary_evolution_frex_50_15.tex
\begin{tabular}{|l|l|l|l|l|l|}
\toprule
\multicolumn{1}{c}{97: 1981--1982} & \multicolumn{1}{c}{98: 1983--1984} & \multicolumn{1}{c}{99: 1985--1986} & \multicolumn{1}{c}{100: 1987--1988} & \multicolumn{1}{c}{101: 1989--1990} & \multicolumn{1}{c}{102: 1991--1992} \\
\midrule
distinguished majority & distinguished majority & distinguished majority & distinguished majority & distinguished majority & distinguished republican\\
distinguished minority & distinguished minority & distinguished minority & distinguished republican & distinguished republican & distinguished majority\\
majority minority & majority minority & majority minority & distinguished friend & reserve remainder & much remaining\\
side aisle & distinguished friend & distinguished friend & majority minority & distinguished friend & reserve remainder\\
much remaining & leadership side & side aisle & next week & majority minority & next week\\
distinguished friend & side aisle & next week & much remaining & next week & distinguished friend\\
minority side & intention leadership & distinguished distinguished & side aisle & addressed chair & addressed chair\\
distinguished distinguished & next week & let indicate & reserve remainder & much remaining & republican side\\
inquire distinguished & much remaining & much remaining & friend distinguished & friend distinguished & let let\\
distinguished senior & distinguished senior & far concerned & distinguished distinguished & republican side & majority minority\\
\midrule
\multicolumn{1}{c}{103: 1993--1994} & \multicolumn{1}{c}{104: 1995--1996} & \multicolumn{1}{c}{105: 1997--1998} & \multicolumn{1}{c}{106: 1999--2000} & \multicolumn{1}{c}{107: 2001--2002} & \multicolumn{1}{c}{108: 2003--2004} \\
\midrule
distinguished republican & distinguished majority & addressed chair & distinguished majority & department homeland & distinguished majority\\
distinguished majority & much remaining & much remaining & next week & much remaining & much remaining\\
much remaining & addressed chair & appreciate much & lot things & distinguished majority & much appreciate\\
addressed chair & next week & much appreciate & appreciate much & appreciate much & appreciate much\\
next week & labor relations & issue issue & much appreciate & next week & couple days\\
republican side & distinguished colleague & members sides & due respect & much appreciate & lot things\\
distinguished friend & national labor & lot things & much remaining & try find & next days\\
labor relations & democratic side & try find & majority minority & lot things & distinguished democratic\\
lot things & distinguished democratic & kind thing & try find & finite list & move along\\
let let & lot things & democratic side & couple days & move along & issue issue\\
\midrule
\multicolumn{1}{c}{109: 2005--2006} & \multicolumn{1}{c}{110: 2007--2008} & \multicolumn{1}{c}{111: 2009--2010} & \multicolumn{1}{c}{112: 2011--2012} & \multicolumn{1}{c}{113: 2013--2014} & \multicolumn{1}{c}{114: 2015--2016} \\
\midrule
worker program & worker program & republican side & change rules & immigration reform & department homeland\\
guest worker & immigration reform & national mediation & labor relations & border security & homeland security\\
immigration reform & distinguished republican & bipartisan way & national labor & department homeland & security governmental\\
comprehensive immigration & temporary worker & appreciate much & together bipartisan & comprehensive immigration & immigration system\\
border patrol & guest worker & mediation board & friend majority & immigration system & democratic side\\
border security & together bipartisan & together bipartisan & relations board & broken immigration & border security\\
much remaining & comprehensive immigration & much appreciate & bipartisan way & security governmental & immigration reform\\
immigration system & immigration system & couple days & appreciate much & deliberative body & deliberative body\\
real id & republican side & democratic side & find common & border patrol & governmental entitled\\
patrol agents & friend majority & much remaining & immigration reform & southern border & fund department\\
\midrule
\end{tabular}